\documentclass[structabstract]{aa}  %A&A version

\usepackage{amstext,amsmath,mathtools}
\usepackage{graphicx}
\usepackage{textcomp}
\usepackage{txfonts}
\usepackage{time}
\usepackage{subfig}
\usepackage{color}
\usepackage{url}
\usepackage{natbib}
\bibpunct{(}{)}{;}{a}{}{,}
\usepackage{overpic}
\usepackage{epic}
\usepackage{fancyvrb}

\usepackage{multirow} 
\usepackage{array} 
\usepackage{booktabs}

\def\gs{}
\def\gg{}
\begin{document}
\title{SST/CRISP Observations of Convective Flows in a Sunspot Penumbra}

\author{G.B. Scharmer\inst{1,2} \and V.M.J. Henriques\inst{1,2}}
\institute{Institute for Solar Physics, Royal Swedish Academy of Sciences,
AlbaNova University Center, 106\,91 Stockholm, Sweden \and
Stockholm Observatory, Dept. of Astronomy, Stockholm University,
AlbaNova University Center, 106\,91 Stockholm, Sweden}
%\authorrunning{L{\"o}fdahl \& Scharmer}

%   \date{Received September 15, 1996; accepted March 16, 1997}
\date{Draft: \now\ \today}
\frenchspacing

\abstract{Recent discoveries of intensity correlated \emph{downflows} in the \emph{interior} of a sunspot penumbra provide direct evidence for overturning convection, adding to earlier strong indications of convection from filament dynamics observed far from solar disk center, and supporting recent simulations of sunspots.}
{Using spectropolarimetric observations obtained at a spatial resolution approaching 0\farcs1 with the Swedish 1-m Solar Telescope (SST) and its spectropolarimeter CRISP, we investigate whether the convective downflows recently discovered in the \ion{C}{i} line at 538.03~nm can also be detected in the wings of the \ion{Fe}{i} line at 630.15~nm}
{We make azimuthal fits of the measured LOS velocities in the core and wings of the 538~nm and 630~nm lines to disentangle the vertical and horizontal flows. To investigate how these depend on the continuum intensity, the azimuthal fits are made separately for each intensity bin. By using spatially high-pass filtered measurements of the LOS component of the magnetic field, the flow properties are determined separately for magnetic spines (relatively strong and vertical field) and inter-spines (weaker and more horizontal field).}
{The dark convective downflows discovered recently in the 538.03~nm line are evident also in the 630.15~nm line, and have similar strength. This convective signature is the same in spines and inter-spines. However, the strong radial (Evershed) outflows are found only in the inter-spines.}
{At the spatial resolution of the present SST/CRISP data, the small-scale intensity pattern seen in continuum images is strongly related to a convective up/down flow pattern that exists everywhere in the penumbra. Earlier failures to detect {\gs the dark convective downflows in the interior penumbra} can be explained by inadequate spatial resolution in the observed data.}


\maketitle

\section{Introduction}
\label{sec:introduction}
The origin of sunspot penumbral filamentary structure and strong horizontal flow \citep{1909MNRAS..69..454E}, their complex magnetic field topology, and the mechanism that transports energy to the visible surface represent some of the most longstanding puzzles in astrophysics. {\gs An early proposal for explaining this filamentary structure was in terms of convection rolls \citep{1961ApJ...134..289D}, based on sunspot photographs recorded during the 1959 Stratoscope flights \citep{1961ApJ...134..275D}. Initial support for this interpretation came from spectroscopic observations at low spatial resolution of a sunspot at 16\degr{} heliocentric distance \citep{1969SoPh...10..384B}, demonstrating a correlation of 29-39\% between the continuum intensity and LOS velocity measured in the line wings of the \ion{Fe}{i} line at 557.6 nm. Later observations did not give consistent support for strong such correlations \citep[e.g.,][]{1989A&A...225..528W,1990ApJ...355..329L}. \citet{1993A&A...273..633J} found a generally weak or non-existent correlation between the continuum intensity and the flow field, with only a tendency for bright structures to correspond to local blue shifts on both sides of the spot, while \citet{1993ASPC...46..192S} claimed clear evidence of such a correlation, based on 21 slit \ion{Fe}{i} 630.15~nm and 630.25~nm spectra recorded with an 0\farcs{5} spectrograph slit. More recent observations with the Swedish 1-m Solar Telescope (SST) at much higher spatial resolution of part of a sunspot at 35\degr{} heliocentric distance \citep{2007ApJ...658.1357S} demonstrated a clear correlation between intensity and vertical velocity in the sense expected for convection. We note, however, that none of the above observations demonstrated evidence that the dark penumbral filaments actually have \emph{absolute downward} flows, as required for these flows to be identified as convective. {\gg We finally note that \emph{indirect} evidence for penumbral convection was found from analysis of proper motions of small-scale intensity structures, obtained by employing local cross correlation techniques to a time series of SST G-band images \citep{2006ApJ...638..553M}. A relevant question is to what extent velocities inferred from intensity patterns reflect true flows.} 

In parallell, attempts were made to explain the Evershed flow, and to characterize its relation to bright and dark filamentary fine structure.} A highly influential explanation {\gs was in terms of} siphon flows in flux tubes \citep{1968MitAG..25..194M}. This apparently triggered the development of models and numerical simulations of flows in flux tubes for a period of over 30 years \citep{1993ApJ...407..398T, 1993A&A...275..283S, 1998A&A...337..897S, 2002AN....323..303S}. {\gs Support for such models came from observations at about 0\farcs{7} {\gg spatial resolution} showing bright upflows in the inner penumbra and dark downflows in the outer penumbra \citep{1999A&A...349L..37S,2000A&A...364..829S}. These authors also found an  overall correlation between brightness and vertical velocity that can be interpreted as an indication of convection. However, throughout the inner two thirds of the penumbra observed, both the locally bright and dark component of the penumbra on the average showed only upflows.}

An important step toward our understanding of the fine structure of penumbral magnetic fields came with the discovery of broadband circular polarization in sunspots \citep{1974A&A....35..327I,1974A&A....37...97I}, implying the simultaneous existence of strong line-of-sight (LOS) gradients in the magnetic field and flow velocity \citep{1992ApJ...398..359S}. The inferred strong gradients appeared to imply extremely strong currents and curvature forces in the visible layers of penumbrae \citep{1992ApJ...398..359S,1993A&A...277..639S,1993A&A...275..283S} {\gs \citep[note however, that strong gradients in the magnetic field do not necessarily imply strong forces: potential fields above nearly field-free gaps give rise to strong LOS gradients above the gaps, but are force-free;][]{2006A&A...447..343S}.  This (apparent) problem} was addressed by the uncombed penumbra model \citep{1993A&A...275..283S}, based on the assumed existence of slender nearly horizontal flux tubes with constant magnetic field within the flux tube (no volume currents). The flux tubes were assumed to be elevated above the penumbral photosphere and carrying the Evershed flow, and to be embedded in a more vertical background magnetic field. Various calculations and inversions \citep{1993A&A...275..283S,2000A&A...361..734M,2002A&A...381..668S,2003A&A...403L..47B,2004A&A...427..319B,2007ApJ...666L.133B,2007ApJ...671L..85T} based on this and similar simplified flux tube models gave consistency with the spectropolarimetric data, at that time rarely reaching better than 1\arcsec{} spatial resolution. {\gg 
An alternative explanation of observed penumbral Stokes spectra was given by \citet{2005ApJ...622.1292S,2006ASPC..358...13S}. He postulated the existence of micro-structured magnetic field and velocity structures (represented as two optically thin magnetic components) and used a special inversion code \citep{1997ApJ...491..993S} to test this conjecture. Fitting his observed spectra required the existence of simultaneous upflows and downflows throughout the penumbra, which can be interpreted as evidence of convection. However, these inversions were constrained by the temperature being equal for the two magnetic components, which excludes any convective energy flux. A surprising result of the inversions is the existence of opposite polarity field everywhere in the penumbra. We note that the inversions constrain the magnetic fields and flows in the two magnetic components to be strictly parallell, such that any mixture of upflows and downflows must be associated with mixed polarity field, and vice versa.} 
\begin{figure}[tbp]
 \centering
\includegraphics[width=0.49\textwidth]{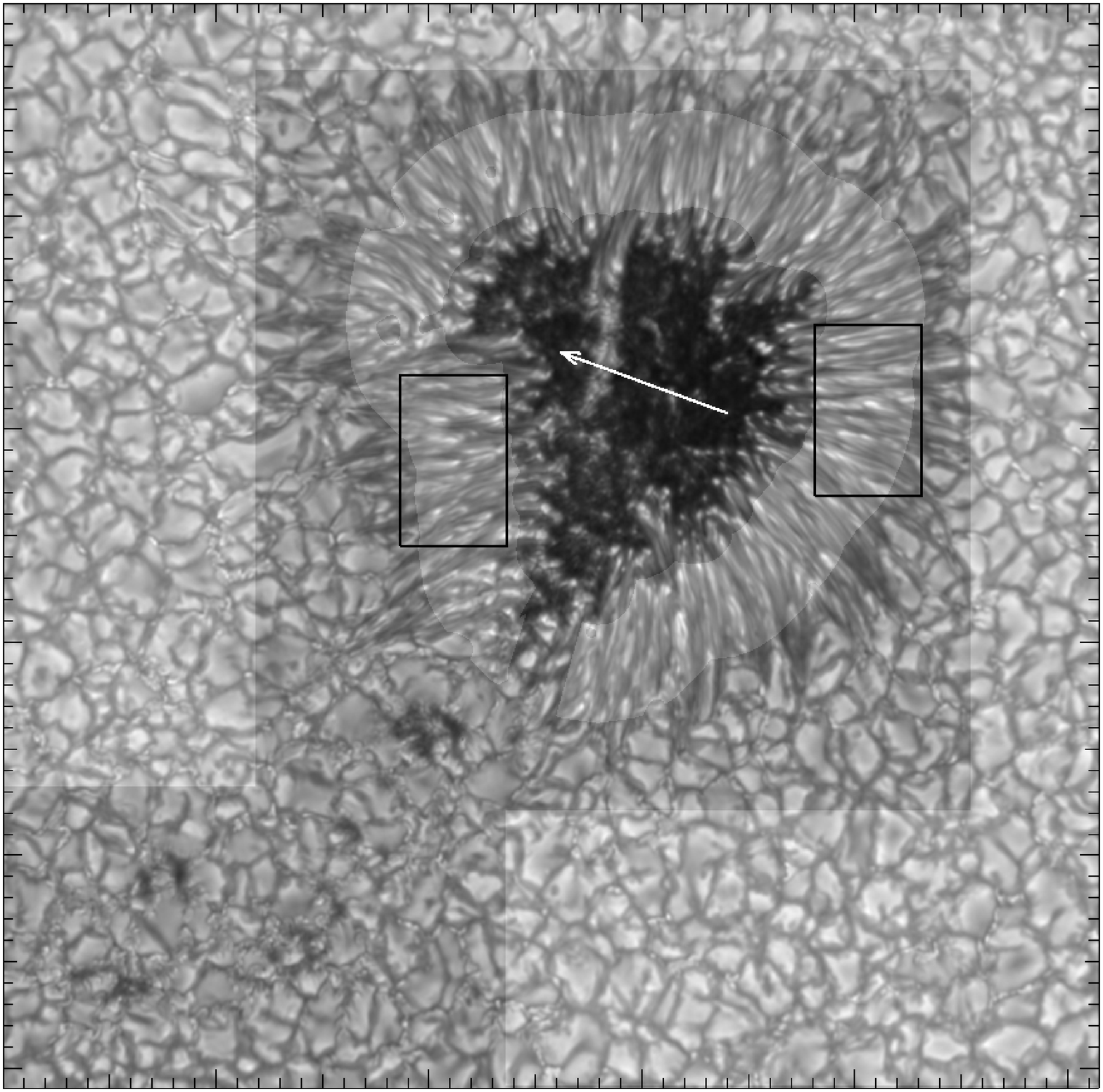}
\includegraphics[width=0.49\textwidth]{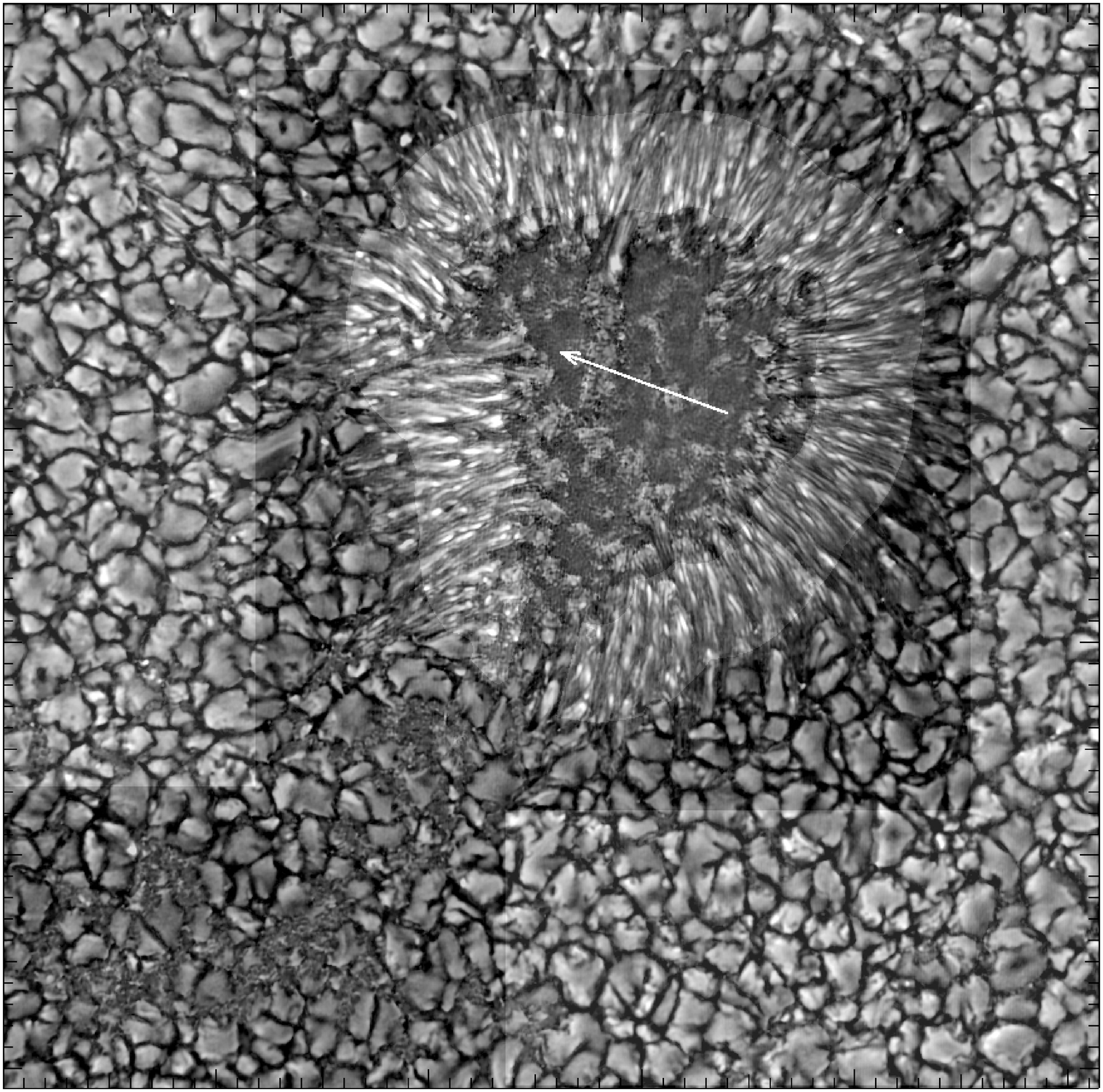}
\caption{Overview of the entire FOV observed. The highlighted contiguous area close to the boundary of the FOV corresponds to the area used as quiet Sun reference, the highlighted area in the penumbra corresponds to what is referred to as the ``interior penumbra''. The top panel corresponds to the 538~nm continuum, the right bottom to the COG velocity measured in the same line. Tick marks are at 1\arcsec{} intervals. The arrow points in the direction of Sun center. The dark rectangles outline sub-fields shown in Figs.~\ref{fig:fig_o} and \ref{fig:fig_o2}.}
\label{fig:fig_i}
\end{figure}

\begin{figure}[tbp]
 \centering
\includegraphics[width=0.49\textwidth]{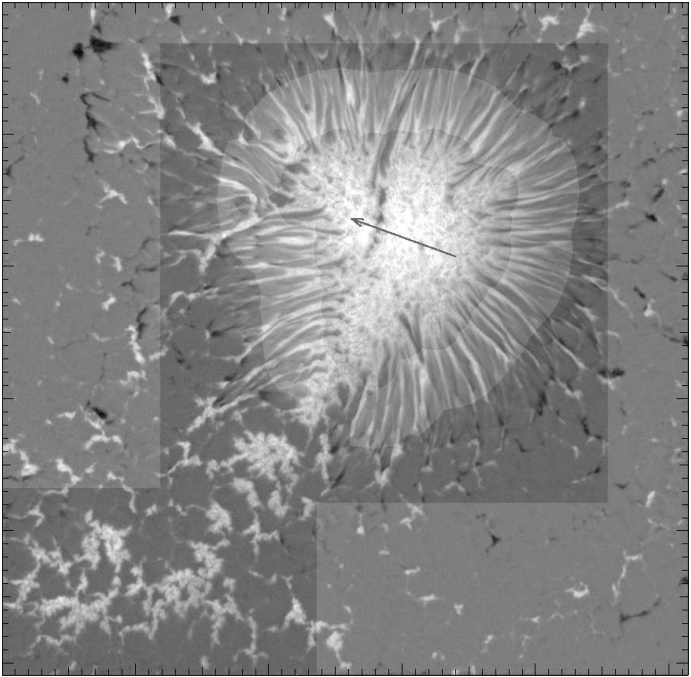}
\caption{The panel shows LOS component of the magnetic field over the same FOV as in Fig.~\ref{fig:fig_i}. Tick marks are at 1\arcsec{} intervals. The arrow points in the direction of Sun center.}
\label{fig:fig_i2}
\end{figure}

It was soon evident that a flow in a slender flux tube cannot carry the energy needed to compensate the energy losses throughout the entire radial extent of a penumbra \citep{2003A&A...411..257S}. This and other problems of flux tube models were high-lighted by \citet{2006A&A...447..343S}. They concluded that the origin of the filamentary structure and the large fluctuations in field strength and magnetic field inclination across penumbral filaments is not embedded flux tubes but must be convection in (nearly) field-free gaps \emph{below} the visible surface\footnote{{\gs We make two remarks. First, as in the quiet Sun, the penumbral photosphere is convectively stable \citep{2011ApJ...729....5R}, and the flows seen in the line forming layers represent the overshoot of this convection. Gaps are thus expected primarily in the invisible layers just below the photosphere, and to what extent they are field-free can probably only be investigated by means of simulations. Second, the gaps are not expected to be strictly field-free, but rather to have sufficiently reduced field strengths to allow the overturning convective flows to occur \citep{2010mcia.conf..243N,2010A&A...521A..72S}.}}, resolving the heat flux problem. Simple potential field and magnetohydrostatic models \citep{2006A&A...460..605S} of such ``gaps'' demonstrated that large fluctuations in magnetic field strength and inclination are \emph{unavoidable} consequences of such gaps, but without giving rise to any magnetic forces, and that the magnetic field should be much more horizontal above than outside the gaps. The model also explained the penumbral dark cores \citep{2002Natur.420..151S} as due to a strongly warped $\tau$=1 surface caused by the strong azimuthal fluctuations in gas pressure (induced by the strong fluctuations in magnetic field strength) combined with an overall drop in temperature with height\footnote{This is very similar to what is seen in faculae and bright points where locations of strong magnetic field correspond to a locally depressed $\tau=1$ surface, leading to enhanced intensity due to the higher temperatures in the deeper layers \citep{2009SSRv..144..229S}.}. While the convective gap model did not at that time explain the \emph{origin} of the Evershed flow, it was evident that this flow can only exist in the convective gaps or immediately above them, where the field is nearly horizontal \citep{2006A&A...447..343S}.

{\gs Theoretical} support for the convective origin of penumbral fine structure came with the first MHD simulation of penumbral fine structure \citep{2007ApJ...669.1390H}. While limited to a thin slice of a quite small modelled sunspot, these simulations demonstrated convection with strongly reduced (but far from zero) field strength, a systematic (but weak) radial \emph{outflow} in gaps, dark cores above short penumbral filaments, an inward migration of the penumbral structure toward the penumbra (as observed in movies), and a moat flow outside the sunspot. Based on these simulations,  the Evershed flow was explained as being \emph{identical} to the horizontal component of penumbral convection \citep{2008ApJ...677L.149S}. Simulations of much larger sunspots and with higher grid resolution \citep{2009Sci...325..171R,2009ApJ...691..640R,2011ApJ...729....5R} demonstrated a higher degree of realism, with much longer filaments and stronger Evershed flows. Detailed analysis of the simulation data has clarified some of the properties and driving mechanisms of the penumbral convection and Evershed flow \citep{2011ApJ...729....5R}. Remarkably, the length scales, mass fluxes and RMS velocities of penumbral convective flows are quite similar to those of field-free convection in the quiet Sun \citep{2009Sci...325..171R}.

{\gs Strong} evidence for penumbral convection came from observations of {\gs filamentary dynamics in sunspots located} well away from disk center. \cite{2007Sci...318.1597I} {\gs found} ``twisting motions'' in filaments from space-time plots along lines crossing filaments in the inner penumbra. Further evidence for convection (interpreted as convective rolls) in penumbral filaments in SST observations of a sunspot at 40\degr{} heliocentric distance was reported by \cite{2008A&A...488L..17Z}. More recently, evidence for penumbral convection was observed in a sunspot located at 58\degr{} heliocentric distance \citep{2010A&A...521A..72S}. These authors also pointed out that the tilted ``striations'', seen in the filaments are similar to those observed in faculae close to the limb and reproduced in MHD simulations \citep{2004ApJ...610L.137C}, and most likely outline magnetic field lines.

Whereas the convective origin of the penumbral structure and Evershed flow thus seems quite clear from a theoretical perspective, and also was receiving strong support from observations, the accumulating evidence for penumbral convection was not readily accepted. The main objection is the absence of direct observational evidence for convective \emph{downflows} inside the penumbra. \citet{2009A&A...508.1453F} and \citet{2011arXiv1107.2586F} analyzed spectropolarimetric data from Hinode, recorded from sunspots close to disk center and found no indications of convection in penumbral filaments. However, \citet{2011arXiv1107.2586F} actually measured a correlation coefficient for small-scale (high-pass filtered) intensity and velocity fluctuations of up to $-53$\% in the inner parts of one penumbra in the \ion{Fe}{i} 630.15~nm line (hereafter referred to as ``6301 line''), but considered this degree of correlation as being too small to be significant. 

While clear evidence for convection thus appears to be missing (or, is claimed to be missing) in spectropolarimetric Hinode data, the Swedish 1-m Solar Telescope (SST) provides clear such evidence. We recently \citep{2011Sci...333..316S} made observations in the \ion{C}{i} 538.03~nm line (hereafter referred to as the ``5380 line''), formed very close to the visible photosphere \citep{1999A&A...349L..37S}, where (overshooting) convective flows are expected to be significantly stronger than at the heights of formation of the 6301 line. The spatial resolution of this SST data is more than 2.3 times higher than that of the spectropolarimetric Hinode data, representing a major improvement in our ability to resolve penumbral filamentary structure and flows. By making azimuthal fits to separately determine the variation of radial (horizontal) and  vertical velocities with intensity, clear evidence of a correlation between intensity and vertical velocity was found in the sense expected for convection. {\gs In addition, the \emph{darkest}} penumbral features in the {\gs\emph{interior}} penumbra\footnote{Here and in the following we use the term ``interior penumbra'' to mean the part of the penumbra that is well away from both the inner (umbral) and outer (quiet Sun) boundaries.} show {\gs\emph{downflows}} up to about 1 km\,s$^{-1}$ while the brightest features show upflows of several km\,s$^{-1}$ \citep{2011Sci...333..316S}. Additional support for the  existence of convective downflows was recently found from analysis of other SST data in the 5380 line \citep{2011ApJ...734L..18J}. Using simulated SST observations based on numerical simulations of \citet{2009ApJ...691..640R}, \citet{2011ApJ...739...35B} recently confirmed that convective downflows should be observable with the 5380 line at the spatial resolution of the SST, while such downflows would be more difficult to observe in the \ion{Fe}{i} 709.1~nm line, explaining in part the failure of \citet{2010ApJ...725...11B} in detecting such flows in this line. The other difficulty in detecting vertical flows from their data is the contamination of the LOS velocity from the strong horizontal Evershed flows, due to the 5\fdg4 heliocentric distance \citep{2011ApJ...739...35B}.

In the present paper we extend the analysis of the observations reported by \citet{2011Sci...333..316S} to include Doppler and magnetic field measurements made in the 6301 line, recorded nearly simultaneously with the recently published 5380 data. We demonstrate that the convective signatures found in the 5380 line also are obvious in the wings of the 6301 line, although with reduced velocity amplitudes, {\gs and that the dark penumbral features in the \emph{interior} penumbra on the average are associated with (convective) downflows also when observed in the 6301 line}. We also find that the convective signatures are the same in magnetic spines and inter-spines \citep{1993ApJ...418..928L}, but the strong radial outflows are seen only in the inter-spines. In Sect. 2 of this paper, we describe the observations, data processing and methods of data analysis. Correlations between continuum intensities and radial and vertical velocities obtained from azimuthal fits of LOS velocities measured in the 5380 and 6301 lines are compared and discussed in Sect. 3. Finally, in Sect. 4 we discuss the results and draw conclusions with respect to models. 

\section{Observations and data processing}
\label{sec:data}

\begin{figure*}[tbp]
 \centering
\includegraphics[width=0.98\textwidth]{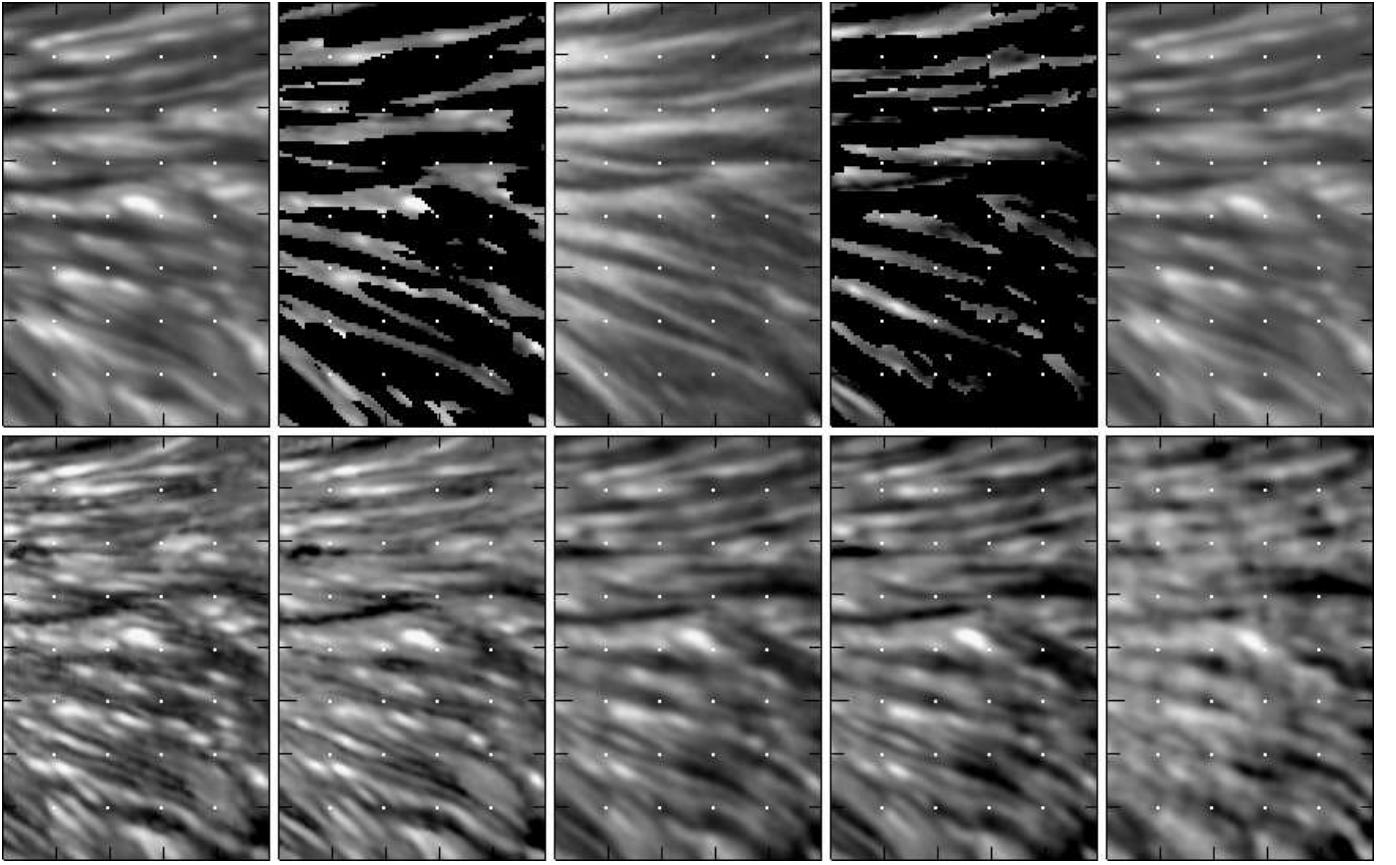}
\caption{The Figure shows a 5\arcsec{}$\times$8\arcsec{} sub-field from the right-hand side (mostly limb-side) penumbra, outlined in Fig.~\ref{fig:fig_i}. The top row shows the 538~nm continuum, the 5380 COG velocity within the spine mask, the LOS magnetic field, the 5380 COG velocity within the inter-spine mask, and the 6301 continuum intensity. The bottom row shows the 5380 COG velocity, the 5380 line core velocity, the 6301 COG velocity, 6301 70\% bisector velocity and the 6301 line core velocity. The velocity maps have been scaled individually to enhance fine structure. {\gs Maps shown represent original quantities and have not been processed with unsharp masking}. Tick marks and small white dots are at 1\arcsec{} intervals.}
\label{fig:fig_o}
\end{figure*}

\begin{figure*}[t!]
 \centering
\includegraphics[width=0.98\textwidth]{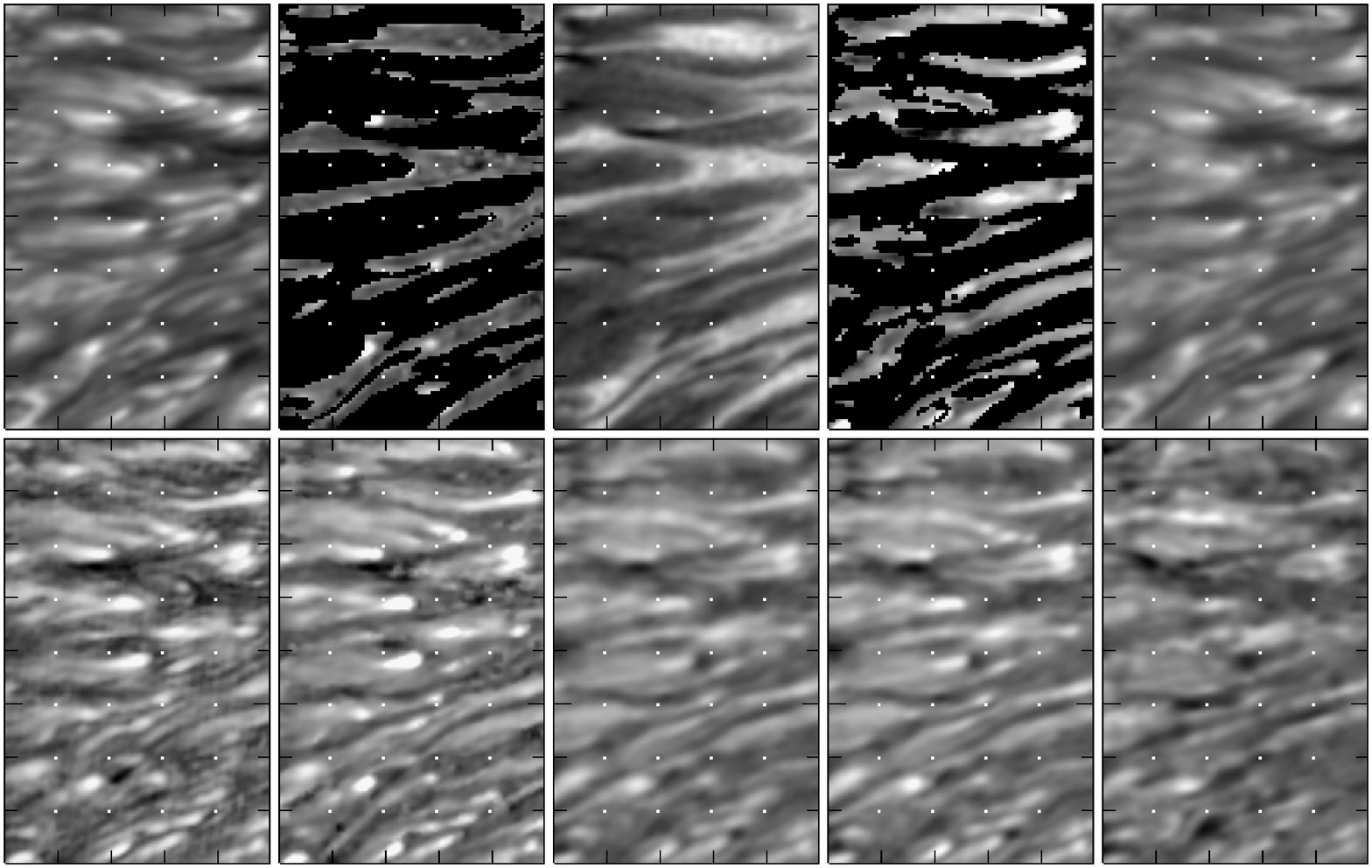}
\caption{The panel shows a 5\arcsec{}$\times$8\arcsec{} sub-field from the left-hand side (mostly disk center-side) penumbra outlined in Fig.~\ref{fig:fig_i}. The arrangement of the panels is the same as shown in Fig.~\ref{fig:fig_o}.}
\label{fig:fig_o2}
\end{figure*}

The data were recorded with the SST \citep{2003SPIE.4853..341S} using the CRisp Imaging SpectroPolarimeter \citep[CRISP;][]{2006A&A...447.1111S,2008ApJ...689L..69S}. The SST is an un-obscured 1-meter aperture telescope that presently employs low-order (37 electrode) adaptive optics \citep{2003SPIE.4853..370S} to achieve nearly diffraction limited performance in excellent seeing at wavelengths as short as 390~nm. Of importance in explaining the excellent data quality is the use of multiple short (17~msec) exposures that are post-processed with an elaborate image reconstruction technique \citep{2005SoPh..228..191V} to compensate residual low-order aberrations left by the AO system. CRISP is a dual Fabry-Perot filter system with the low-resolution etalon having reduced reflectivity in order to mitigate the effects of cavity errors (variations in separation between the reflecting surfaces of the etalons). The telecentric mounting of the etalons gives wavelength shifts of the transmission peak over the FOV, but the low reflectivity of the low-resolution etalon gives a passband that is wide enough to accommodate the relative wavelength shifts between the two etalons, such that the wavelength transmission profile shows only small variations in its shape over the FOV. The spectral resolution of CRISP is modest, about 4.4~pm at 538~nm and 6~pm at 630~nm, in order to allow fast tuning of spectral lines at high signal-to-noise and without {\gs spectral} undersampling. To support image reconstruction and co-alignment between images recorded at different wavelengths, the setup includes one CCD camera that records broad-band images through the CRISP pre-filter used to select a particular spectral line. To enable polarimetric measurements with low levels of seeing-induced cross-talk from Stokes $I$ to $Q$, $U$, and $V$ there are two cameras at the CRISP final focal plane, with the light divided between the two cameras by a polarizing beam splitter. Polarization modulation is made with two tunable liquid crystals (LC's) with their fast axes at 0\degr{} and 45\degr{} with respect to the vertical axis on the optical table. The LC's have sufficient stroke to allow full Stokes measurements over the entire wavelength range of CRISP (from about 510~nm to 860~nm).

The present data were recorded on 23 May 2010 at 14:11~UT from a reasonably regular sunspot at 15\degr{} heliocentric distance. The data consists of three nearly simultaneous (the time difference between the 5380 and 6301 line scans is 13~s) spectral scans of the 5380 line, the 6301 line, and the \ion{Fe}{i} at 630.25~nm (hereafter the ``6302 line''). Due to the telluric blend in the red wing of the 6302 line preventing measurements of the bisector wavelength shifts close to the continuum, we will not discuss the 6302 data in the present paper. The images were recorded at an image scale of 0\farcs059 per pixel. After image processing, co-alignment of the images recorded at the three different wavelengths and removing the apodized part of the FOV near the edges, the science data consisted of 881$\times$881 pixels covering 52\arcsec$\times$52\arcsec{}, shown in Figs.~\ref{fig:fig_i} and \ref{fig:fig_i2}. The 5380 data consist of a total of 960 images sampled with CRISP at 23 wavelengths in typical steps of 2.8~pm plus 480 broad-band images. The 6301 (and 6302) data consist of a total of 600 images sampled at 15 wavelengths in steps of 4.4~pm and 320 broadband images. 

Figure~\ref{fig:fig_i} shows the 538~nm continuum image (top) and the 5380 center-of gravity (COG) line-of-sight (LOS) velocity of the entire field-of-view (FOV) observed, Fig.~\ref{fig:fig_i2} shows the LOS magnetic field obtained with the COG method \citep{1979A&A....74....1R, 1993SoPh..146..207C}. The high-lighted region around most of the outer boundary of the FOV defines the area used as quiet Sun reference (although about a few per cent of this area obviously is covered with strong field), the high-lighted area in the sunspot outlines what we refer to as the ``interior penumbra'', which is the main target of our analysis. Figures~\ref{fig:fig_o} and \ref{fig:fig_o2} show all measured quantities within sub-fields (marked with dark rectangles in Fig.~\ref{fig:fig_i}) close to the limb-side and disk center-side parts of the penumbra.

{\gs We note that Fig.~\ref{fig:fig_i2} shows very little evidence of the opposite polarity fields observed \emph{well inside} the penumbra in a sunspot located at small heliocentric distances \citep{2007PASJ...59S.593I,2011arXiv1107.2586F}. However, Fig.~\ref{fig:fig_o2} (upper row, mid panel) shows several small-scale patches where the LOS magnetic field is very weak at the center-side penumbra. The absence of opposite polarity patches in our LOS magnetic map may be related to the associated strong downward flows and the corresponding abnormal Stokes $V$ profiles that are not well captured with our simple (COG) method for estimating the LOS magnetic field. However, inspection of the Stokes $V$ images at  +26.3~pm and +30.7~pm confirms that opposite polarity patches are found primarily in the outer penumbra and that only a few such patches are seen well inside the penumbra.}

\subsection{MOMFBD image reconstruction and alignment}
The images were corrected for gain and offset, using flats at a fixed continuum wavelength. The gain and dark corrected images from {\gs the 5380 and 6301/6302 lines} and their corresponding broadband images were processed separately {\gs as two data sets} with the Multi-Object Multi-frame Blind Deconvolution (MOMFBD) method \citep{2005SoPh..228..191V}. %After image reconstruction, the restored images were multiplied with the (dark corrected) flats previously used and divided by the flat obtained at a fixed continuum wavelength, in order to remove the ``imprint'' of the quiet Sun average line profiles in the restored images. This is equivalent to initial correction of all images with a fixed continuum flat, except that small (field dependent) RMS errors in the alignment as well as time- and space variable blurring compensated for in the MOMFBD processing will spatially smear the flat-field information over distances corresponding to 1--2 pixels. 
In the MOMFBD image reconstruction process we conservatively compensated for only 36 Karhunen-Loeve aberrations. This leaves the small-scale wavefront errors uncompensated for at scales roughly 15 cm or smaller. With sufficiently good seeing, as in our case, the effect of such truncation  is to leave residual straylight in the wings of the point-spread-function (PSF), whereas the core is well corrected, but with reduced peak strength (reduced Strehl) \citep{2010A&A...521A..68S,2011A&A...XX..YYL}. {\gs We expect the restored images to have a spatial resolution close to the diffraction limit of 0\farcs{14} at 538~nm and 0\farcs{16} at 630~nm.} 

After image reconstruction, the recorded Stokes images were demodulated for telescope polarization using the SST polarization model developed by \citet{selbing05sst}
%\citet{2010arXiv1010.4142S}. 
Finally, the 6301 and and 5380 data were co-aligned by destretching the 6301 continuum image onto the 5380 continuum image over 16$\times$16 sub-fields. The so-obtained matrix of geometrical distortions was then applied to all measured 6301 quantities. This process compensates to some extent also for any small effects of large scale flows and evolution of the structure during the 13~sec time elapsed between the recording of the 5380 and 6301 data.

\subsection{Cavity error compensation and wavelength calibration}
\label{sec:cavity}
%The location of the two Fabry-Perot etalons close to the conjugate of the science focal plane necessitates the compensation for field dependent wavelength shifts of the CRISP transmission profile, these are referred to as cavity errors. 
In principle, it would be sufficient to measure the wavelength shifts from the cavity errors and to use these to correct the measured LOS velocities. {\gs However, straylight compensation of the science data also redistributes spatially the wavelength shifts from any cavity errors. Therefore, we first shifted the line profiles to a common wavelength scale by compensating for the cavity errors, and then applied the straylight compensation to the data.}  

The cavity error wavelength shifts were measured from the averaged flat-field images by fitting second-order polynomials to the core of {\gs the Stokes I profile for} each line. The wavelength shifts were determined from the coefficients of the fits. The broad 5380 line is well sampled (at 2.8~pm wavelength steps) and we used linear interpolation to shift the line profiles to a common wavelength scale and at the same time applying a first-order correction to compensate for the transmission profile of the CRISP pre-filters. The 6301 line has coarser wavelength sampling (4.4~pm) and we used cubic splines to resample the line profiles to 1.1~pm steps before shifting the line profiles with linear interpolation according to the measured cavity errors. After compensation for cavity errors, we calculated average 5380 and 6301 line profiles from the high-lighted region surrounding most of the sunspot, shown in Figs.~\ref{fig:fig_i} and \ref{fig:fig_i2}. {\gg We calibrated the wavelength scales such the line core velocities measured from the spatially averaged 5380 and 6301 line profiles correspond to $-917$~m\,s$^{-1}$ \citep{2011Sci...333..316S} resp. $-200$~m\,s$^{-1}$ (negative velocities correspond to blue-shifted profiles), in agreement with the convective blue-shifts obtained from simulations \citep{2011A&A...528A.113D}}. The 6301 convective blue-shift adopted is somewhat larger than the shift obtained from convection simulations ($-188$~m\,s$^{-1}$) at disk center for the line core, when the averaged quiet Sun profile is convolved with the CRISP transmission profile \citep{2011A&A...528A.113D}. As a check of this calibration, we calculated the average quiet Sun LOS velocity from the \emph{spatially resolved} spectra and obtained $-130$~m\,s$^{-1}$ and $-220$~m\,s$^{-1}$ for the COG and line core 5380 velocities, and $150$~m\,s$^{-1}$, $120$~m\,s$^{-1}$ and $-100$~m\,s$^{-1}$ for the 6301 COG, 70\% bisector and line core velocities. Averaging over the entire penumbra, the corresponding values are $-200$~m\,s$^{-1}$ and $-290$~m\,s$^{-1}$ for the 5380 line, and $50$~m\,s$^{-1}$, $30$~m\,s$^{-1}$ and $-130$~m\,s$^{-1}$ for the 6301 line. The remaining convective blue-shifts in the 5380 line are similar to those expected from convection simulations (see SOM, Table S2). The values for the 6301 line are reasonably consistent and indicate that the systematic errors are not larger than $\pm150$~m\,s$^{-1}$ for this line. The weak blend in the red wing of the 6301 line most likely has a small influence on the velocities measured from the wings (the COG and 70\% bisector measurements) of the 6301 line, but we made no attempt to compensate for that.

{\gs The sunspot is supposedly surrounded by a radially outward moat flow \citep[e.g.,][] {2008ApJ...679..900V} of up to about 0.3--0.5~km\,s$^{-1}$ that will have an influence on our wavelength calibration. No obvious evidence of such a flow can be seen in our Doppler maps, even with heavy spatial smearing of the individual Doppler maps. Since the region used for wavelength calibration surrounds most of the sunspot and because of the relatively small heliocentric distance, we expect the systematic errors in the wavelength calibration from the moat flow to be small.}

\subsection{Straylight calibration and compensation}
\label{sec:straylight}
Recently, the RMS contrast of solar granulation was measured from Hinode data, combined with a detailed characterization of the straylight properties of the telescope and instrumentation used. \citet{2008A&A...487..399W} determined the straylight PSF from images recorded during both a Mercury transit and a partial solar eclipse. After deconvolution of the granulation images for the measured stray-light PSF, the corrected granulation contrasts are close to those obtained from two independently developed MHD simulation codes \citep[the CO$^5$BOLD and Stein-Nordlund codes;][]{2009A&A...503..225W}. \citet{2008A&A...484L..17D} degraded synthetic continuum images at 630~nm calculated from simulations with the MURAM code with the PSF of Hinode. Taking into account only the PSF of a perfect unobscured 50~cm telescope, the RMS contrast is reduced from 14.4\% to 10.9\%. Including the effects of the large central obscuration and spider plus the CCD, reduces the RMS contrast to 8.5\%. This is still higher than the observed value of 7\%. Danilovic et al. interpret this difference as the result primarily of small focus errors and/or other residual aberrations, but the work of \citet{2009A&A...503..225W} as well as the Gaussian fits of \citet{2009A&A...501L..19M} clearly indicate significant contributions also from the far wings of the PSF. We conclude that there are no longer scientific grounds for questioning the RMS granulation contrast values obtained from numerical 3D simulations, and that the measured granulation contrast in highly resolved granulation images recorded with the SST and other solar telescopes therefore can be used to constrain the amount of straylight. We also note that the large central obscuration, the wide spider and aberrations and/or other imperfections combine to reduce the observed RMS contrast of granulation to about 64\% of that theoretically possible at wavelengths around 630~nm for a telescope with the diameter (but without the spider and central obscuration) of Hinode. 

For the SST, the observed RMS contrast is around 6.5--7\% at 630~nm in the best ``raw'' images, but since the spatial resolution is higher than for Hinode, we could in principle reach 14\% contrast with a perfect SST operating outside the Earth's atmosphere. MOMFBD image reconstruction increases the RMS contrast somewhat to about 7.5\% at 630~nm and 9\% at 538~nm \citep{2010A&A...521A..68S}. Clearly, the SST data must contain on the order of 50\% straylight, operating at scales corresponding to granulation or larger. At the same time, ``conventional'' straylight, often characterized by a PSF having very wide (e.g., Lorentzian) wings, must be low in this and other SST data: the minimum umbral intensity in the sunspot presently discussed is 15\% at 538~nm and 18\% at 630~nm. Measurements of straylight made during the 2004 Venus transit suggests on the order of 4\% at 390~nm and 9\% at 700~nm \citep{2008PhST..133a4016K}. Measurements of straylight from the primary focal plane just outside the SST focal plane and including all relay optics as well as CRISP and the CCD itself suggests levels of ``conventional'' straylight of at most a few percent \citep{2011A&A...XX..YYL}. The origin of the dominant source of straylight is largely unknown but is now suspected to be mostly from aberrations, including the high-order aberrations truncated in MOMFBD processing, high-altitude seeing and/or effects from the relatively long integration times (17 ms) used \citep{2010A&A...521A..68S,2011A&A...XX..YYL}. {\gs This conclusion receives at least partial support from a recent comparison between speckle and MOMFBD image restorations of spectropolarimetric data recorded at the German VTT. This comparison demonstrated that speckle restorations and MFBD processing (identical to MOMFBD processing, but with a single wavelength channel) gives significantly higher granulation contrast than with MOMFBD \citep{2011A&A...533A..21P}.}

\begin{figure*}%[t!]
 \centering
\includegraphics[width=0.33\textwidth,clip]{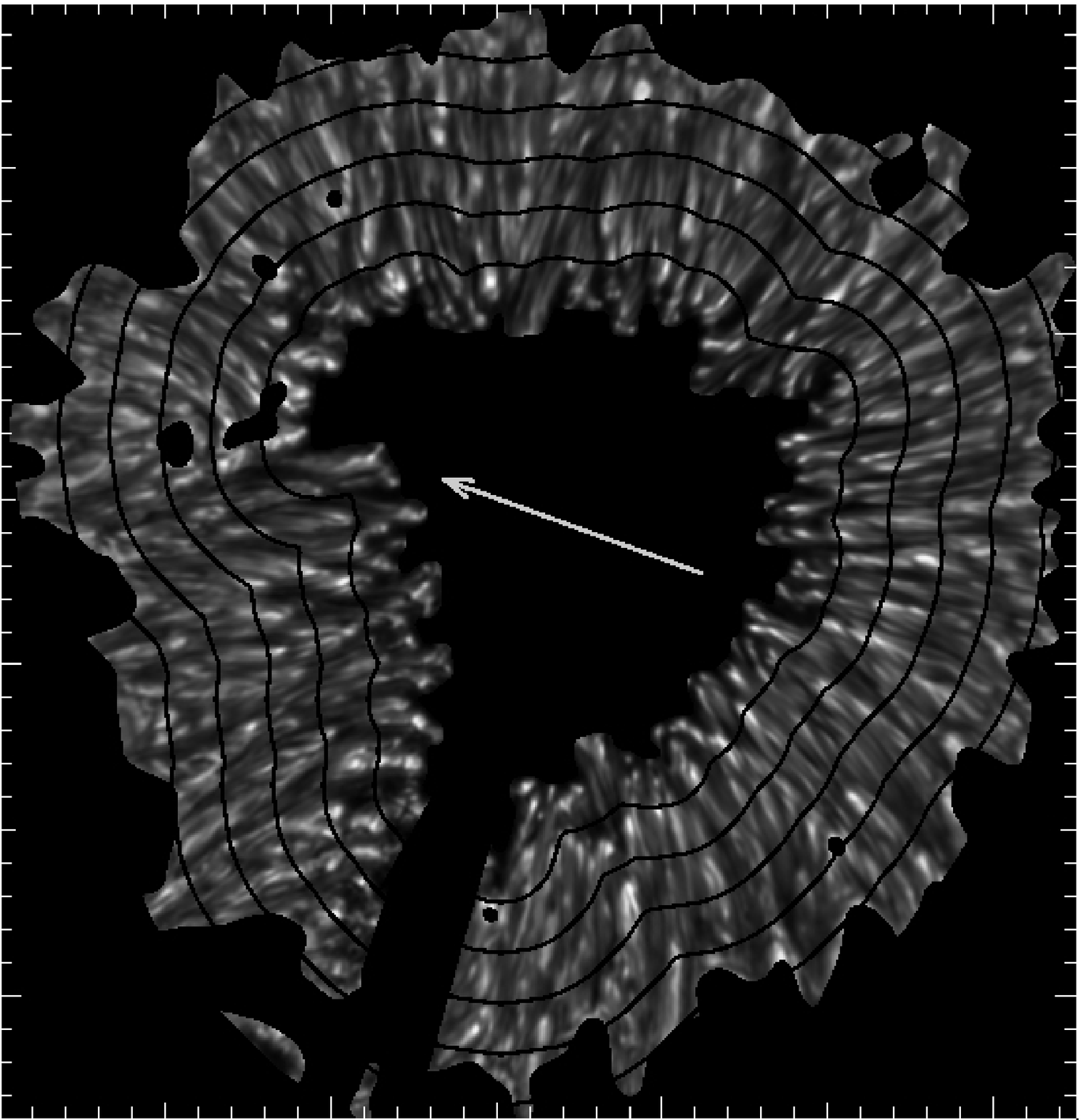}
\includegraphics[width=0.33\textwidth,clip]{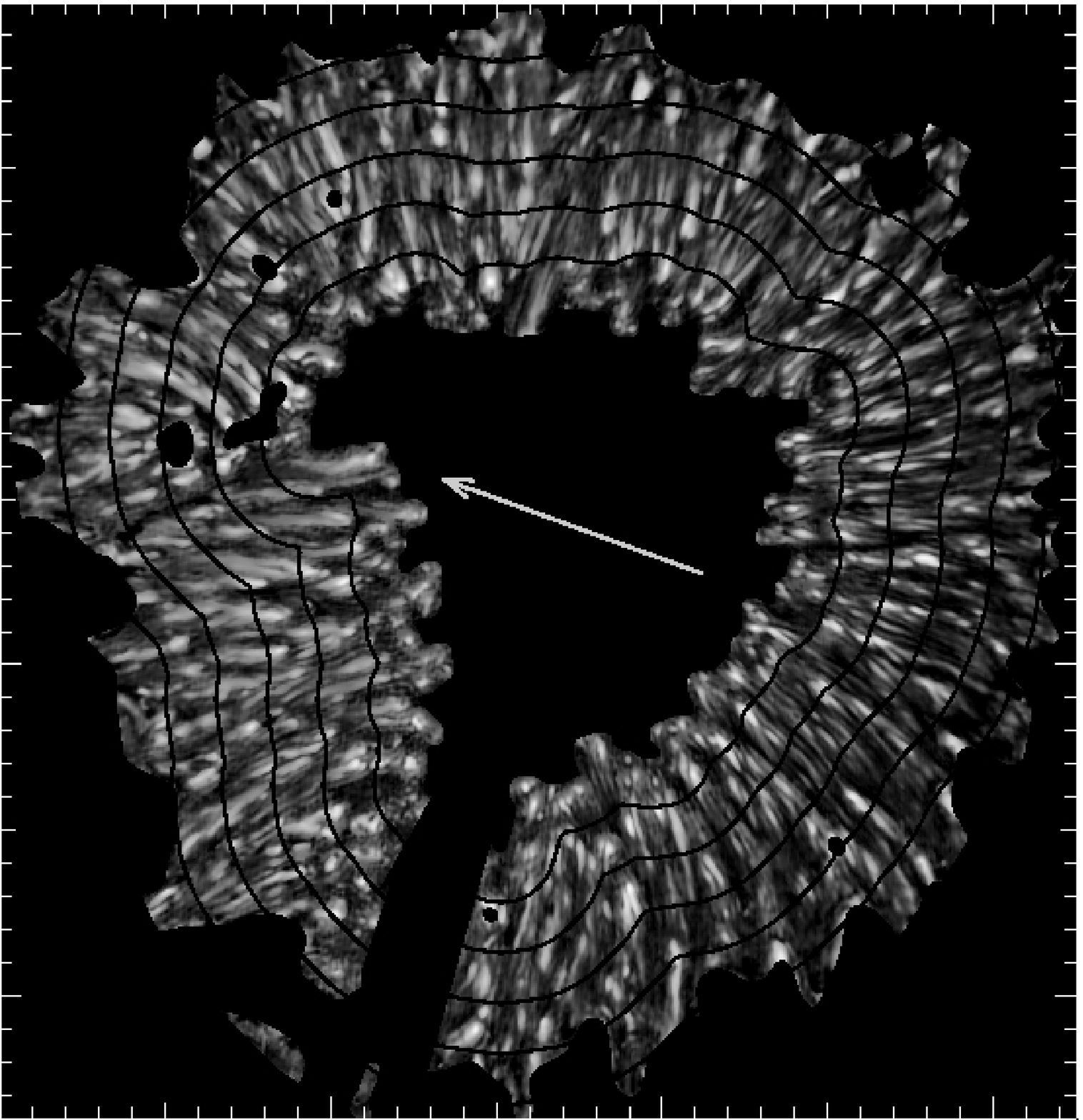}
\includegraphics[width=0.33\textwidth,clip]{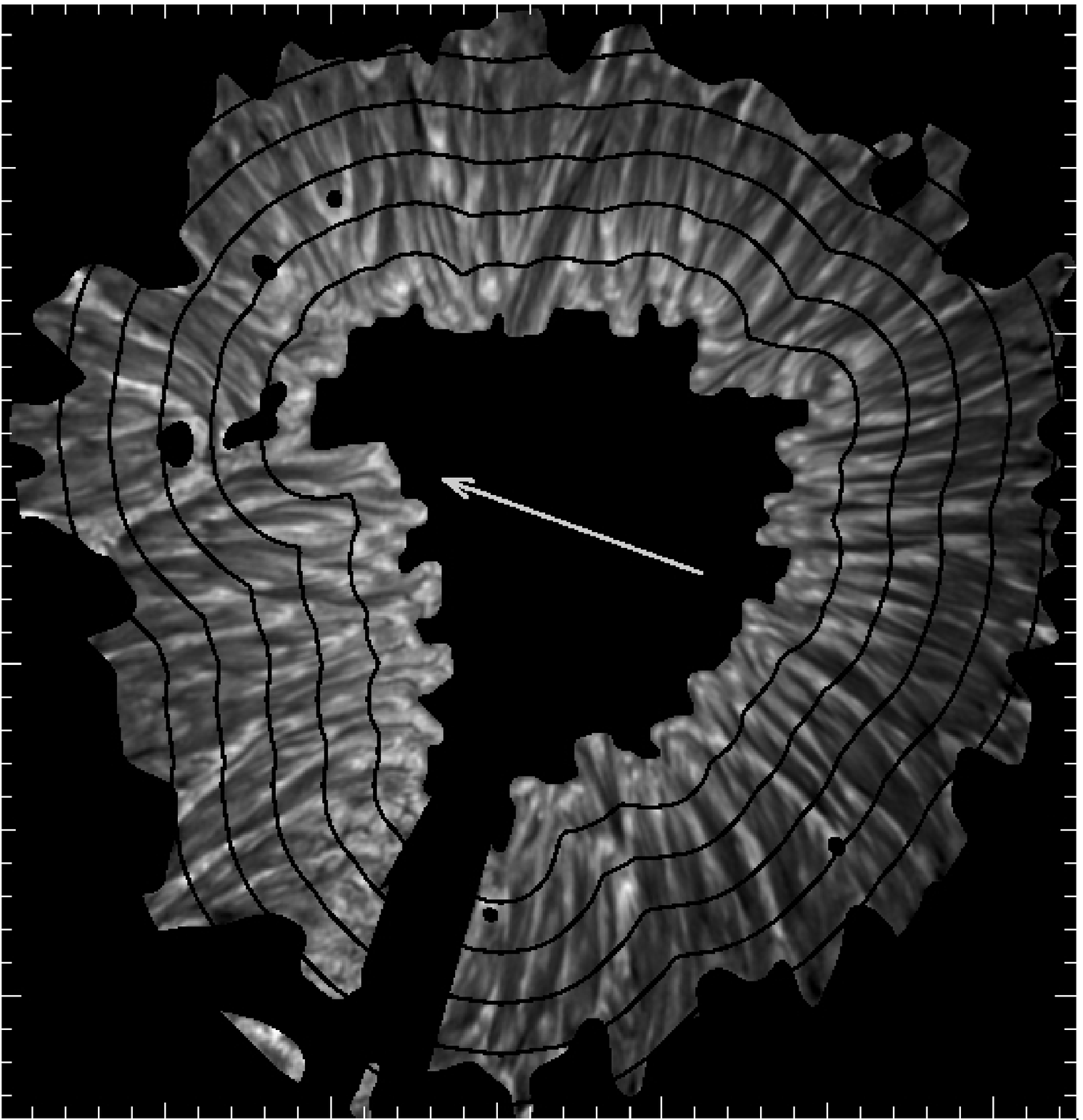}
\includegraphics[width=0.33\textwidth,clip]{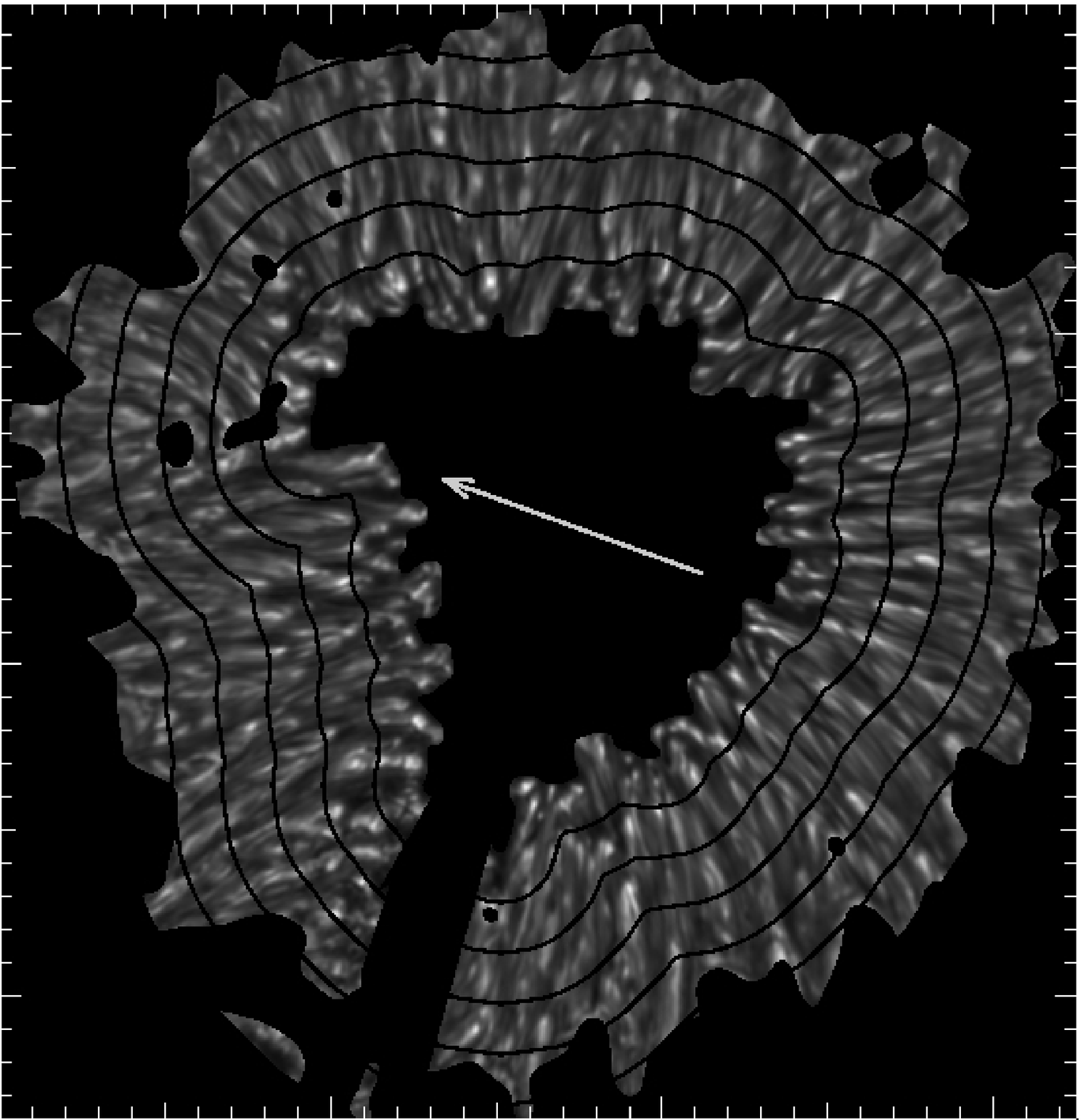}
\includegraphics[width=0.33\textwidth,clip]{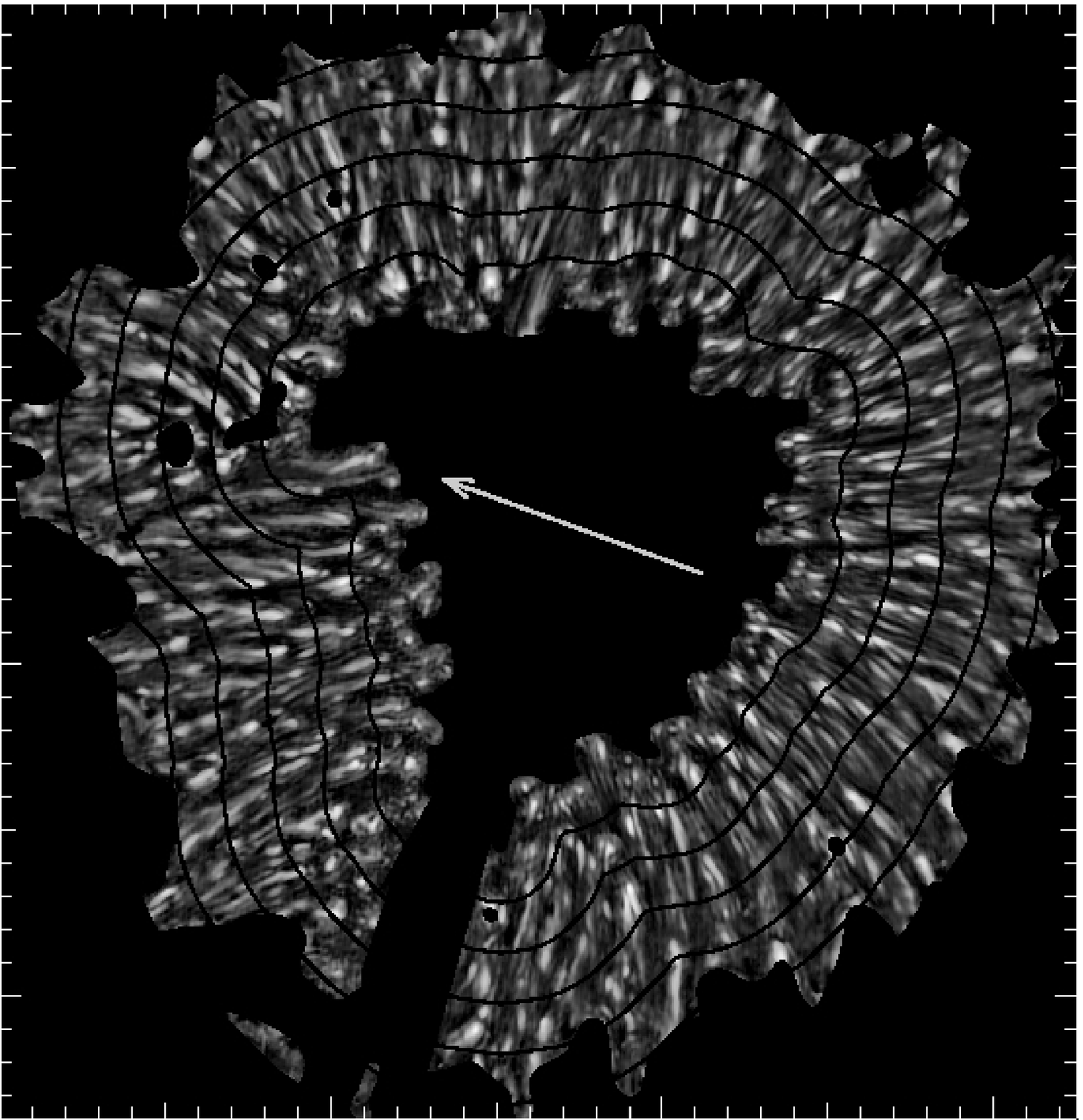}
\includegraphics[width=0.33\textwidth,clip]{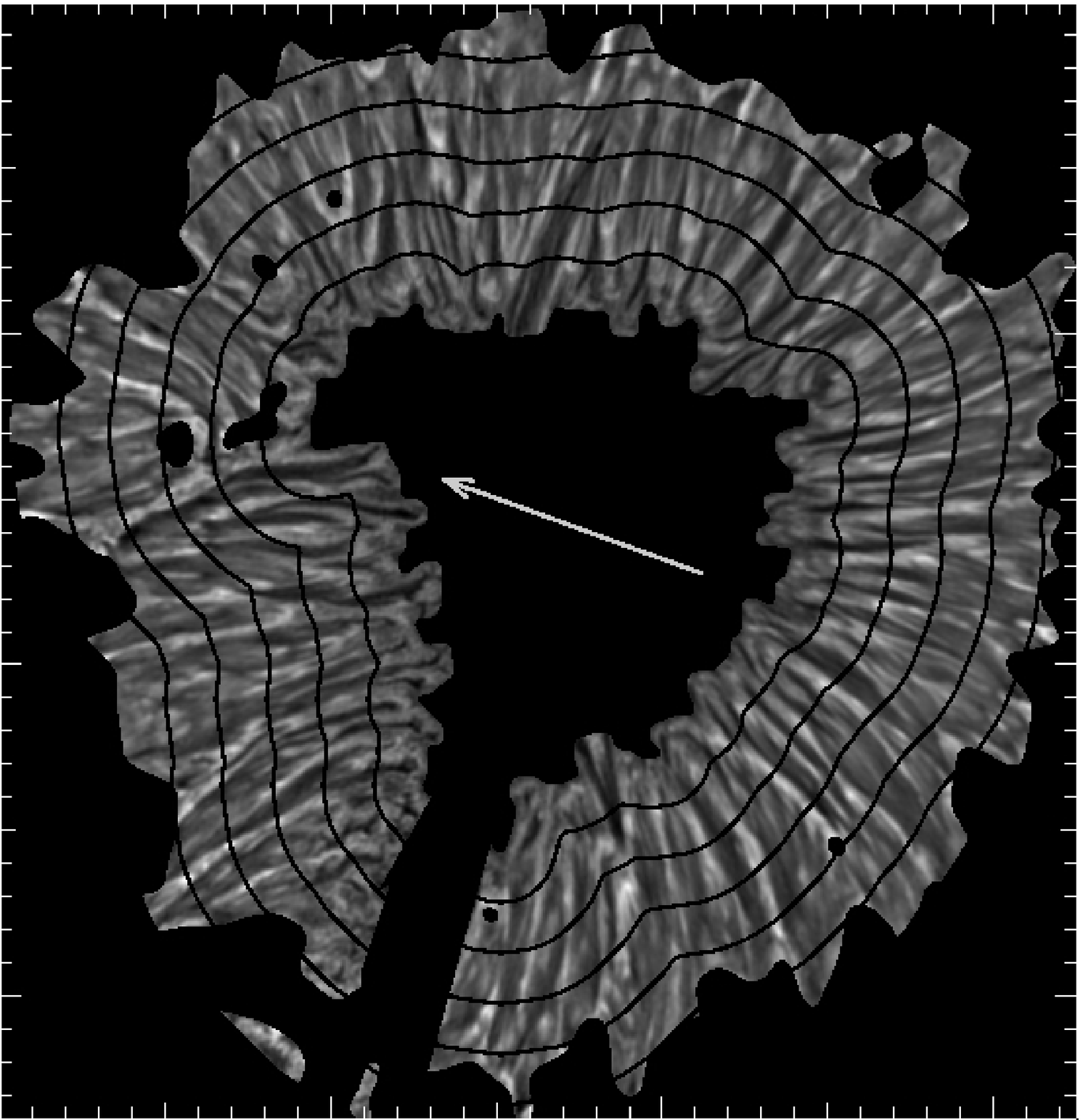}
\caption{Examples of measured quantities with superimposed contours outlining the six radial zones in the penumbra, numbered 1 (adjacent to the umbra) to 6 (the outermost penumbra). Radial zones 2--4 constitute what is referred to as the ``interior penumbra''. In the top row are shown (left to right) the 538~nm continuum intensity, the COG velocity in the same line and the LOS component of the magnetic field, obtained with the COG method. In the lower row are shown the same quantities but high-pass spatially filtered to remove large-scale azimuthal and radial variations. Tick marks are at 1\arcsec{} intervals. The arrow points in the direction of Sun center.}
\label{fig:fig_j}
\end{figure*}

An important component of our data processing is therefore straylight compensation. In the Supplementary Online Material (hereafter SOM) of \citet{2011Sci...333..316S} (see, \url{http://www.sciencemag.org/content/333/6040/316/suppl/DC1}), we discuss this in detail for the 5380 line, here we summarize the analysis made. We used snapshots from field-free simulations obtained with the \citet{1998ApJ...499..914S} code, degraded with the PSF of a perfect 1-m solar telescope. This constitutes our ``true'' theoretical data. We assumed that the actual PSF of the MOMFBD restored science data corresponds to a diffraction limited core with reduced peak strength (reduced Strehl) and enhanced wings due to straylight as follows
\begin{equation}
  \label{eq:Stray}
  I_o=(1-\alpha) I_t + \alpha I_t * S(W) ,
\end{equation}
where $I_o$ and $I_t$ are the observed and ``true'' intensities at any wavelength and polarization state, $\alpha$ the straylight fraction, ``*'' denotes convolution and $S$ is the straylight PSF, having a full width at half maximum (FWHM) $W$. This corresponds to a total PSF $P$ of 
\begin{equation}
 P = (1-\alpha) \delta + \alpha S(W) ,
\end{equation}
where $\delta$ is Dirac delta function. Note that this PSF corresponds to assuming that \emph{the core of the PSF is perfectly compensated} for by MOMFBD {\gs processing; this} is consistent with our approach of first degrading the synthetic simulation data with the diffraction limited PSF of the SST. When using Eq. (2) to deconvolve the data we thus compensate \emph{only} for straylight. We used this PSF to both degrade the simulated data and to deconvolve the observed data. As basis for our comparison, we calculated both synthetic continuum images and line profiles corresponding to a heliocentric distance of 15\degr. We then calculated 5380 line core LOS velocities from both the synthetic and the observed data. The calculated quantities compared were the RMS continuum intensity, the 5380 line core RMS velocity and the spatially averaged line core velocity\footnote{The \ion{C}{i} 5380 line is very temperature sensitive and weakens considerably in dark granular and penumbral lanes. This makes it very sensitive to straylight in dark (cool) lanes, explaining the large convective blueshift ($-917$~m\,s$^{-1}$ at 15\degr{} heliocentric distance). This also makes the spatially averaged 5380 velocity from spatially \emph{resolved} line profiles a sensitive indicator of straylight. Indeed, even the straylight compensated SST data leaves a residual convective blueshift of about $-240$~m\,s$^{-1}$ when averaging LOS velocities from the spatially \emph{resolved} line profiles for quiet Sun (this residual is consistent with what is obtained from simulations). A similar blue-shift is obtained when averaging the 5380 line core velocities over the penumbra.}. 
%The line profiles were also averaged and from the averaged line profile we established the wavelength scale by comparing with the convective blueshift obtained by \citet{2011A&A...528A.113D}. 
This comparison does not give much guidance about the shape of the straylight PSF. However, the {\gs low umbra intensity observed implies} that the straylight PSF cannot be very wide. In particular, {\gs a PSF having extended Lorentzian wings} would lead to observed umbra intensities on the order of 50\%, {\gs given that the level of straylight must be on the order of 50\% to explain the observed granulation contrast}. If we assume that the true minimum umbra intensity is not less than half of the measured values, then up to 7--9\% of the straylight could be {\gs from a Lorentzian PSF. A similar problem with the umbra intensity arises even with a Gaussian PSF, if it is too wide. Based on extensive tests, we conclude that} the majority of the straylight must be contained inside a PSF {\gs with a FWHM that is less than about 2\arcsec{}, else the restored umbra intensity is close to zero or negative \citep{2011Sci...333..316S}}.

Our best agreement with granulation data, taking into account the constraints from the umbra intensity, is a Gaussian PSF with a FWHM of about only 1\farcs2 and a straylight contribution of $\alpha$=58\% at 538~nm.\footnote{A narrow straylight PSF is consistent with main contributions from high-order aberrations, such as from seeing or possibly from the adaptive mirror \citep{2010A&A...521A..68S,2011A&A...XX..YYL}.} We have adopted these values for the 5380 line. We use the same FWHM for the 6301 line, but with lower straylight contribution, $\alpha$=50\% at 630~nm, {\gs since our present observations as well as those discussed by \citet{2009A&A...503..225W} suggest that the SST straylight increases at shorter wavelengths}. This leads to a quiet Sun RMS granulation contrast of about 13.3\% at 630~nm, which is somewhat lower than from simulations \citep{2008A&A...484L..17D, 2010A&A...521A..68S}, but the ``quiet'' region we observed does contain some small fraction of strong field that is likely to reduce the overall contrast. Admittedly, this determination of the straylight PSF is not precise, but the main effect of the straylight is present already when its PSF includes about one granule. Compensation with a wider straylight PSF will require somewhat smaller values of $\alpha$, but the overall results do not change qualitatively. Whereas the \emph{accuracy} of our measured intensities and LOS velocities thus is limited by the uncertainties about the precise shape of the straylight PSF, we are confident that the conclusions drawn about the clear existence of correlations between measured continuum intensities and LOS velocities, {\gs as well as about the existence of dark convective \emph{downflows} in the interior penumbra}, are robust. 

\subsection{LOS velocities and magnetic field measurements}
When measuring spectral line properties at low temperatures (low continuum intensities) there is always a concern about the possible influence of blends from (mostly) molecular lines. The 5380 line weakens considerably at reduced temperatures and must disappear in the umbra, strongly aggravating any such influence of blends. In SOM, we established that the influence of blends is small near the core of the 5380 line at continuum intensities corresponding to granulation and penumbral filaments. {\gs However}, there is evidence of weak blends both in the far blue and red wings (see Fig.~S5 in SOM) in the penumbral \emph{dark} structures and in the dark intergranular lanes. {\gs These} blends are obvious in the umbra (where also a strong molecular blend, shifted by about 3--3.5~pm{} to the blue of the \ion{C}{i} line, is present -- see SOM\footnote{This blend has now been identified to be from \ion{Mg~H}{} by Uitenbroek, Dumont and Tritschler (in prep.).}). To mitigate the influence of the blends in the wings, we initially choose to estimate LOS velocities from the line core only, by fitting a second-order polynomial to 5 wavelengths closest to the line minimum \citep{2011Sci...333..316S}. In the present work, we estimate also 5380 COG velocities, expected to be dominated by flows even closer to the photosphere than from the line core. In order to remove influence from the above mentioned blue/red blends, we crop the 5380 line profiles by excluding the first 6 and last 2 wavelengths shown in Fig.~S3 in SOM. This leads to underestimates of the strengths of the strongest blue- and redshifted COG velocities, as is the case when the COG method is applied to LOS magnetic field measurements \citep{1993SoPh..146..207C}, but these data nevertheless add support to the earlier line core measurements and also provide interesting information about flows in the deepest observable layers of the penumbra. 

We measured the 6301 line core shifts from the 3 wavelengths nearest to the line minimum, the COG velocity from the entire line profile and (using the spline interpolated line profiles) the 70\% bisector velocity, where the 70\% intensity level is relative to the line minimum intensity at each pixel. We estimated the LOS magnetic field $B_\text{LOS}$ by applying the COG method to Stokes $I+V$ and $I-V$ \citep{1979A&A....74....1R,1993SoPh..146..207C,2010A&A...518A...2O} and from the difference in the two COG wavelengths {\gs  $\delta \lambda$ we obtained  
$B_\text{LOS} = \delta \lambda~(9.334~10^{-13} \lambda^2 g_L)^{-1} $, where wavelengths are given in units of Angstrom, and $g_L$ is the Land\'e splitting (=1.667 for the 6301 line). }
%ANA: lande_g=[1.6667,2.5] c_b=1.e-3/(2.*4.667e-13*6302.^2*lande_g) ;from mA to Gauss
%We also estimated the zero-$V$ crossing velocities from the spline interpolated Stokes $V$ profiles, but do not discuss these data here.

\subsection{High-pass spatial filtering and masks}
\label{sec:filter}
In order to investigate relations between \emph{small-scale} fluctuations in intensity, LOS velocity and magnetic field, we make repeated use of high-pass filtered (unsharp masked) quantities. In particular, such filtering strongly reduces radial variations of the intensity in the penumbra, large-scale radial and azimuthal variations in the LOS component velocities from the horizontal (Evershed) flow and the horizontal component of the magnetic field. For example, the high-pass filtered continuum intensity $\delta I_c$ is calculated as 
\begin{equation}
 \delta I_c = I_c - G(W_2)*I_c
\end{equation}
where G is a Gaussian profile having a FWHM width $W_2$ of 20 pixels, or approximately 1\farcs2. The same high-pass filter is used with the measured LOS velocities and LOS magnetic field. {\gs The FWHM used here is coincidentally the same as used for the straylight compensation, but the precise value of the FWHM is not critical for the analysis.} It might be argued that a filter kernel with a larger FWHM would be more appropriate, but we prefer a narrow filter in order to reduce the spatial mixing of information across the borders to the umbra and the surrounding quiet Sun. 

{\gs An obvious effect of the spatial filtering is that the absolute zero point of any filtered quantity is lost and that all filtered quantities are referenced only to a \emph{local} mean, defined by the filter kernel. In the following, we therefore analyze both the filtered velocities to show their relation to \emph{local} intensity fluctuations, and the \emph{absolute} (unfiltered) velocities to demonstrate that \emph{the darkest penumbral structures are indeed associated with downflows}.}

In Fig.~\ref{fig:fig_j} are shown in the top row the 538~nm continuum image, the 5380 COG velocity and the LOS component of the magnetic field. In the bottom row are shown the same quantities after applying the previously defined high-pass filter. Clearly, the high-pass filter strongly reduces both systematic radial variations as well as differences between the limb and center sides of the penumbra, but leaves the small-scale structures intact.

\begin{figure}%[t!]
 \centering
\includegraphics[width=0.33\textwidth]{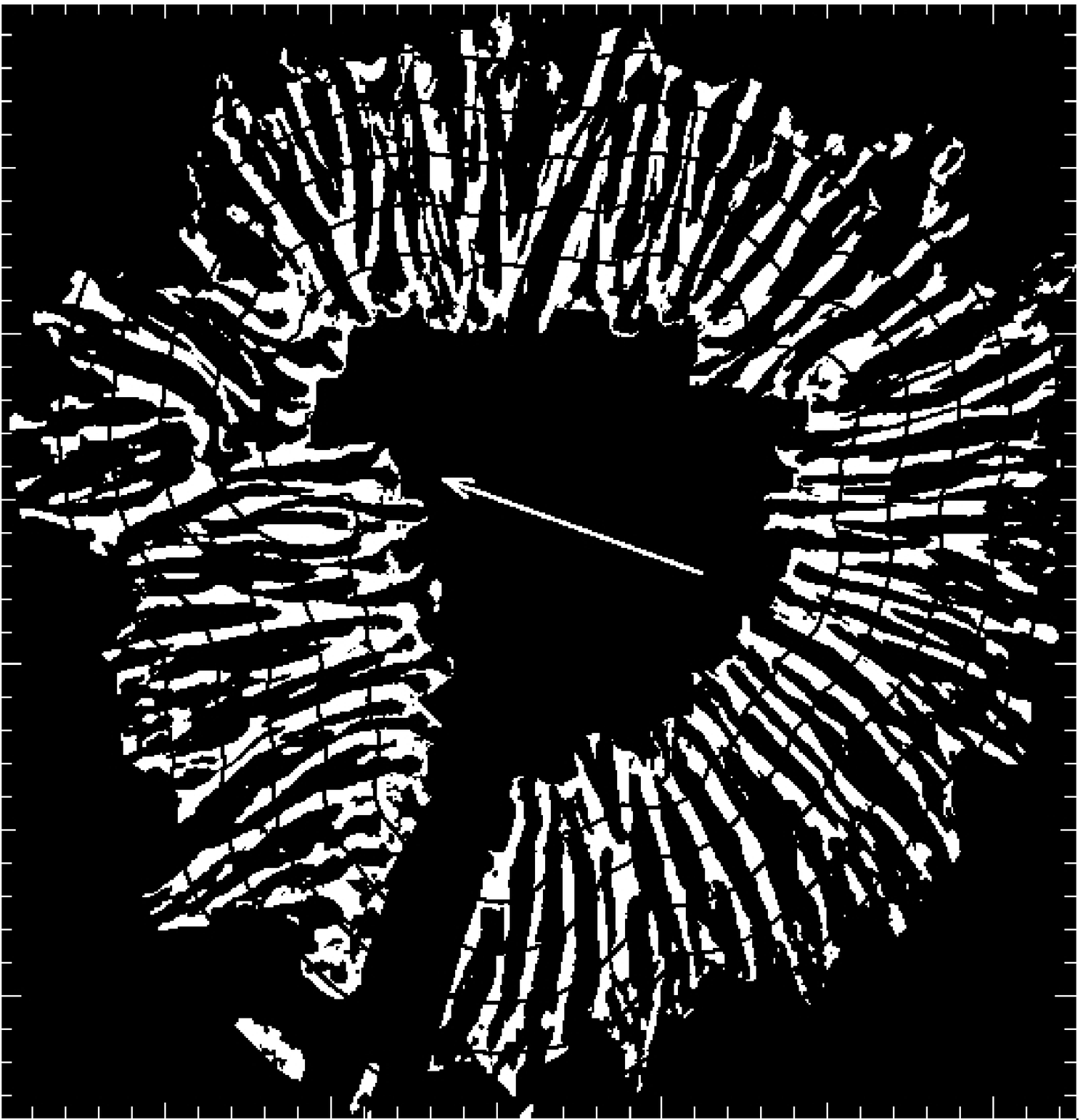}
\includegraphics[width=0.33\textwidth]{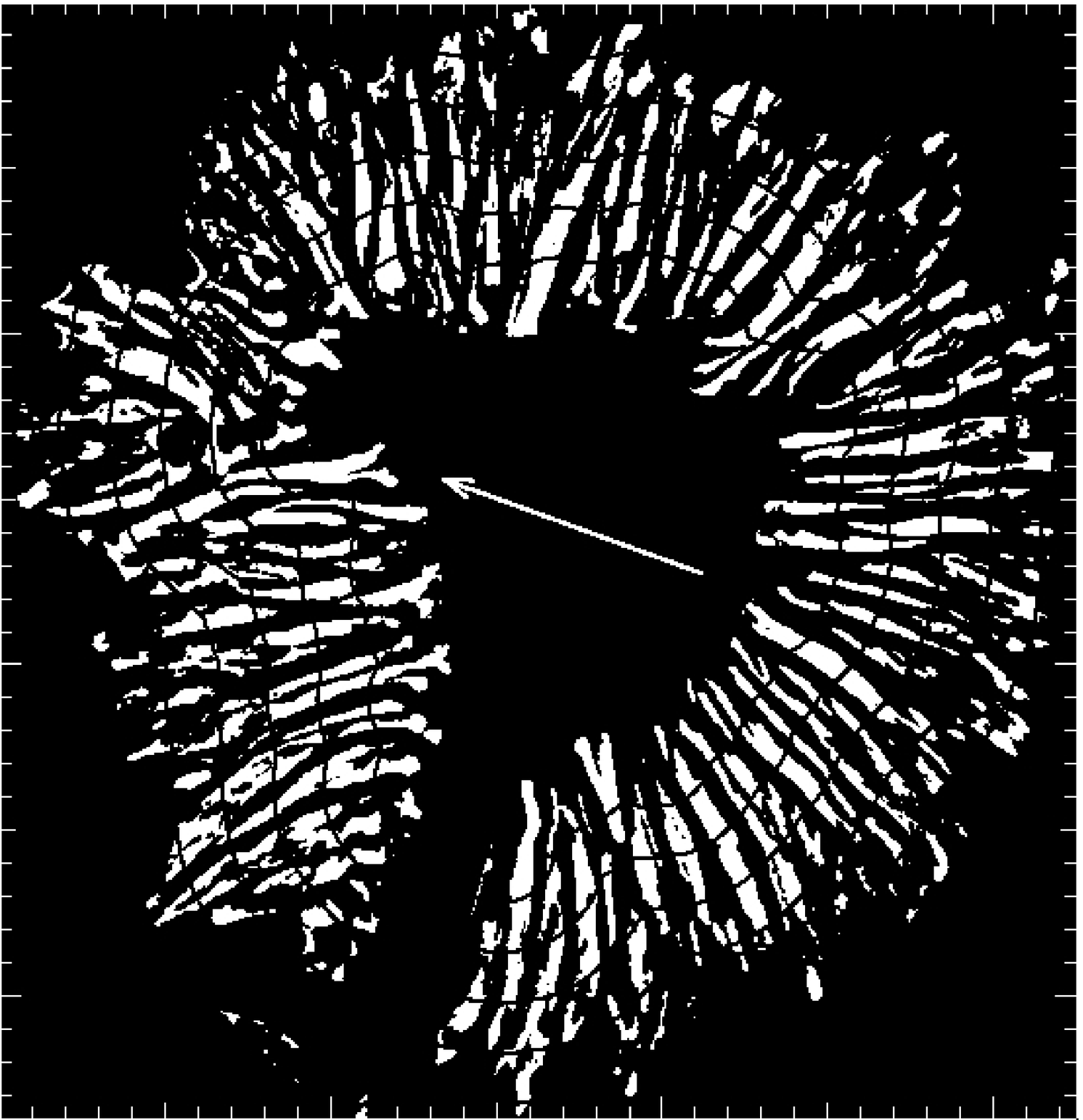}
\caption{Spine (upper panel) and inter-spine (lower panel) masks, obtained by thresholding the spatially filtered LOS magnetic field map, shown in the lower-right panel of Fig.~\ref{fig:fig_j}. {\gs The thresholds have been set such that roughly 33\% of the penumbra is contained in each of the two masks.}}
\label{fig:fig_k}
\end{figure}

For some of our fits (see Sect.~\ref{sec:azfit}), we use the spatially filtered quantities to define spatial masks, isolating certain measured parameters in a given range. In particular, we use the filtered 538~nm continuum image to define masks that have intensities in a specific range to estimate the relation between intensity and vertical and radial flows (see Sects.~\ref{sec:ivcorr} and \ref{sec:vmcorr}). We also use the filtered LOS magnetic field map to identify penumbral structure where the LOS magnetic field is locally strong (spines) or weak  and nearly horizontal (inter-spines). {\gs These masks, shown in Fig.~\ref{fig:fig_k}, are based on the high-pass filtered $B_\text{LOS}$ map, with thresholds set such that roughly 33\% of the penumbra is contained in each of the two masks.}

\subsection{Azimuthal fits}
\label{sec:azfit}

\begin{figure}%[bt!]
 \centering
\includegraphics[bb=18 42 578 702,angle=-90,width=0.242\textwidth]{fig_e2_2.eps}
\includegraphics[bb=18 42 578 702,angle=-90,width=0.242\textwidth]{fig_e2_42.eps}
\includegraphics[bb=18 42 578 702,angle=-90,width=0.242\textwidth]{fig_e2_6.eps}
\includegraphics[bb=18 42 578 702,angle=-90,width=0.242\textwidth]{fig_e2_46.eps}
\includegraphics[bb=18 42 578 702,angle=-90,width=0.242\textwidth]{fig_e2_18.eps}
\includegraphics[bb=18 42 578 702,angle=-90,width=0.242\textwidth]{fig_e2_58.eps}
\includegraphics[bb=18 42 578 702,angle=-90,width=0.242\textwidth]{fig_e2_22.eps}
\includegraphics[bb=18 42 578 702,angle=-90,width=0.242\textwidth]{fig_e2_62.eps}
\includegraphics[bb=18 42 578 702,angle=-90,width=0.242\textwidth]{fig_e2_34.eps}
\includegraphics[bb=18 42 578 702,angle=-90,width=0.242\textwidth]{fig_e2_74.eps}
\includegraphics[bb=18 42 578 702,angle=-90,width=0.242\textwidth]{fig_e2_38.eps}
\includegraphics[bb=18 42 578 702,angle=-90,width=0.242\textwidth]{fig_e2_78.eps}
\caption{Examples of azimuthal fits of the {\gs absolute (unfiltered)} LOS velocities {\gs to Eq. (4), measured} with different methods in the 5380 and 6301 lines, for relatively dark ($\delta I_c\approx -0.19$) and bright ($\delta I_c\approx 0.19$) structures. The left column shows the data (dark dots) and fits (red curves) for spines, the right column for inter-spines.}
\label{fig:fig_e2}
\end{figure}

The sunspot observed is at 15\degr{} heliocentric distance, such that the measured LOS velocity contains a mixture of contributions from both vertical and (strong) radial flows\footnote{A sunspot located precisely at disk center would be allow clean measurements of vertical velocities. However, even a few degrees heliocentric distance leads to significant contributions from radial Evershed flows that nearly always must be accounted for in such measurements, as demonstrated by \citet{2011ApJ...739...35B}.}. %Furthermore, it is of considerable interest to be able to simultaneously estimate radial (Evershed) flow velocities, now believed to be closely related to the same convective process that drives the vertical flows beneath the visible surface \citep{2008ApJ...677L.149S,2009Sci...325..171R,2009ApJ...691..640R,2011ApJ...729....5R}. 
We must resort to statistical methods to separately determine average properties of the radial and vertical flows. The presently observed penumbra has boundaries to the umbra and quiet Sun that are rather irregular, but the penumbral filaments and the radial extent of the penumbra appear similar on the disk center and limb sides. We assume that the small-scale velocity field and magnetic field fluctuations do not vary in amplitude with azimuth angle in order to separate the observed LOS component of the velocity into its radial and vertical components. 

Ignoring any azimuthal components of the flow field, the relation between the LOS velocity $v_\text{LOS}$, the radial (horizontal) velocity $v_r$, the vertical velocity $v_z$, the heliocentric distance $\theta$ (=15\degr) and the azimuthal angle $\phi$ is then\footnote{The same relation, but with different sign conventions, is used to resolve $B_\text{LOS}$ into its radial and vertical components.} 
\begin{equation}
 v_{\rm LOS} = -v_r \cos \phi~\sin \theta + v_z \cos \theta 
\end{equation}
\citep{1952MNRAS.112..414P}. Fits based on this relation have been used to disentangle the radial and vertical components of penumbral velocity and/or magnetic fields from LOS measurements by numerous earlier investigators \citep[e.g.,][]{1964ApNr....8..205M,1993ApJ...403..780T,2000A&A...358.1122S,2003A&A...403L..47B,2004A&A...415..717T,2005A&A...436.1087L,2006A&A...453.1117B,2007ApJ...658.1357S,2011Sci...333..316S}. The sign conventions here are that $\phi$ is zero in the disk center direction, radial outflows are counted positive, and vertical velocities as well as velocities in the direction toward the observer are counted as negative. Our procedure for defining the azimuth angle and radial distances within the penumbra is described in SOM. The zero-point azimuth direction was determined from full disk images recorded by {\gs the Solar Dynamics Observatory (SDO)} close to the time of our observations and is indicated by the arrows shown in, e.g., Figs.~\ref{fig:fig_i} and \ref{fig:fig_i2}. Guided by the orientation of the penumbral filaments, we determined an approximate center of the sunspot from which straight lines line up as well as possible with the penumbral filaments. The umbra/penumbra boundary was identified by applying an intensity threshold and removing the light bridge and innermost part of a few filaments. From that boundary six radial zones were {\gs defined by successively} applying a dilate operation using a circular kernel of 1.5\arcsec{} radius. The boundaries of the six radial zones are shown as dark contours in Fig.~\ref{fig:fig_j}. Examples of azimuthal fits made at different high-pass filtered 538~nm continuum intensities can be found in Fig.~\ref{fig:fig_e2}. These show fundamental differences between the spines and inter-spines: in the inter-spines the amplitude of the azimuthal variation of $v_\text{LOS}$ implies a strong radial component of the LOS velocity that is much weaker in the spines and virtually absent at low continuum intensities. 

\section{Results}
Figures~\ref{fig:fig_o} and \ref{fig:fig_o2} show all measured quantities within two 5\arcsec$\times$8\arcsec{} subfields on (mostly) the limb and disk center sides, with tick marks separated at 1\arcsec{} intervals. The first and last panels in the top row show the 538~nm and 630~nm continuum images. These appear nearly identical but {\gs the continuum image} at 538~nm shows somewhat higher spatial resolution, as expected. The first two panels in the bottom row show the 5380 COG and line core velocities. These are quite similar, but with significant differences for the smallest flow features shown. Although not so obvious in Fig.~\ref{fig:fig_o2}, there is an overall tendency, also outside the sunspot, for the COG velocity map to show stronger small-scale flows than for the 5380 line core velocity map. This is interpreted to mean that the flows with the smallest scales change rapidly with height, whereas the large-scale flows show weaker gradients with height. 

The 6301 flow maps in general show much less fine structure than the 5380 flow maps. This is not a question of spatial resolution, as the continuum images demonstrate, but most likely is due to either or both of two effects: the first is that small-scale upflows do not extend to the same height as large-scale strong upflows. {\gg The second effect} is the relatively large formation height range of the 6301 line, aggravated by the limited spectral resolution of CRISP, causing smearing of flow topologies that change with height within the formation range. We note that the 6301 line core velocity (last panel) is the most ``fuzzy'' of these flow maps on the limb side, but not on the disk center side. We also note the 6301 COG and 70\% bisector flow maps (third and fourth panels in the bottom row of Figs.~\ref{fig:fig_o} and \ref{fig:fig_o2}) appear almost identical. {\gg In the following,} we do not discuss these flow maps separately. We attribute this similarity to relatively low spectral resolution of CRISP, smearing out the height variations of the flow field. 

We emphasize the small azimuthal scale of almost all the flow structures, in particular the (dark) red-shifted lanes in between the (bright) blue-shifted features. \emph{Many of these flow structures appear to have azimuthal scales close to the SST diffraction limit} (0\farcs14 at 538~nm). This clearly illustrates the difficulties of earlier (failed) attempts aimed at detecting convective downflows {\gg (i.e., that \emph{dark} features are associated with \emph{downflows})} in the interior penumbra at much lower spatial resolution \citep{2010ApJ...725...11B,2009A&A...508.1453F,2011arXiv1107.2586F}, as demonstrated recently also from analysis of synthetic spectra calculated from simulation data and degraded to the spatial resolution of SST \citep{2011ApJ...739...35B}. In the following, we characterize the complicated relations between the observed small-scale variations in the continuum intensity, flow field and magnetic field, and attempt to explain some of the observed differences between the limb and disk center sides of the penumbra.

Finally, we note that the LOS magnetic field (mid panel, upper row in Figs.~\ref{fig:fig_o} and \ref{fig:fig_o2}), shows only weak evidence for variations that can easily be linked to the small-scale velocity field. However, the LOS magnetic field in Fig.~\ref{fig:fig_o2} shows {\gs several} narrow dark structures, where either the polarity reverses or where the LOS magnetic field is nearly zero (according to the present COG measurements). These can in most cases be linked to locally red-shifted flow features. {\gs We will discuss these features in more detail in a forth-coming paper.}
\begin{figure*}%[t!]
\centering
%\includegraphics[bb=18 42 578 702,angle=-90,width=0.245\textwidth]{FIGS_C/fig_c_1.eps}
%\includegraphics[bb=18 42 578 702,angle=-90,width=0.245\textwidth]{FIGS_C/fig_c_2.eps}
%\includegraphics[bb=18 42 578 702,angle=-90,width=0.245\textwidth]{FIGS_C/fig_c_3.eps}
%\includegraphics[bb=18 42 578 702,angle=-90,width=0.245\textwidth]{FIGS_C/fig_c_4.eps}\\[2mm]
%\includegraphics[bb=18 42 578 702,angle=-90,width=0.245\textwidth]{FIGS_C/fig_d_1.eps}
%\includegraphics[bb=18 42 578 702,angle=-90,width=0.245\textwidth]{FIGS_C/fig_d_2.eps}
%\includegraphics[bb=18 42 578 702,angle=-90,width=0.245\textwidth]{FIGS_C/fig_d_3.eps}
%\includegraphics[bb=18 42 578 702,angle=-90,width=0.245\textwidth]{FIGS_C/fig_d_4.eps}
%bb=32 78 600 730, 
\includegraphics[bb=32 78 600 730,clip,angle=-90,width=0.245\textwidth]{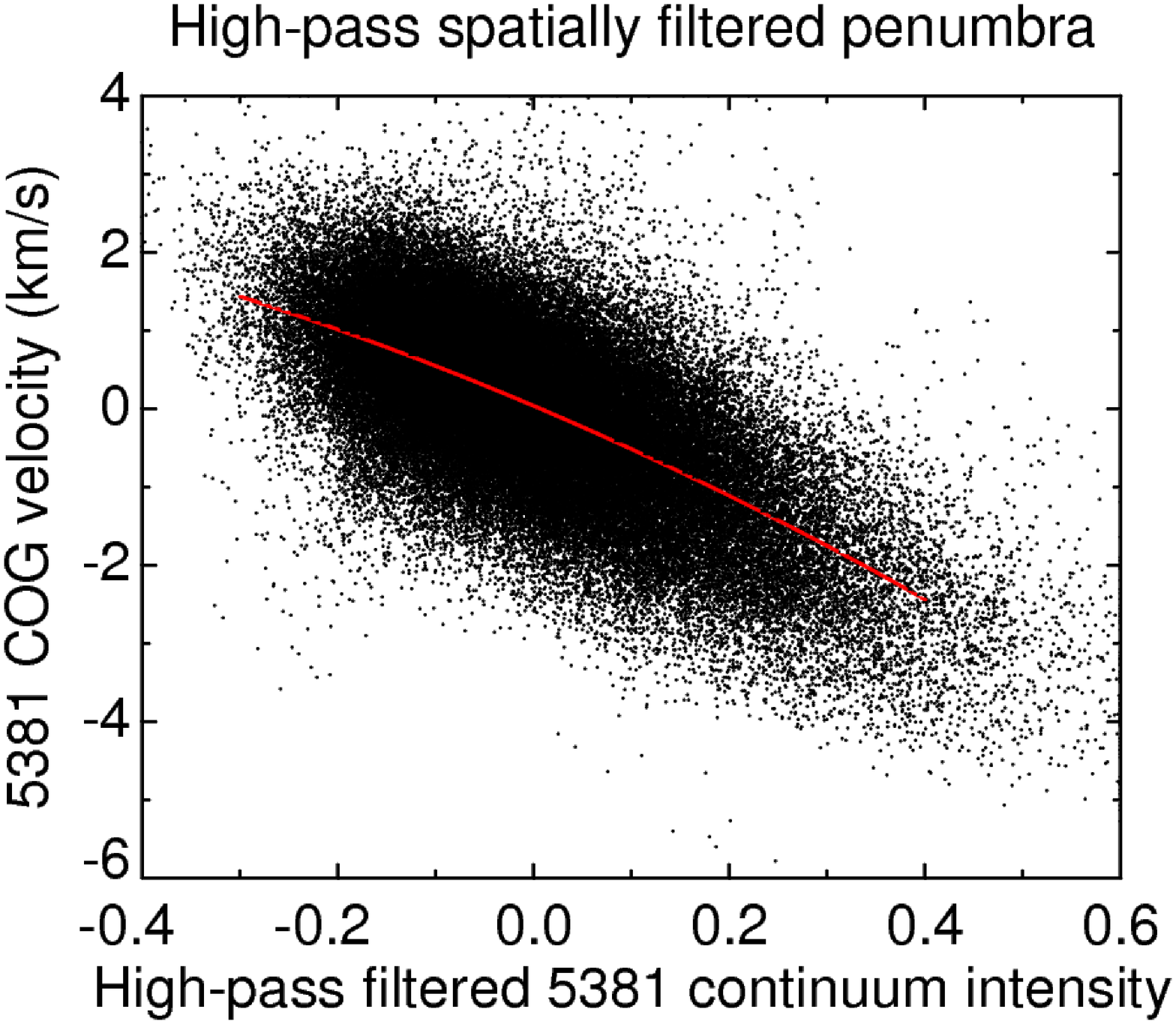}
\includegraphics[bb=32 78 600 730,clip,angle=-90,width=0.245\textwidth]{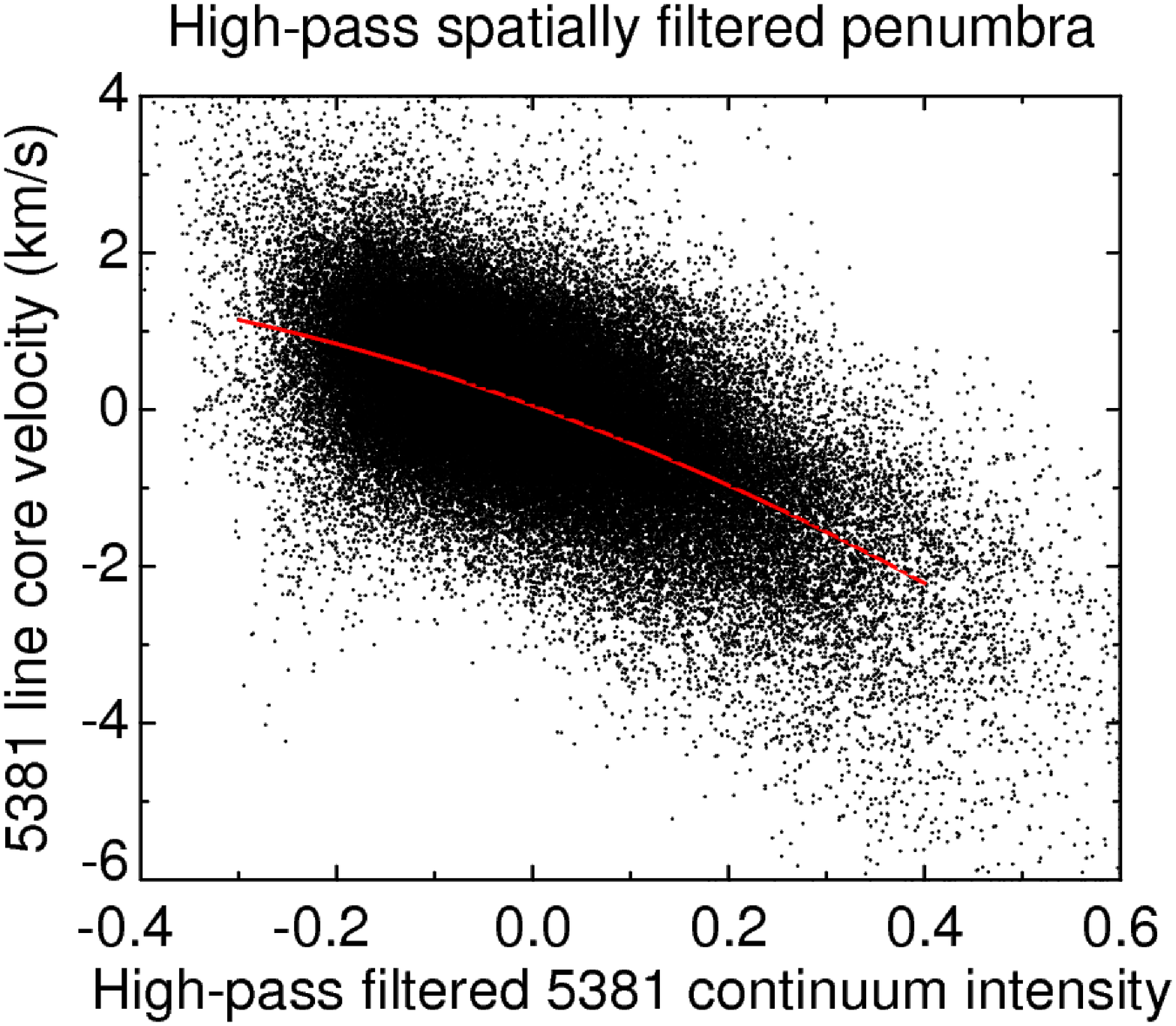}
\includegraphics[bb=32 78 600 730,clip,angle=-90,width=0.245\textwidth]{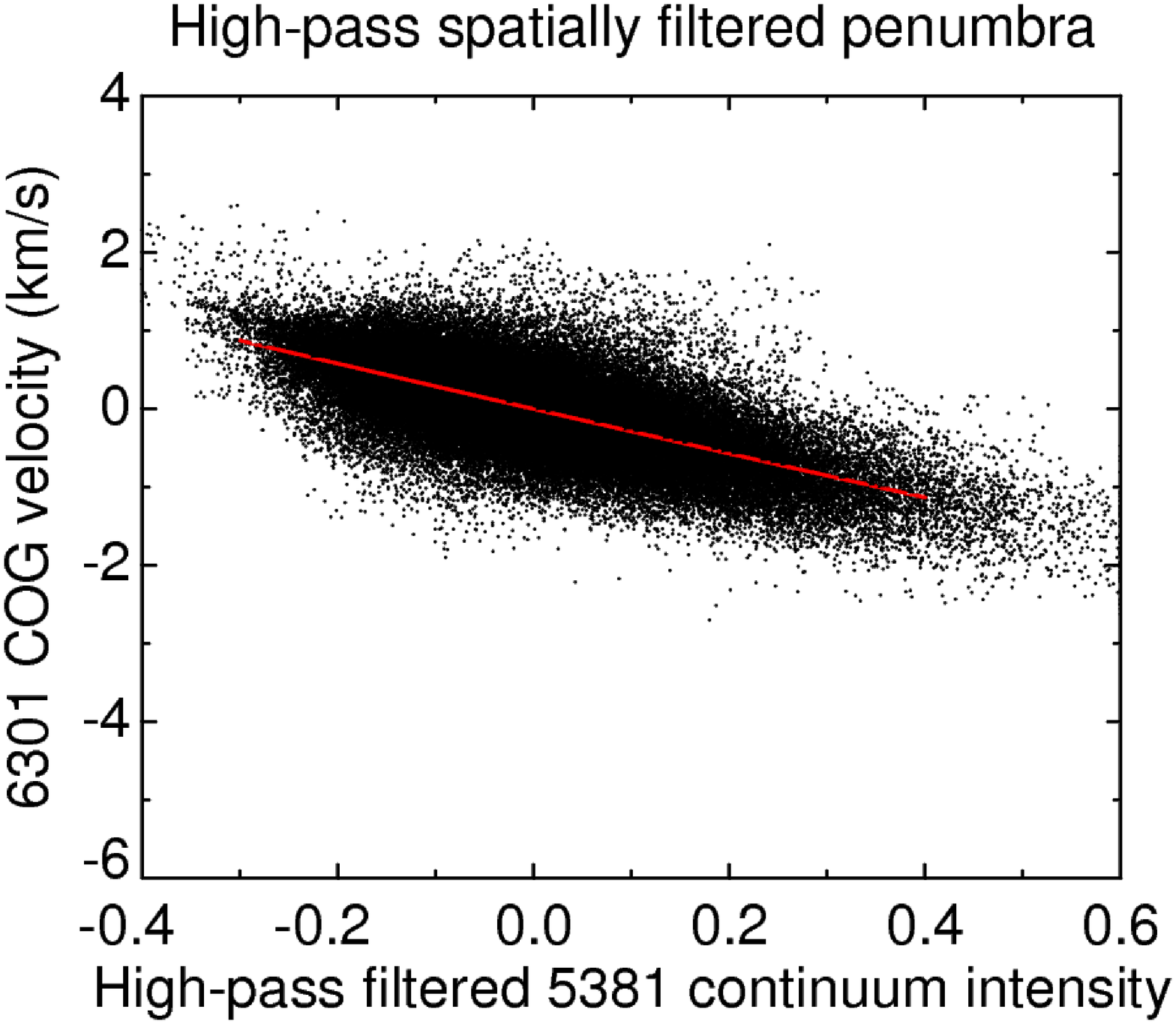}
\includegraphics[bb=32 78 600 730,clip,angle=-90,width=0.245\textwidth]{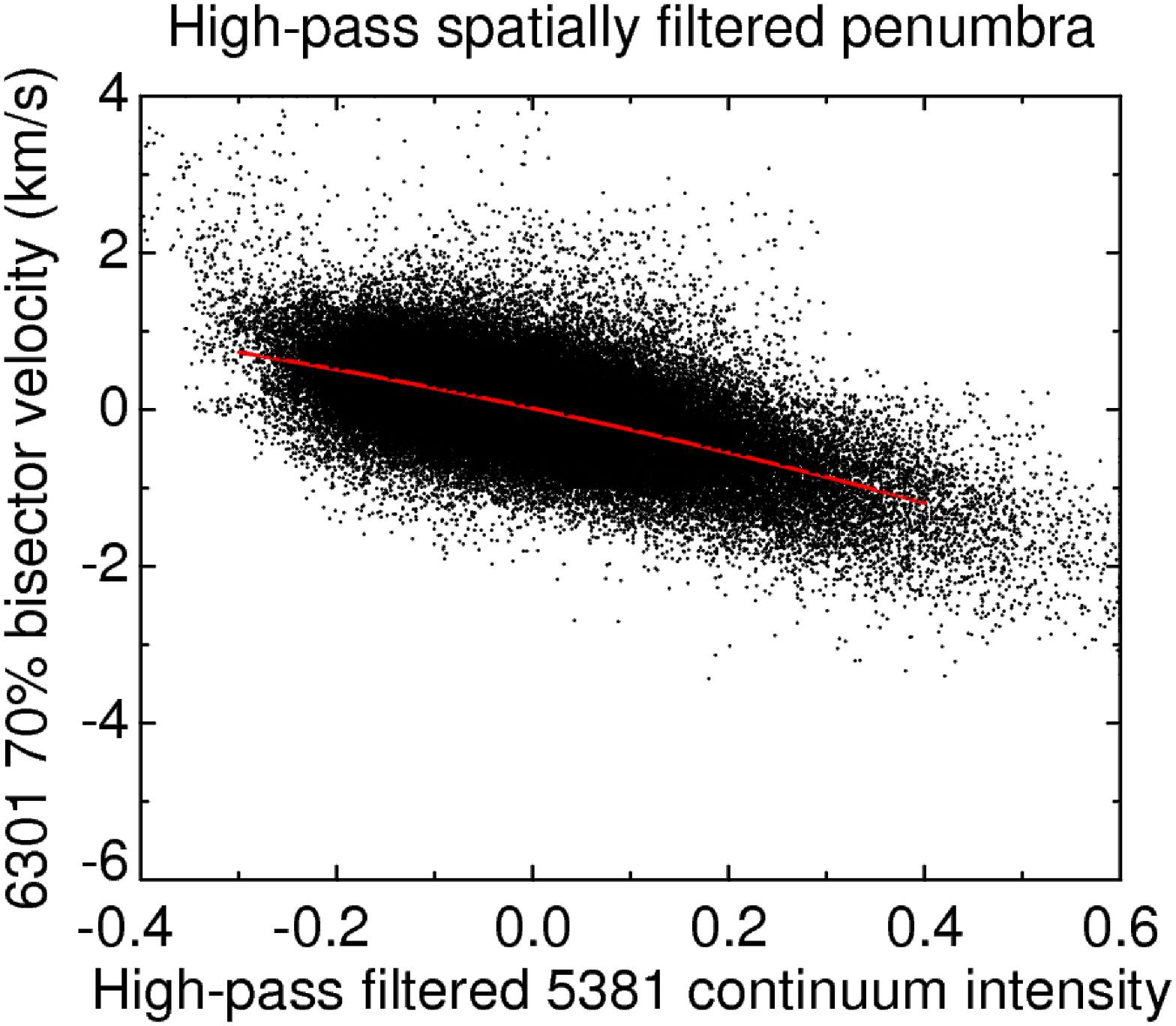}\\[2mm]
\includegraphics[bb=32 78 600 730,clip,angle=-90,width=0.245\textwidth]{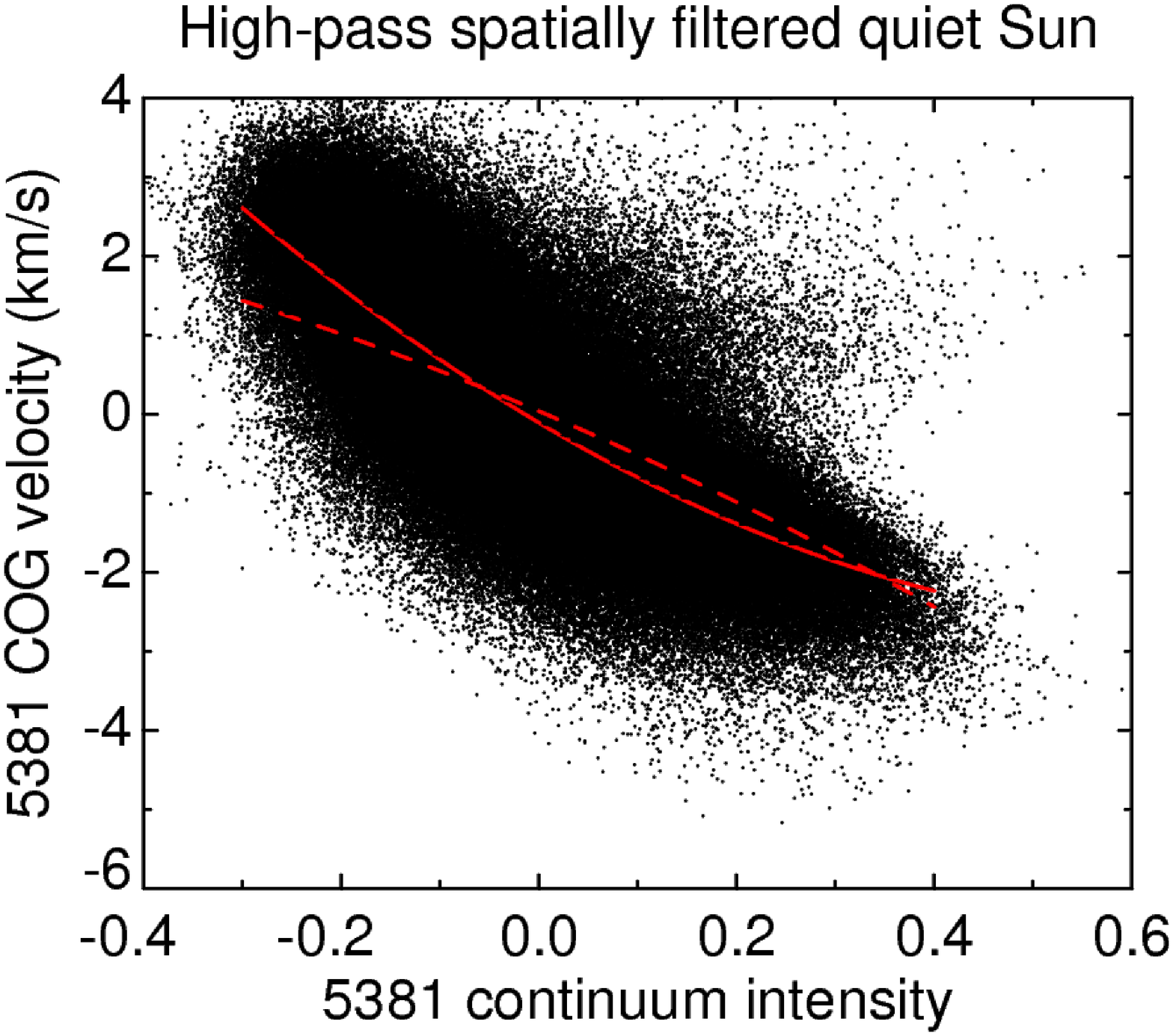}
\includegraphics[bb=32 78 600 730,clip,angle=-90,width=0.245\textwidth]{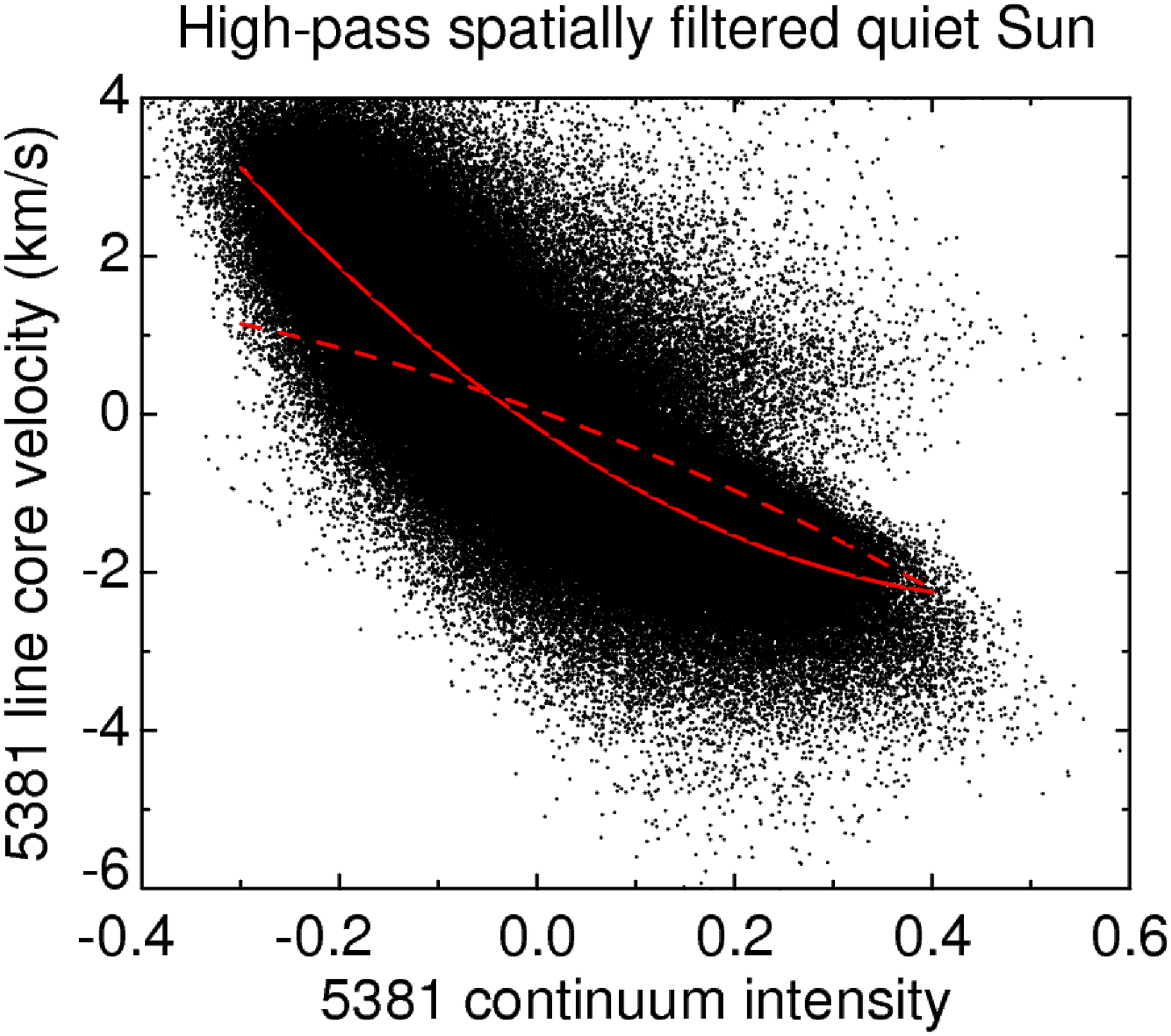}
\includegraphics[bb=32 78 600 730,clip,angle=-90,width=0.245\textwidth]{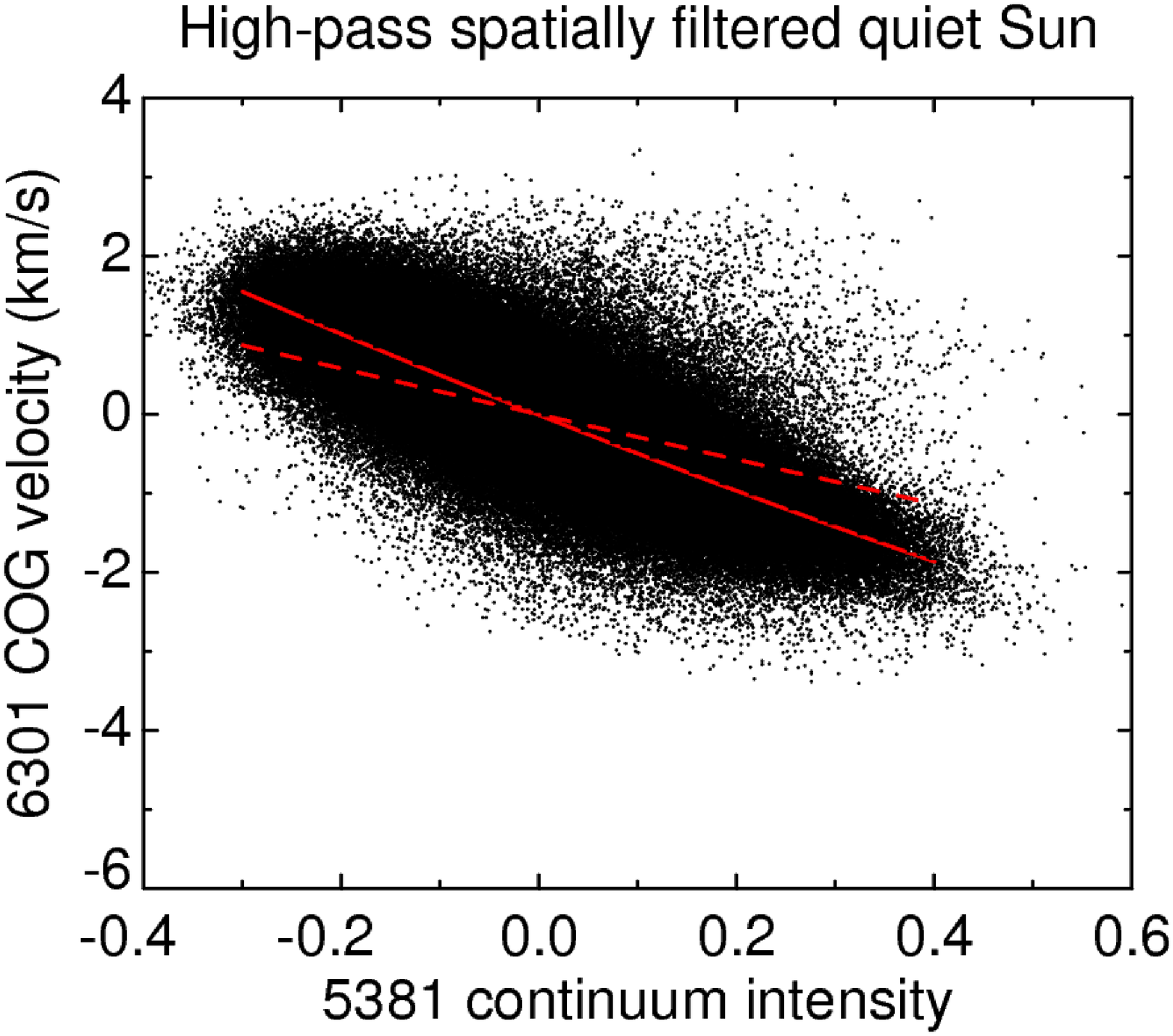}
\includegraphics[bb=32 78 600 730,clip,angle=-90,width=0.245\textwidth]{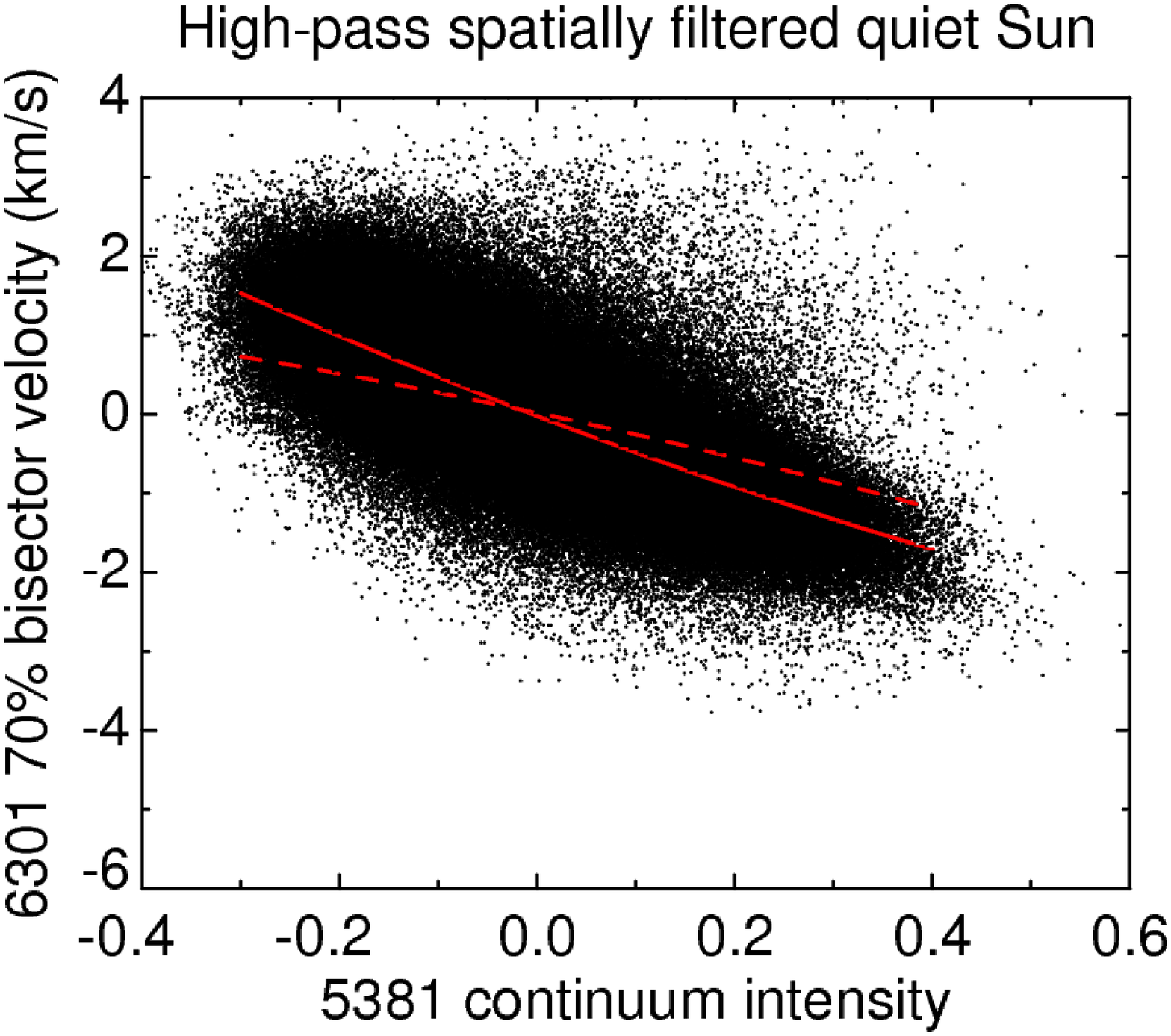}
\caption{Correlation between the high-pass filtered 538~nm continuum intensity and the {\gs high-pass filtered} LOS velocity, measured with different methods in the 5380 and 6301 lines. The top row shows plots for the ``interior penumbra'', the bottom row corresponding plots for quiet Sun. For easier comparison, the red curves show in the upper panels are shown as dashed curves in the lower panels.{\gs Note that the velocities plotted are referenced to \emph{local} averages by spatial filtering (unsharp masking).}}
\label{fig:fig_cd}
\end{figure*}

\begin{table*}%[t!]
  \centering
  \caption{Correlations and RMS velocities.}
  \begin{tabular}{ l r r r r r}
    \hline\hline\noalign{\smallskip}
    Quantity & 5380 COG & 5380 core & 6301 COG & 6301 70\% bis. & 6301 core \\
    \hline\noalign{\smallskip}
    $R$ interior penumbra & $-0.686$ & $-0.610$ & $-0.693$ & $-0.602$ & $-0.404$ \\
    $R$ quiet Sun & $-0.772$ & $-0.796$ & $-0.804$ & $-0.734$ & $-0.588$ \\
    \hline\noalign{\smallskip}
    $v_\text{RMS}$ interior penumbra (km\,s$^{-1}$) & 1.08 & 1.06 & 0.58 & 0.64 & 0.32 \\
    $v_\text{RMS}$ quiet Sun (km\,s$^{-1}$) & 1.38 & 1.50 & 0.91 & 0.96 & 0.56 \\
%    RMS velocity interior penumbra (km/s) & 1.08 & 1.06 & 0.58 & 0.64 & 0.32 \\
%    RMS velocity quiet Sun (km/s) & 1.38 & 1.50 & 0.91 & 0.96 & 0.56 \\
    \hline
  \end{tabular}
  \tablefoot{Summary of RMS velocities ($v_\text{RMS}$) and correlation coefficients ($R$)  between high-pass spatially filtered 538~nm continuum intensities and LOS velocities ($v_\text{LOS}$) measured with the 5380 and 6301 lines for quiet Sun and the interior penumbra. The different methods used for measuring the velocities are indicated above.}
\label{tab:table_cd}
\end{table*}
%\bigskip

\subsection{Intensity and LOS velocity correlations}
\label{sec:ivcorr}

A quick glance at Figs.~\ref{fig:fig_o} and \ref{fig:fig_o2} makes it evident that \emph{the measured small-scale LOS velocities are strongly related to the continuum intensity:} e.g., blue-shifted (shown bright) features can in almost all cases be associated with a locally bright continuum feature. This does not necessarily mean that there exist simple one-to-one relations between these quantities: it is easy to find examples where a relatively strong blue-shifted feature shows only a marginal brightening in the continuum and vice-versa. In other cases, the flow features and continuum structures appear somewhat displaced spatially relative to each other. There is no reason to expect a perfect correlation between intensity (temperature) and flow velocities: even for quiet Sun granulation, this correlation is less than about 80\%. We first quantify these correlations in the simplest possible manner. Figure~\ref{fig:fig_cd} shows the correlation between the high-pass filtered 538~nm continuum intensity and the \emph{high-pass filtered} LOS velocity for the interior penumbra (top row) and quiet Sun (bottom row). As shown in Fig.~\ref{fig:fig_j}, the main effect of high-pass filtering the LOS velocity map is to remove the {\gs systematic} differences in LOS velocity between the disk center and limb sides -- these differences are from the radial (horizontal) flow. 

After high-pass filtering the LOS velocity map, the remaining small-scale flow field structures appear similar on the disk center and limb sides, suggesting a dominance of \emph{vertical} flows {\gs \citep[c.f., ][ for similar arguments]{1969SoPh...10..384B}}. The correlations between intensity and LOS velocity in Fig.~\ref{fig:fig_cd} are obvious, the corresponding correlation coefficients are given in Table~\ref{tab:table_cd} and range from $-60$\% to $-69$\% (except in the 6301 line core, where it is $-40$\%) for the penumbra. For the quiet Sun, the corresponding correlations are stronger and range from $-73$\% to $-80$\% ($-59$\% at the 6301 line core). The values for the quiet Sun are similar to those found from bisector measurements close to the continuum in the 6301 line by \citet{2011arXiv1107.2586F}, $-78$\%, the lower values found by him for the penumbra are likely to be explainable by the much lower spatial resolution of the Hinode data and the different methods of analysis used by Franz and us. Irrespective of these differences, our conclusion about the existence of strong relations between the continuum intensity and the flow field does not rely on the {\gs precise} values of the correlation coefficients found; as stated earlier they are obvious in Figs.~\ref{fig:fig_o} and \ref{fig:fig_o2}. The correlation coefficients calculated here only {\gs quantify this relation in a simple way} that is easily testable against simulations. In the following we will analyze this in more detail, using an independent technique. We finally note (Table~\ref{tab:table_cd}) that the RMS velocities obtained for the penumbra are on the order of 65--70\% of those for the quiet Sun, such that they suffice to explain the penumbral heat flux \citep[c.f.,][for a more detailed discussion]{2011Sci...333..316S}.
 
\subsection{LOS velocity and magnetic field correlations}
\label{sec:vmcorr}

\begin{figure}%[bt!]
 \centering
%\includegraphics[bb=18 42 578 702,angle=-90,width=0.242\textwidth]{FIGS_C/fig_g_1c.eps}
%\includegraphics[bb=18 42 578 702,angle=-90,width=0.242\textwidth]{FIGS/fig_g_2.eps}
%\includegraphics[bb=18 42 578 702,angle=-90,width=0.242\textwidth]{FIGS/fig_g_4.eps}
%\includegraphics[bb=18 42 578 702,angle=-90,width=0.242\textwidth]{FIGS/fig_g_5.eps}
%bb=55 22 586 700,clip,
\includegraphics[bb=55 22 586 700,clip,angle=-90,width=0.242\textwidth]{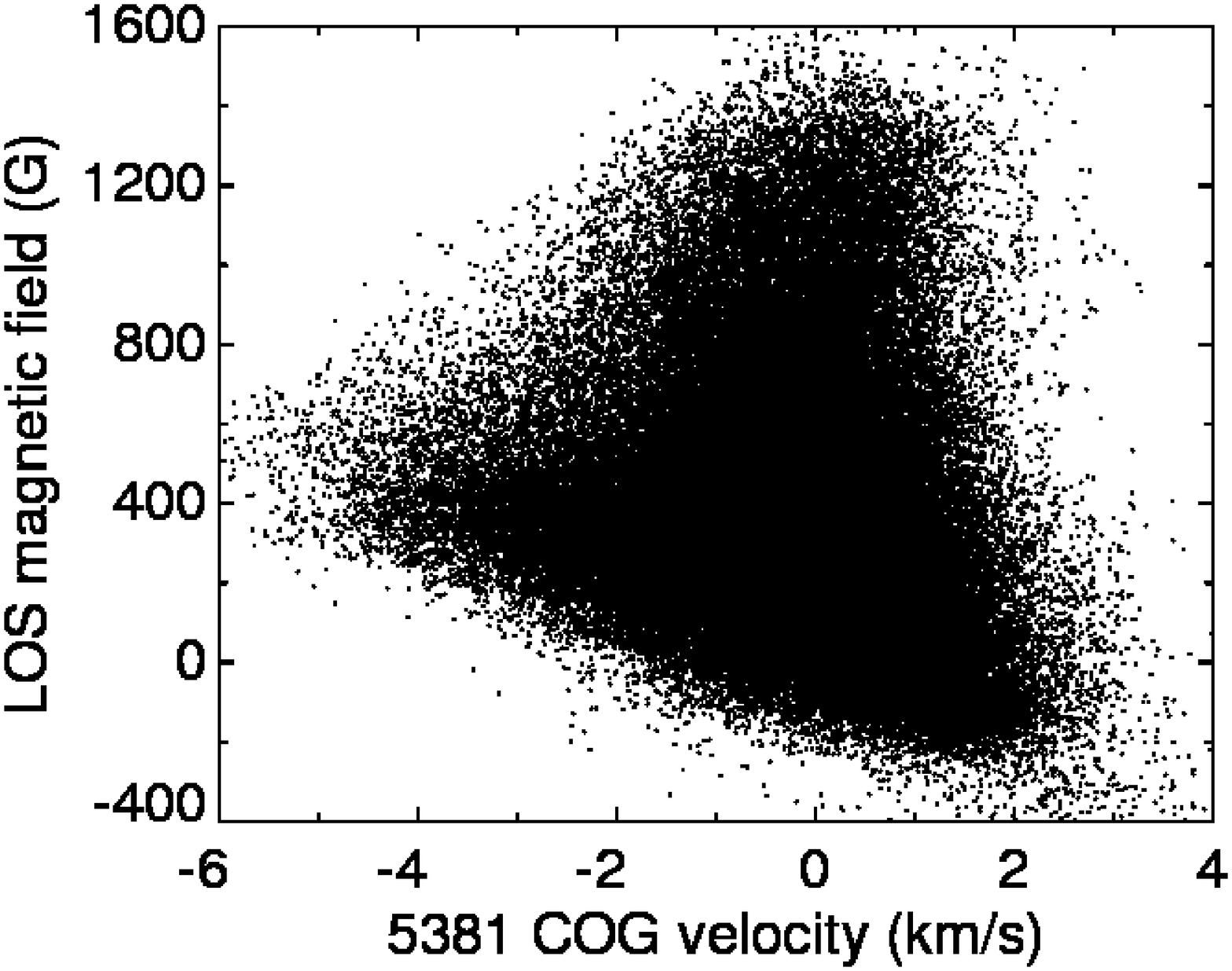}
\includegraphics[bb=55 22 586 700,clip,angle=-90,width=0.242\textwidth]{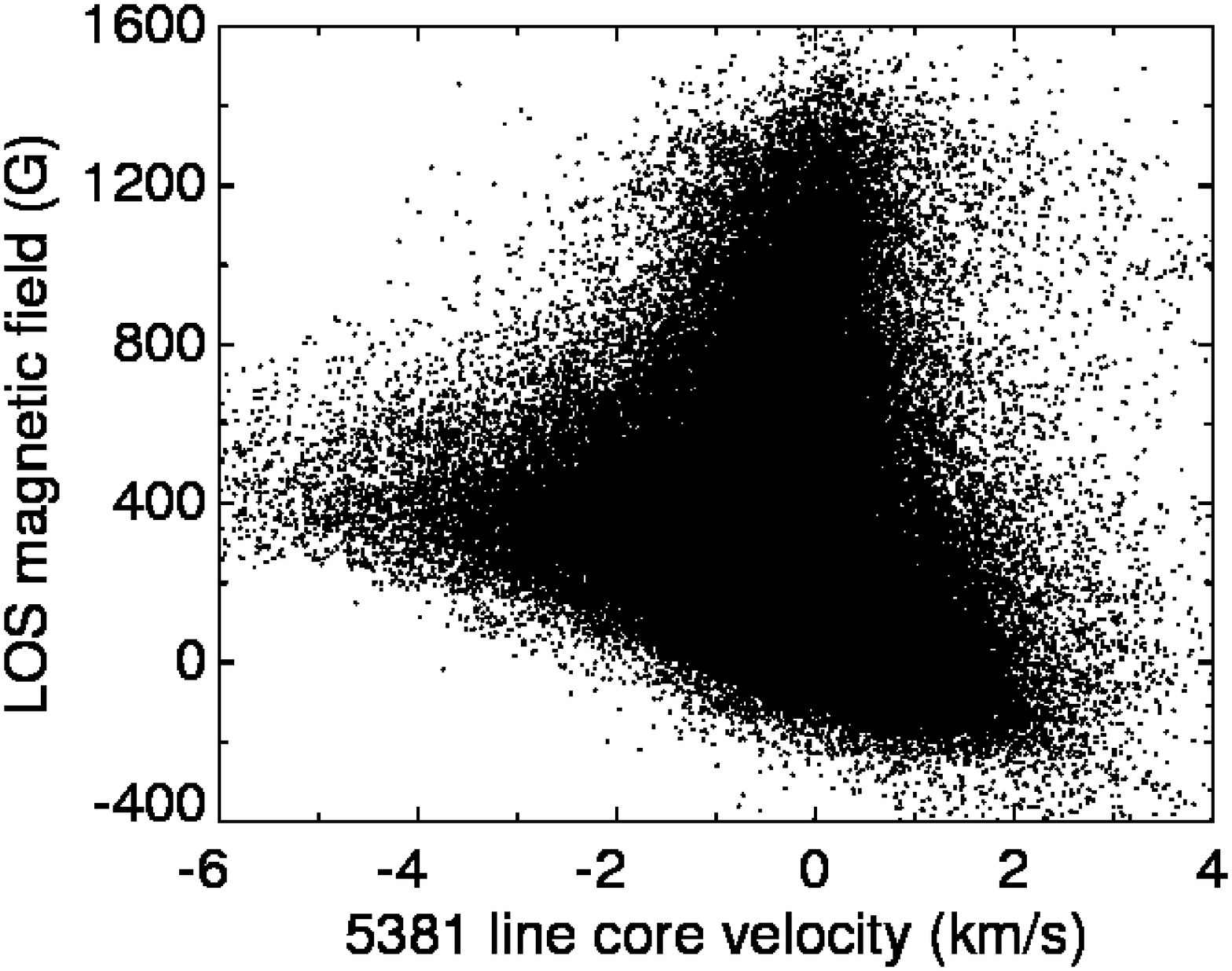}
\includegraphics[bb=55 22 586 700,clip,angle=-90,width=0.242\textwidth]{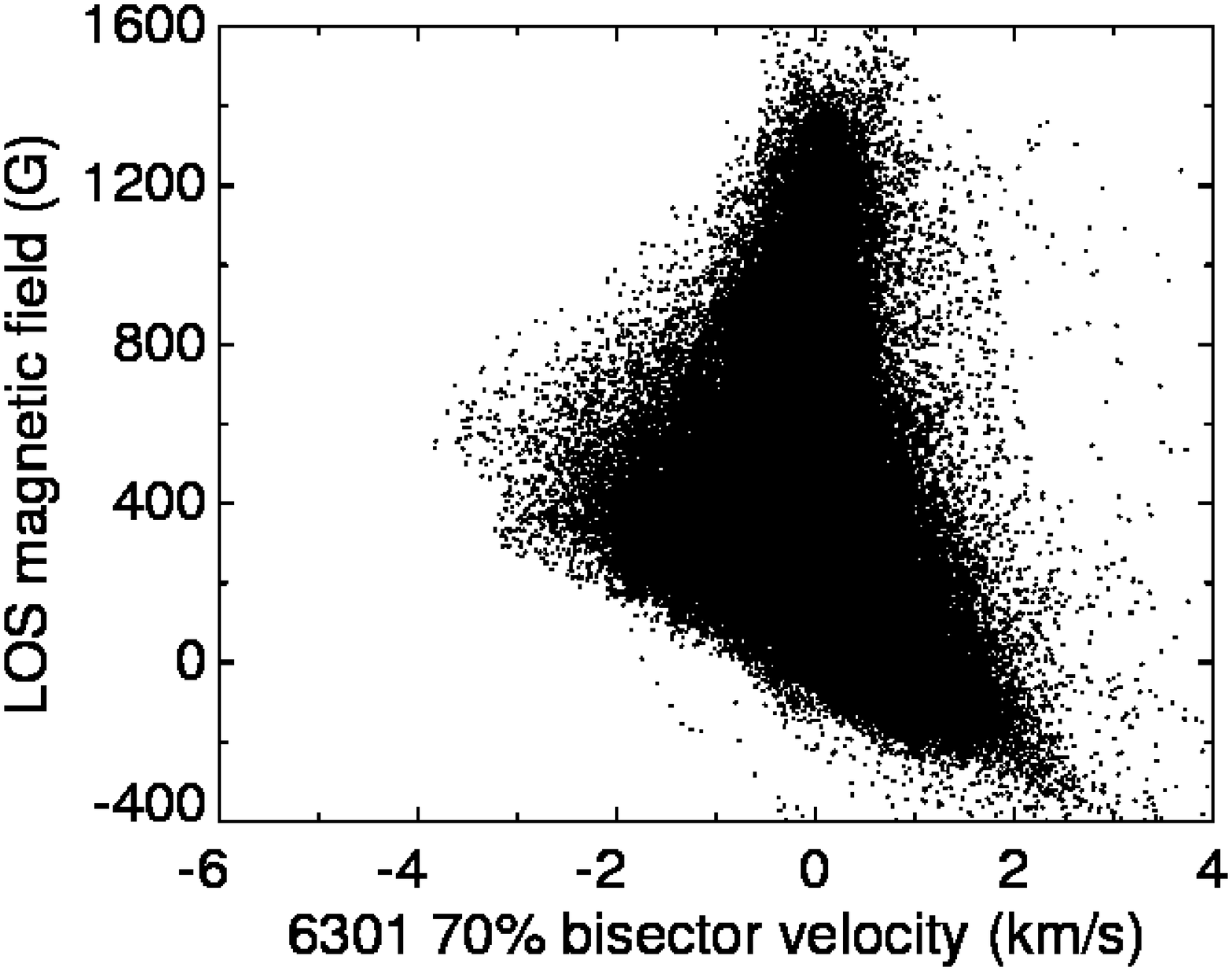}
\includegraphics[bb=55 22 586 700,clip,angle=-90,width=0.242\textwidth]{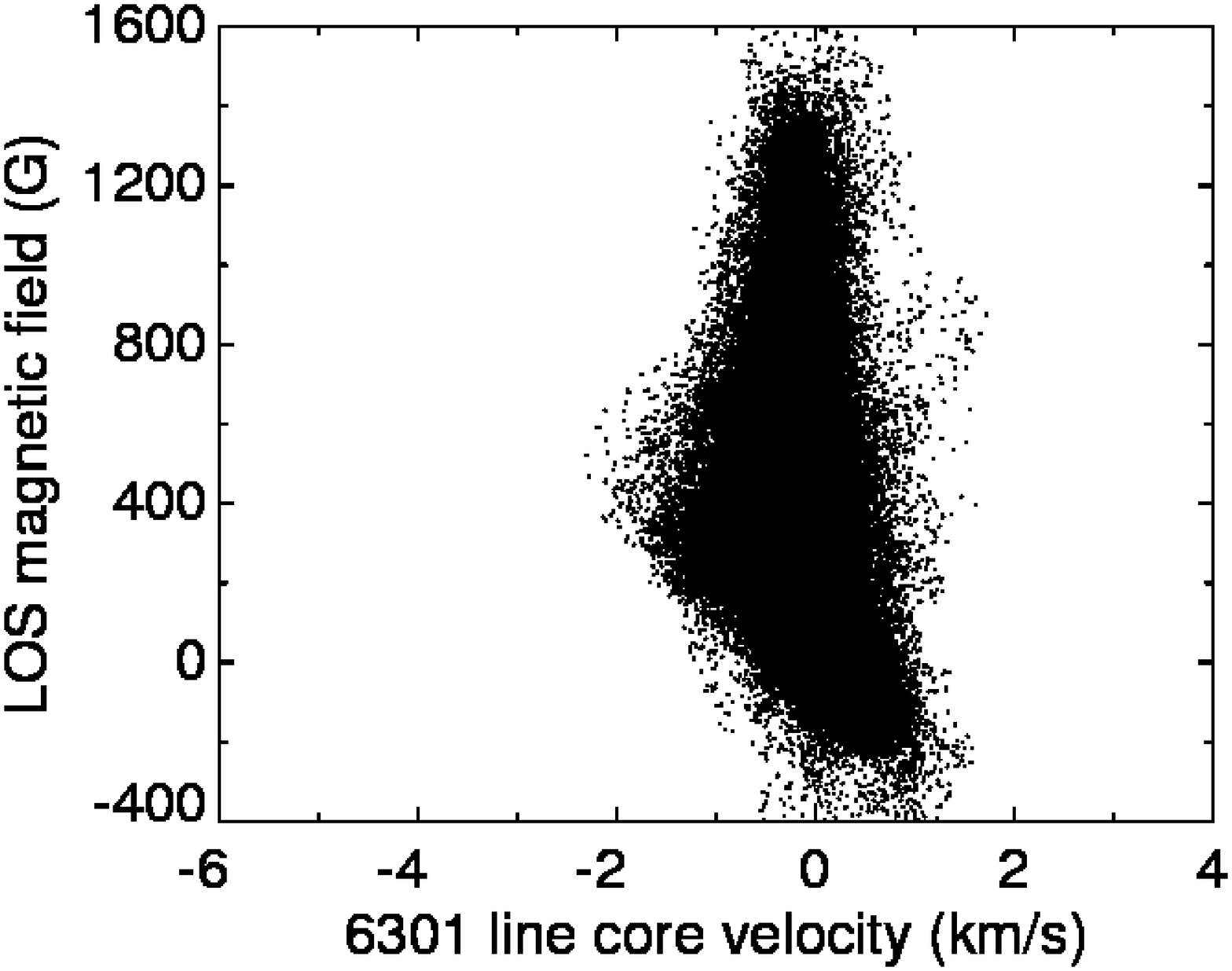}
\caption{Correlation between LOS velocity, measured with different methods in the 5380 and 6301 lines, and the measured LOS magnetic field. {\gs Neither of the two quantities plotted have been high-pass filtered.}}
\label{fig:fig_g}
\end{figure}

Figure~\ref{fig:fig_g} shows the correlation between the measured LOS magnetic field and LOS velocities {\gs in the interior penumbra}, measured with the COG method and from the line core of the 5380 line, and the 70\% bisector and from line core of the 6301 line. Note that the quantities shown in this figure have \emph{not} been spatially filtered. All correlations shown (and that for the 6301 COG velocity) have a roughly triangular shape with the ``corners'' characterized as follows: The strongest blueshifts corresponds to weak ($\sim$400 G) LOS magnetic field with the same polarity as the leading polarity of the spot. The strongest redshifts also correspond to weak LOS magnetic fields but with a polarity that is opposite to that of the spot. {\gs Almost all this opposite polarity field is in the limb-side penumbra, but a few such patches can also be found in the center-side penumbra.} Finally, the strongest LOS magnetic field (up to about 1600~G) systematically corresponds to small blue- and redshifts. %This correlation is explained by the influence of the magnetic field inclination and/or its strength on primarily the \emph{radial} flow component. 

In the lower-right panel of Fig.~\ref{fig:fig_h} we show the variation of the radial velocity with the strength of the high-pass filtered LOS component of the magnetic field. This plot was obtained by creating LOS magnetic masks as explained in Sect.~\ref{sec:filter} and making azimuthal fits of the measured LOS velocities (Sect.~\ref{sec:azfit}) to separately determine the vertical and radial velocities. This plot shows a dramatic variation of the radial velocity with the filtered LOS magnetic field: where $B_\text{LOS}$ is {\gs \emph{locally}} very weak, the radial velocity reaches up to 5~km\,s$^{-1}$ at the height of formation of the 5380 line, where it is very strong the radial velocity drops to less than 0.5~km\,s$^{-1}$! At heights corresponding to the formation of the 6301 line, the radial velocities are systematically {\gs lower}, but the overall trends are the same as for the 5380 line. {\gs The lower-left plot in Fig.~\ref{fig:fig_h} shows that the strength of the spatially filtered LOS magnetic field is closely related to the inclination of the magnetic field: where this quantity is small, the magnetic field is nearly horizontal and where it is strong, the magnetic field is inclined by about 50\degr{} to the vertical. Assuming that most of the variation in the (unfiltered) LOS magnetic field comes from inclination changes, rather than from field strength variations, these plots explain the triangular shape of the correlations shown in Fig.~\ref{fig:fig_g}. Where the LOS magnetic field is weak (nearly horizontal field), strong radial flows produce considerable spread in the measured LOS velocities, with the extremes corresponding to the lower-left and lower right ``corners'' of the triangles. Where the LOS magnetic field is very strong, corresponding to more vertical field (upper ``corner''), radial flows are virtually absent, such that the spread in the measured LOS velocities is limited to that of the (convective) vertical flows.}

%\cleardoublepage

\begin{figure}%[bt!]
 \centering
\includegraphics[bb=18 42 578 702,angle=-90,width=0.242\textwidth]{fig_h_3.eps}
\includegraphics[bb=18 42 578 702,angle=-90,width=0.242\textwidth]{fig_h_5.eps}
\includegraphics[bb=18 42 578 702,angle=-90,width=0.242\textwidth]{fig_h_6.eps}
\includegraphics[bb=18 42 578 702,angle=-90,width=0.242\textwidth]{fig_h_1.eps}
\caption{Variation of ``interior penumbra'' (radial zones 2--4) magnetic field and radial flow speed with the strength of the high-pass filtered LOS magnetic field, obtained from azimuthal fits. The top two plots show the variation of the vertical and horizontal components of the magnetic field, the lower-left the variation in magnetic field inclination. The lower-right plot shows the (remarkable!) variation of the radial flow velocity, measured with five different methods from the 5380 and 6301 lines. Plus symbols correspond to measurements in the 5380 line, squares to the 6301 line, lines drawn full to COG measurements, short dashes to line core, and long dashes to 70\% bisector (6301 line only)  LOS velocities. The vertical dashed lines correspond to the upper and lower thresholds respectively, for ``inter-spines'' and ``spines''.}
\label{fig:fig_h}
\end{figure}

\begin{figure*}%[bt!]
 \centering
\includegraphics[width=0.33\textwidth,clip]{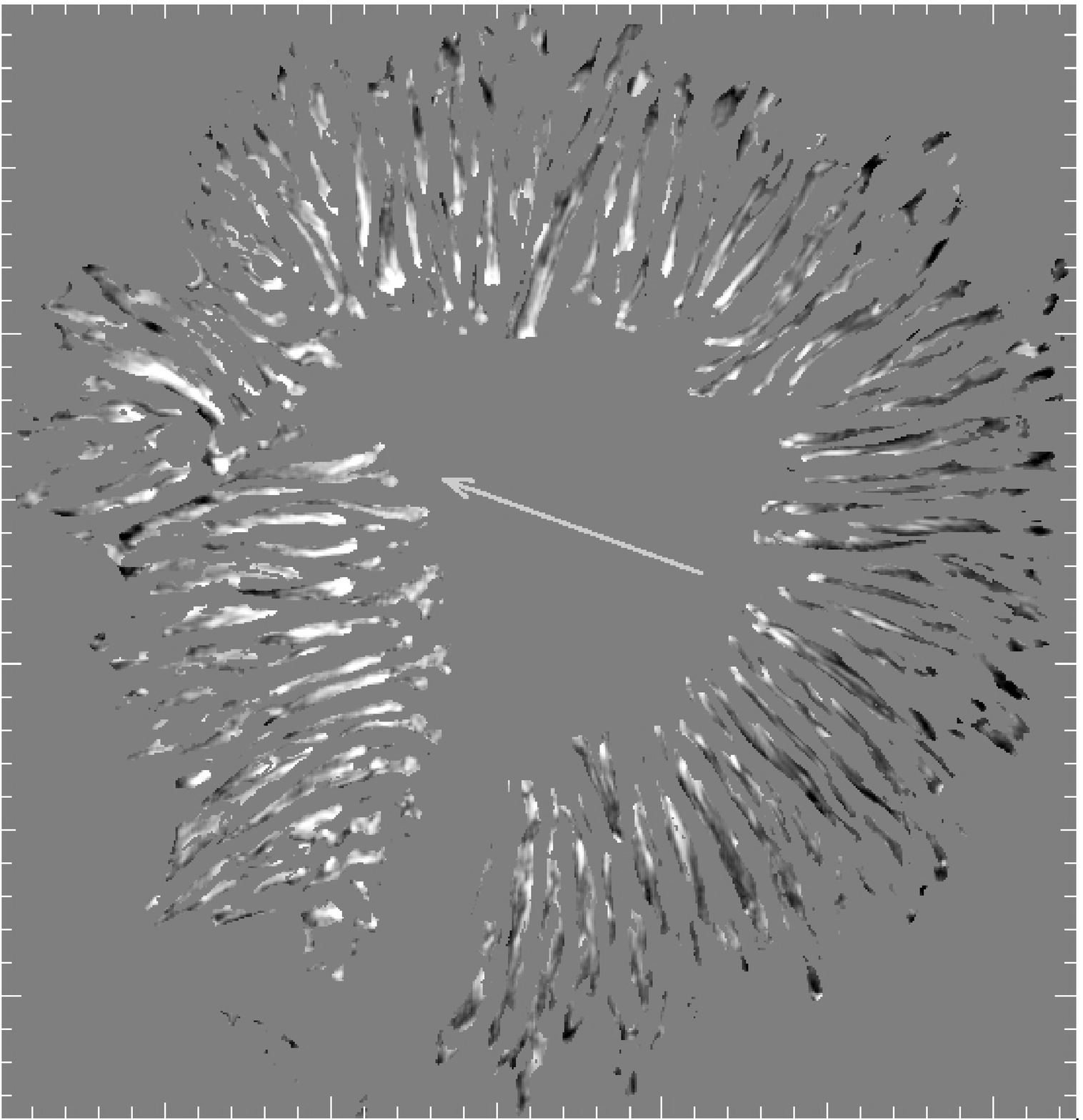}
\includegraphics[width=0.33\textwidth,clip]{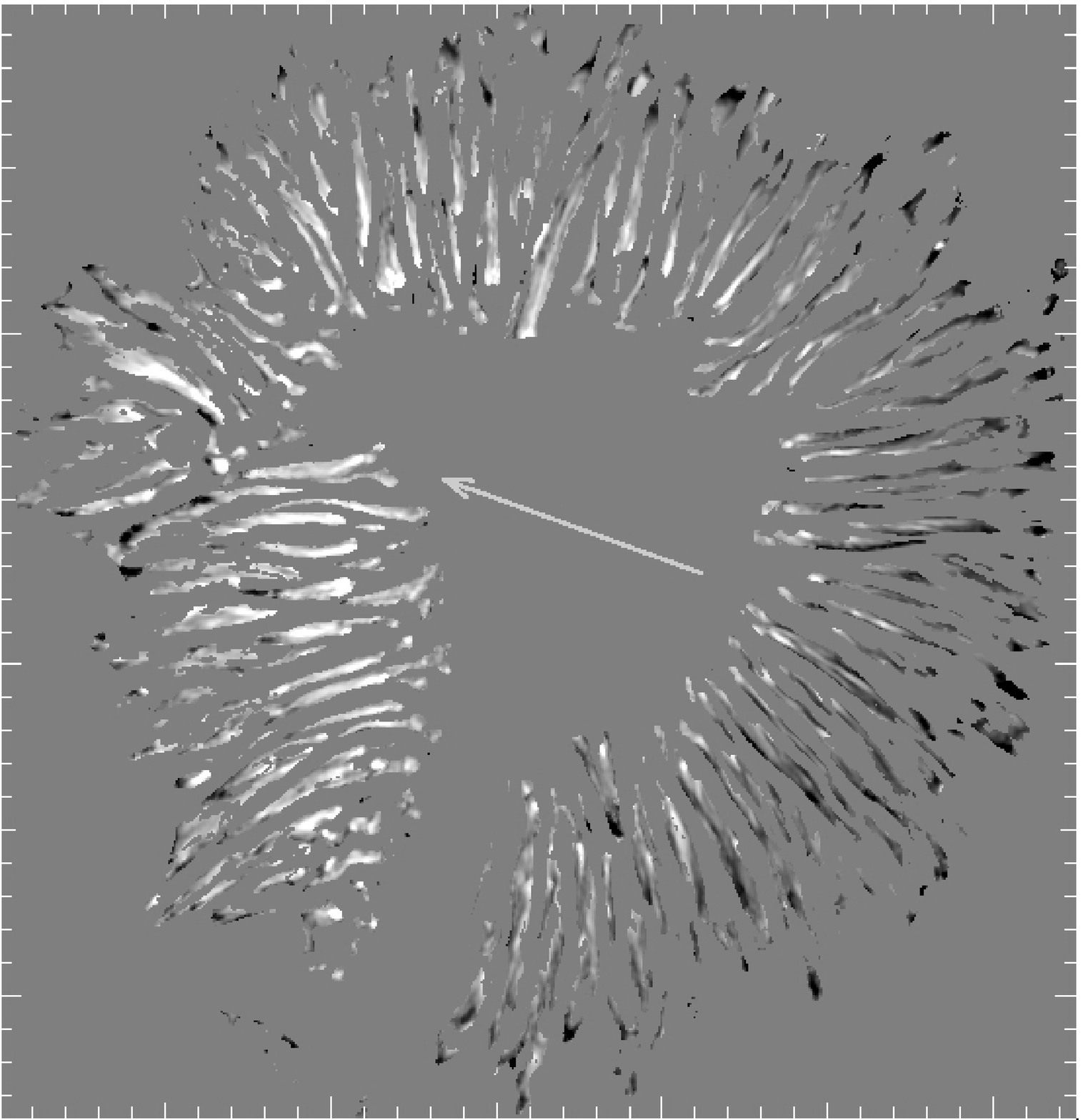}
\includegraphics[width=0.33\textwidth,clip]{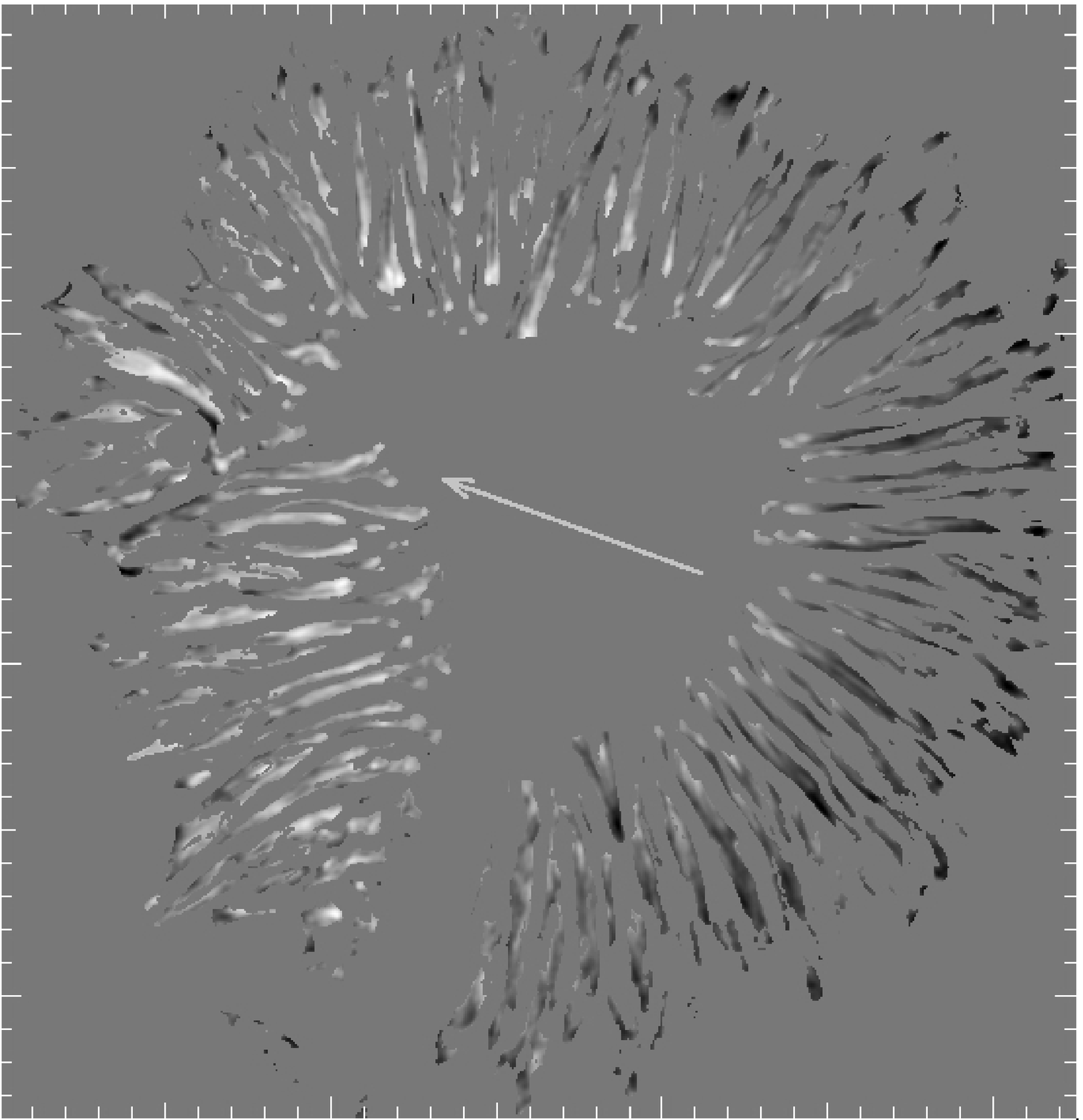}
\includegraphics[width=0.33\textwidth,clip]{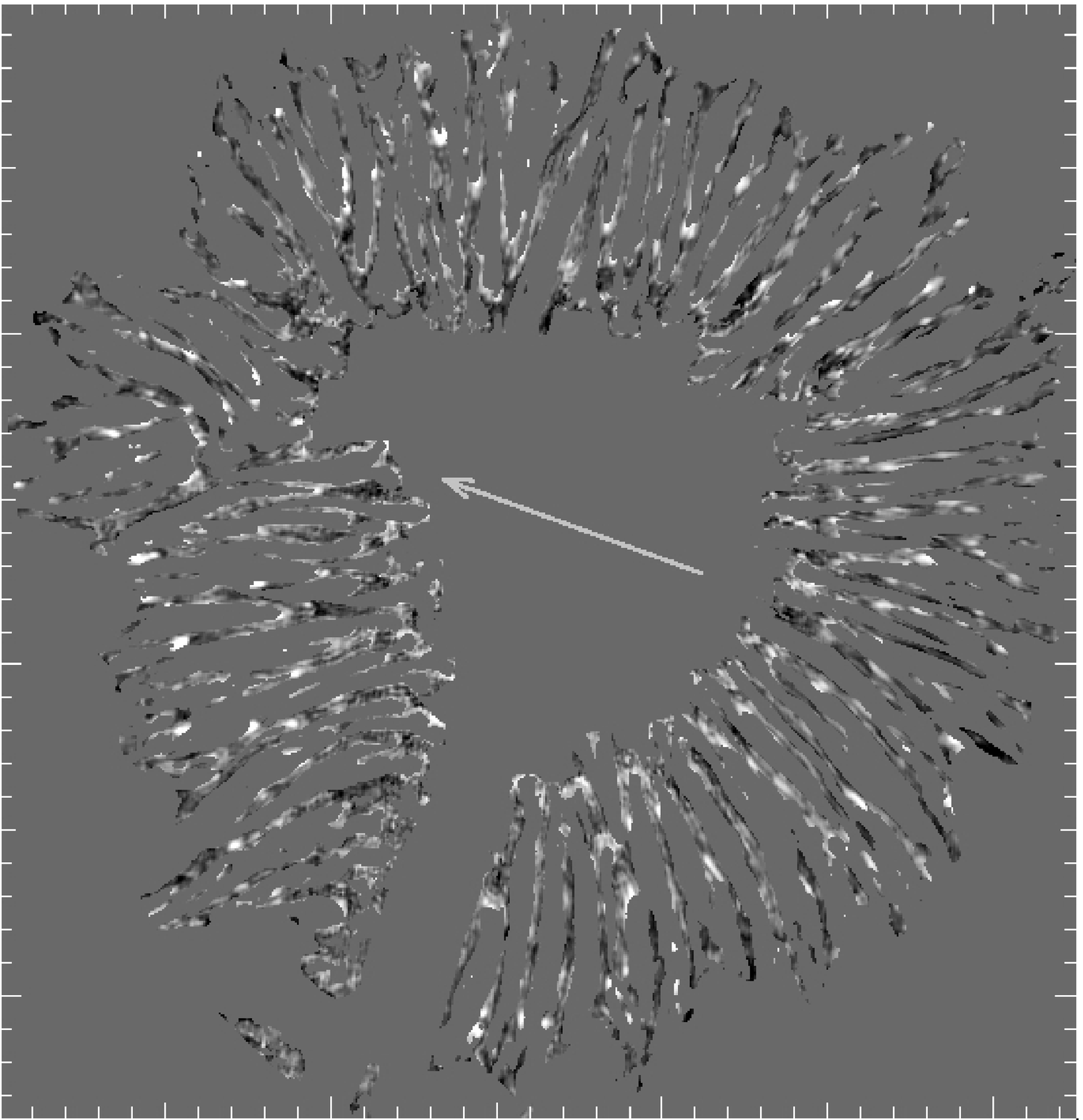}
\includegraphics[width=0.33\textwidth,clip]{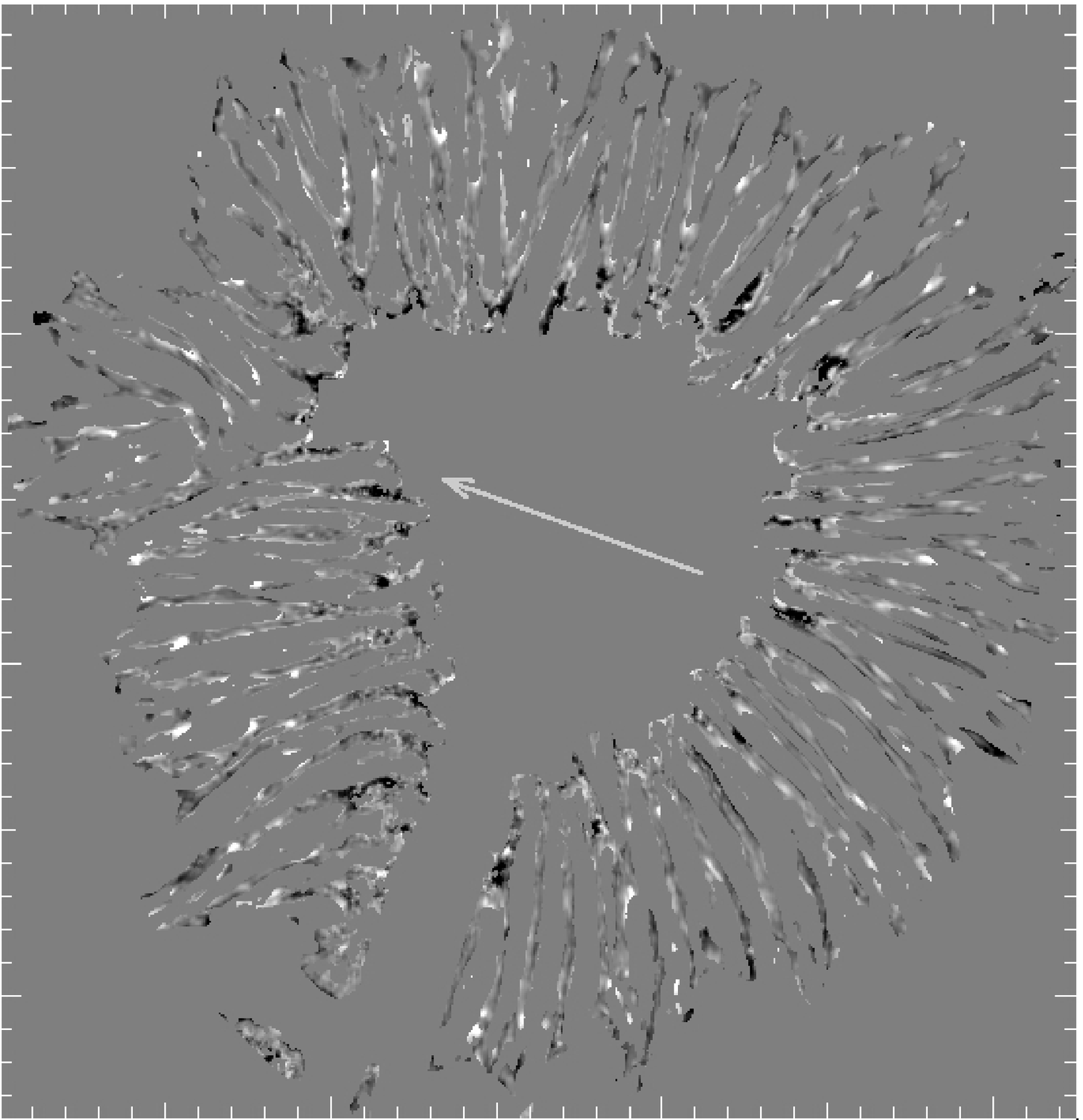}
\includegraphics[width=0.33\textwidth,clip]{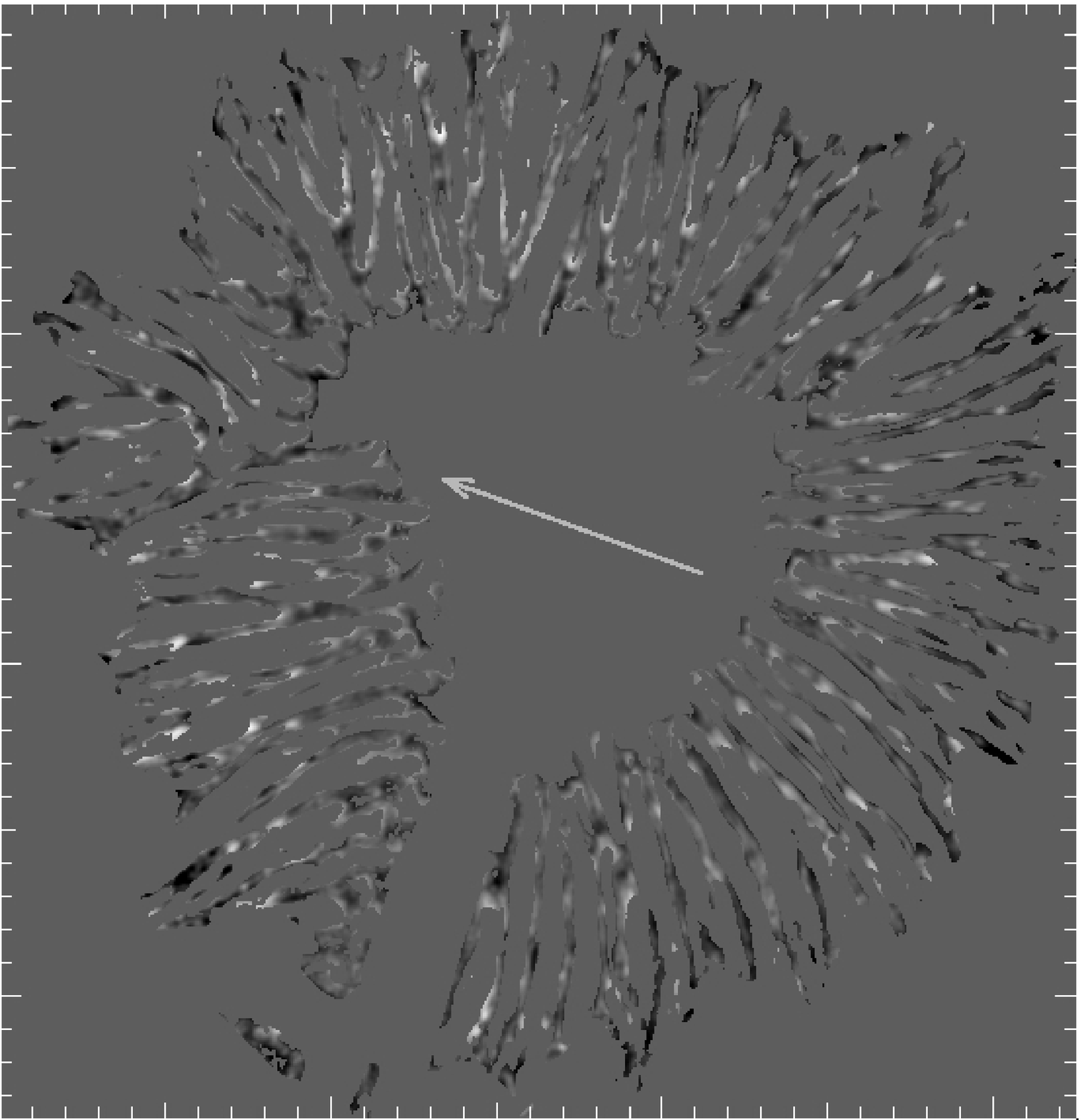}
\caption{Measured {\gs absolute (unfiltered)} LOS velocities within the inter-spine (upper row) and spine masks (lower row) shown in Fig.~\ref{fig:fig_k}, as obtained in the line core and with the COG method in the 5380 line and with the COG method in the 6301 line. The LOS velocities in the inter-spine mask show strong differences between the disk center and limb sides, demonstrating strong radial flows in the inter-spines, whereas no such differences are seen in the spines. Tick marks are at 1\arcsec{} intervals. The arrow points in the direction of Sun center.}
\label{fig:fig_l}
\end{figure*}

\begin{figure}%[!t]
 \centering
\includegraphics[bb=18 42 578 702,angle=-90,width=0.49\linewidth]{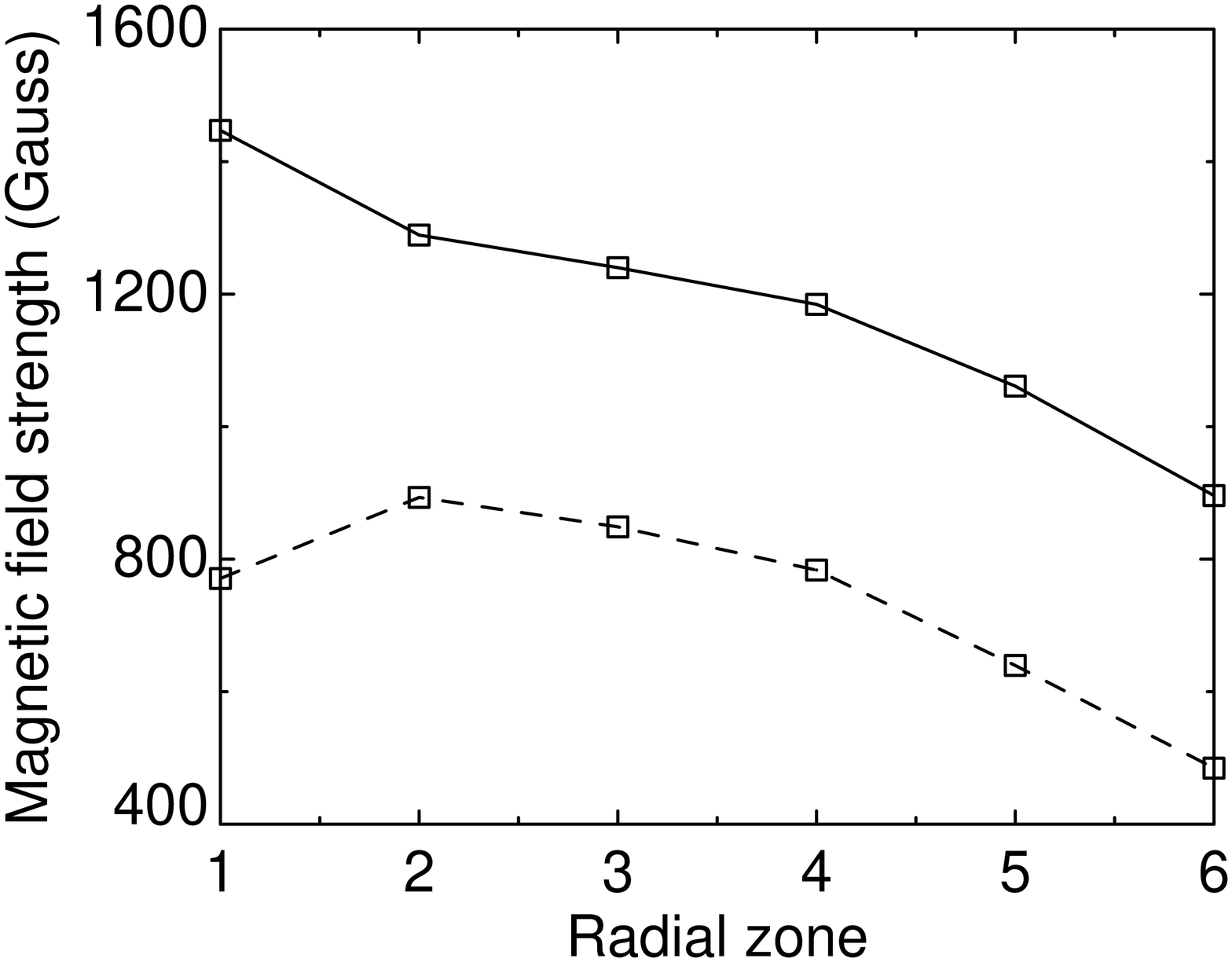}
\includegraphics[bb=18 42 578 702,angle=-90,width=0.49\linewidth]{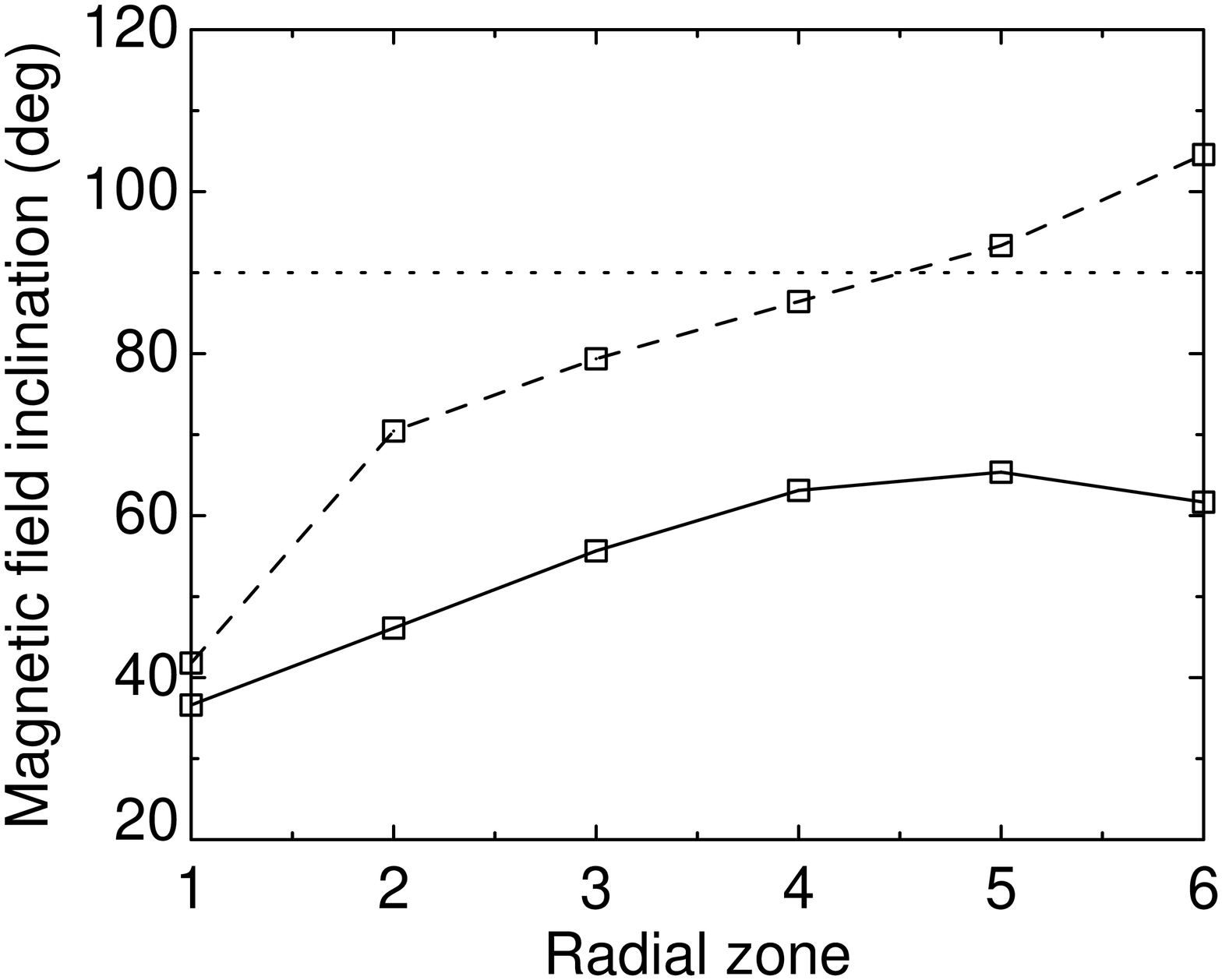}\\[2mm]
\includegraphics[bb=18 42 578 702,angle=-90,width=0.49\linewidth]{fig_a_1.eps}
\includegraphics[bb=18 42 578 702,angle=-90,width=0.49\linewidth]{fig_a_5.eps}\\[2mm]
\includegraphics[bb=18 42 578 702,angle=-90,width=0.49\linewidth]{fig_a_2.eps}
\includegraphics[bb=18 42 578 702,angle=-90,width=0.49\linewidth]{fig_a_6.eps}
\caption{Radial variation of properties of spines and inter-spines, as defined by the LOS magnetic masks, shown in Fig.~\ref{fig:fig_k}. The plots in the top row show field strengths and inclinations for spines (full) and inter-spines (dashed). Symbols in the lower four plots are as in Fig.~\ref{fig:fig_h}.}
\label{fig:fig_ab}
\end{figure}

In the remaining panels of Fig.~\ref{fig:fig_h} we show the results of fits of $B_\text{LOS}$ to the high-pass filtered $B_\text{LOS}$ map. These plots show that \emph{selecting} locally very weak LOS magnetic field corresponds to \emph{selecting} a component of the penumbral field that is relatively weak (field strength of about 850~G) and (not surprisingly) almost horizontal (inclination approximately~80\degr). Selecting the locally strongest LOS magnetic field corresponds to selecting the part of the magnetic field that is stronger (about 1200~G) and more vertical (inclination approximately 50\degr). These properties agree with those of spines and inter-spines \citep{1993ApJ...418..928L}, {\gs although the much lower spatial resolution of their data compared to ours means that it is not obvious that we are identifying the same penumbral structures. Nevertheless}, we can establish that for the observed sunspot, the strong horizontal (Evershed) flow prevails in the inter-spines and is virtually absent in the spines. This division of the penumbra into two main components with quite different magnetic and flow properties is consistent with numerous results obtained from inversions based on spectropolarimetric data at much lower spatial resolution than here \citep{1993A&A...275..283S,2000A&A...361..734M,2002A&A...381..668S,2003A&A...403L..47B,2004A&A...427..319B,2007ApJ...666L.133B,2007ApJ...671L..85T}, and is at the center of interpretations in terms of a ``flux tube'' {\gs (at a smaller scale than our spines)} and a ``background component''. The existence of these two fairly distinct populations is clearly illustrated in Fig.~\ref{fig:fig_l}, showing the 5380 COG and line core and the 6301 COG LOS velocity field within the inter-spine mask (top row) and within the spine-mask (bottom row). In the top row (inter-spines), we see clearly the signatures of the Evershed flow: strong blue-shifts on the disk center side and strong red-shifts on the limb side. In the spines, these signatures are missing entirely, showing that the Evershed flow is virtually absent. 

Figure~\ref{fig:fig_ab} shows the radial variation of the average properties determined for spines and inter-spines, indicating a magnetic field that dips down (inclination larger than 90\degr) in the inter-spines in the outer part of penumbra, consistent with earlier findings. {\gs In comparison, the plot of the average vertical velocity for the inter-spines shows a cross-over to positive velocities (downflows) further toward the center of the sunspot, clearly indicating that the {\gg average} magnetic field and flow field are not precisely aligned (a similar result was obtained from the 5380 Stokes $V$ and LOS velocities, see SOM Fig.~S10, {\gg lower-left panel). %In view of the simplistic COG estimates of the LOS magnetic field that neglects gradients with depth, we attach no particular significance to this descrepancy. 
\emph{Systematic} vertical} and radial flows are small in the spines, but the inter-spines show strong radial flows up to 5~km\,s$^{-1}$ in the interior penumbra and systematic upflows in the inner and downflows in the outer penumbra. Ignoring the existence of the small-scale intensity correlated vertical (convective) flows demonstrated earlier \citep{2011Sci...333..316S} and here, these flow patterns \emph{would be consistent with flows in arched flux tubes}, as concluded in numerous earlier interpretations of observations made at much lower spatial resolution \citep{1993A&A...275..283S,2000A&A...361..734M,2002A&A...381..668S,2003A&A...403L..47B,2004A&A...427..319B,2007ApJ...666L.133B,2007ApJ...671L..85T, 2009A&A...508.1453F} and numerical simulations \citep{1998A&A...337..897S,2002AN....323..303S}. {\gs As pointed out earlier \citep{2007ApJ...658.1357S, 2011Sci...333..316S} and in Sect.~\ref{sec:cavity}, such apparent large-scale penumbral flow patterns are likely to be at least partly the result of inadequate spatial resolution, giving effects similar to convective blue-shifts.}
%However, the present data leads to an entirely different interpretation.

In Fig.~\ref{fig:fig_e} we show the results from azimuthal fits (Sect.~\ref{sec:azfit})  of the measured LOS velocities made at different high-passed filtered 5380 continuum intensities, separately for spines and inter-spines. The fits include pixels only from the ``interior penumbra'', defined here as radial zones 2--4 (see Figs.~\ref{fig:fig_i}--\ref{fig:fig_i2} and Fig.~\ref{fig:fig_j}). We find {\gs \emph{strong vertical velocities that are roughly proportional to $\delta I_c$ both in the spines and inter-spines, with the darkest structures showing downflows of up to about 1~km\,s$^{-1}$ (except in the line core of the 6301 line)}. The brightest structures show upflows up to about 3~km\,s$^{-1}$ in the 5380 line and up to 1.5~km\,s$^{-1}$ in the 6301 line. \emph{Strong radial flows are found only in the inter-spines.}} The 5380 COG measurements corroborate the earlier line core measurements made in the same line \citep{2011Sci...333..316S}; here {\gs we detect similar clear} convective signatures also in the 6301 line. The top four panels in Fig.~\ref{fig:fig_e2} show considerable scatter around the fitted curves. This scatter shows only small variation with the azimuth angle and does not tend to disappear at $\phi=90\degr$ and $\phi=270\degr$, where contributions from radial flows should be absent. This clearly indicates that the scatter comes from fluctuations primarily in the \emph{vertical} flows that do not correlate with intensity. As concluded already {\gs from our data}, the correlation between continuum intensity and vertical velocity is {\gs somewhat} weaker in the penumbra than in the quiet {\gs Sun. The} same tendency is seen in simulations \citep{2011ApJ...729....5R}.

We make a few additional remarks relating to Fig.~\ref{fig:fig_e}: First, the \emph{radial} velocities obtained in the \emph{spines} show a clear tendency to increase with brightness for the brightest structures. We note that Figs.~\ref{fig:fig_o} and \ref{fig:fig_o2} show several examples of bright strong flows that are located just at the boundaries of the spines. Our threshold for $\delta B_\text{LOS}$ set to define spines is ``generous'' compared to that used by \citet{2011Sci...333..316S} based on the integrated 5380 Stokes $V$ data, where the spines occupied only 13.5\% of the interior penumbra. Here, we set the threshold such that spines occupy 33\% of the penumbra. Increasing this threshold decreases the radial flow speed for the the brightest structures in the spines, and also reduces the strength of the radial flows at lower intensities. Second, the roughly linear relation between $\delta I_c$ and the average \emph{vertical} velocity found will not change significantly if the zero-point calibrations of LOS velocities are changed by one or two hundred m\,s$^{-1}$. This will change the value of the strongest downflows and upflows by the same amount but much larger errors would be needed to ``remove'' the penumbral dark downflows. Such large errors would also imply that the penumbra has an average upward velocity of well over 0.5~km\,s$^{-1}$, in strong disagreement with earlier findings \citep[c.f,][]{2011arXiv1107.2586F}.

We finally note that the finding that the convective downflows are of similar strength in the 5380 and 6301 lines, while the upflows are much stronger in the 5380 line than in the 6301 line, suggests significant differences in their height variations. This is consistent with the inversions of \citet{2011arXiv1107.2586F}, based on 6301 line data obtained with Hinode from a sunspot close to disk center. He found that a typical upflow extend only up to heights corresponding to $\log(\tau)\approx-0.5$ whereas typical downflows extend up to heights corresponding to $\log(\tau)\approx-1.2$.  

\begin{figure}[!]
 \centering
\includegraphics[bb=18 42 578 702,angle=-90,width=0.49\linewidth]{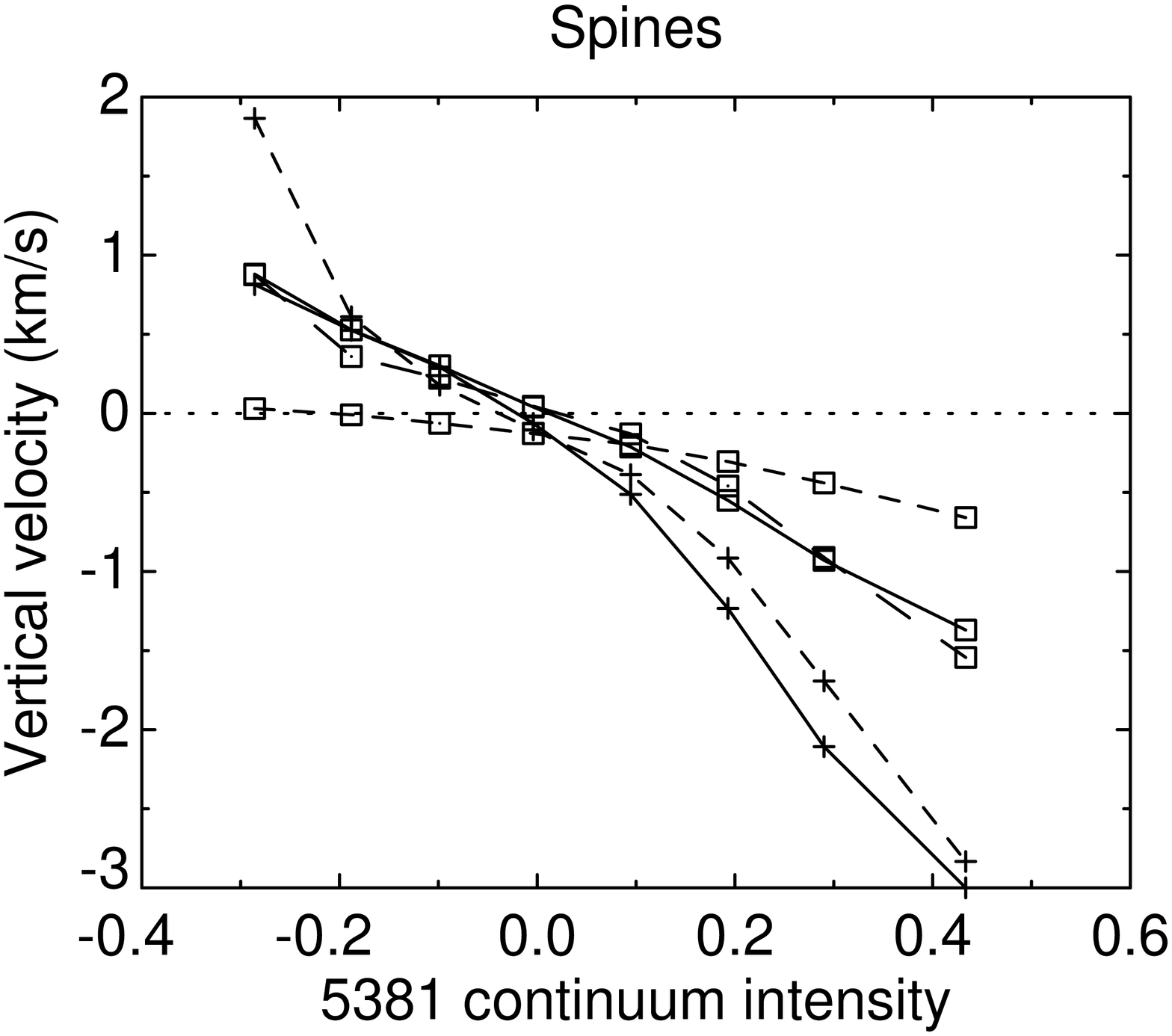}
\includegraphics[bb=18 42 578 702,angle=-90,width=0.49\linewidth]{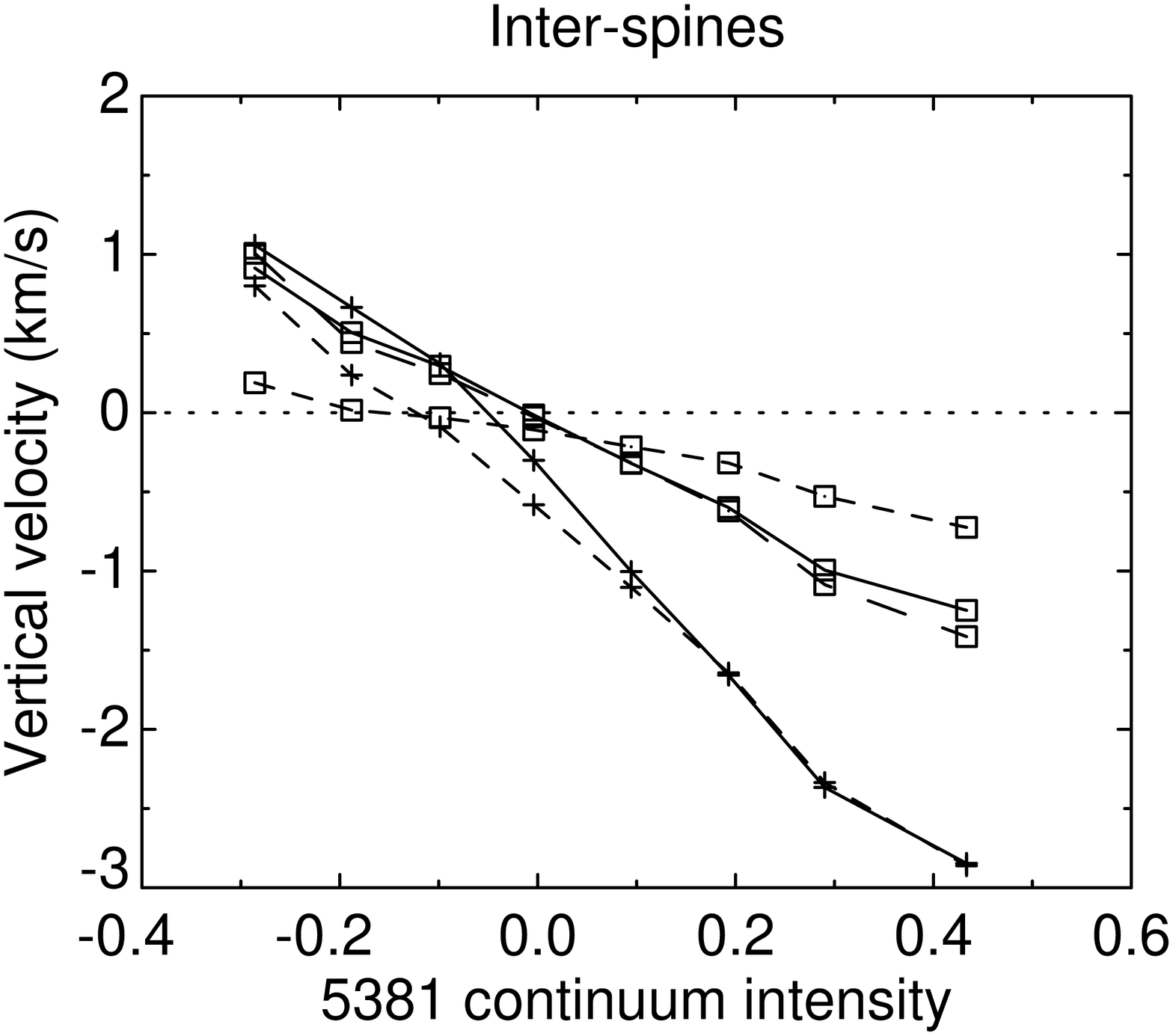}\\[2mm]
\includegraphics[bb=18 42 578 702,angle=-90,width=0.49\linewidth]{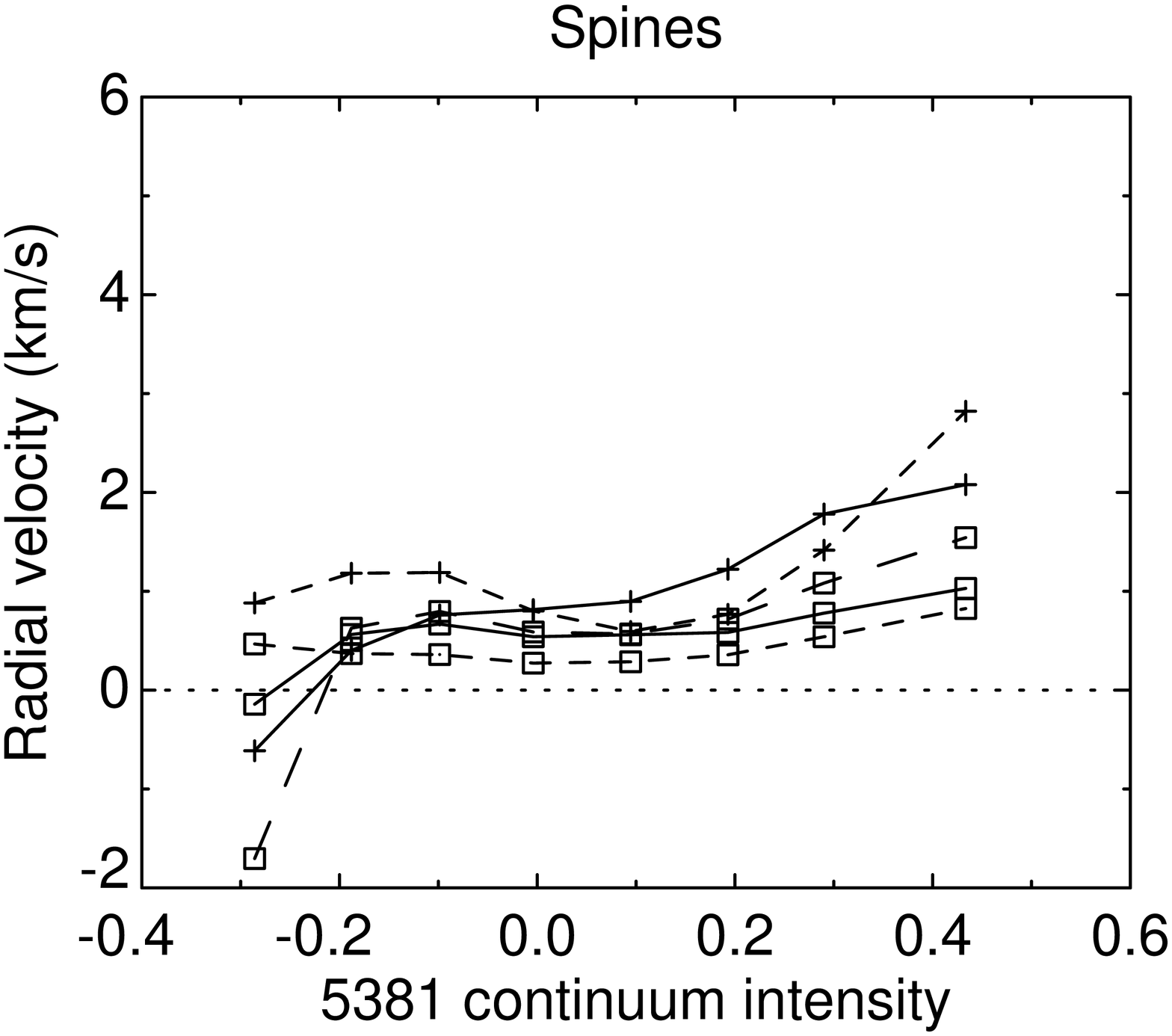}
\includegraphics[bb=18 42 578 702,angle=-90,width=0.49\linewidth]{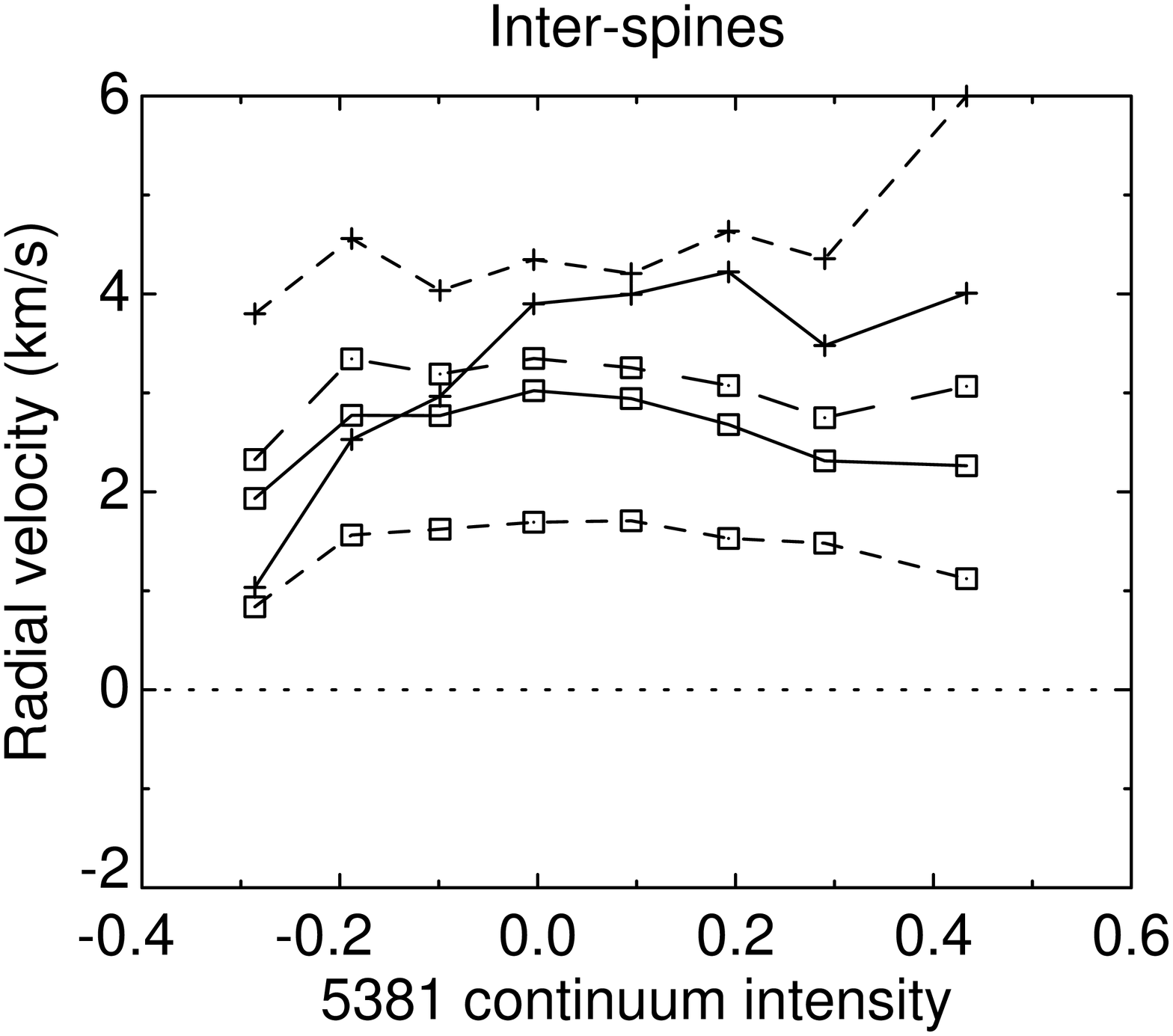}
\caption{Variation of vertical (top) and radial (bottom) velocities with high-pass filter \ion{C}{i} 5380 continuum intensity for spines (left) and inter-spines (right) in the interior penumbra (radial zones 2:4). Plus symbols correspond to measurements in the 5380 line, squares to the {\gs 6301 line}, lines drawn full to COG measurements, short dashes to line core, and long dashes to 70\% bisector (6301 line only) LOS velocities.}
\label{fig:fig_e}
\end{figure}

{\gs \subsection{Effects of straylight on measured vertical/radial flows}
Due to the strong weakening of the \ion{C}{i} 5380 line in dark granular and penumbral structures, straylight compensation is crucial for detecting dark convective downflows in the penumbra in this line (see SOM). The effects of straylight are less dramatic for measurements of LOS velocities in the \ion{Fe}{i} 6301 line. Figure~\ref{fig:fig_e3} compares the vertical and radial velocities obtained in this line with and without straylight compensation. For these fits, we included the entire interior penumbra (spines, inter-spines and the intermediate population). Except for the darkest structures, the radial velocities are similar with and without straylight compensation and (as expected) intermediate to those of spines and inter-spines, shown in Fig.~\ref{fig:fig_e}. The strongest vertical bright upflows are reduced in strength by only a small amount (from 1.5~km\,s$^{-1}$ to 1.3~km\,s$^{-1}$) without straylight compensation, but the darkest downflows are reduced in strength by a factor of 2 to 0.4--0.5~km\,s$^{-1}$. Given our estimated uncertainty of 150~m\,s$^{-1}$, we can thus establish the existence of the dark downflows from the 6301 line even without straylight compensation, but with reduced margin for error in the zero-point calibration. We conclude that both the high spatial resolution of the SST and compensation for straylight is needed for reasonably accurate estimates of the strengths of the dark convective downflows in the interior penumbra.}
\begin{figure}[!]
 \centering
\includegraphics[bb=18 42 578 702,angle=-90,width=0.49\linewidth]{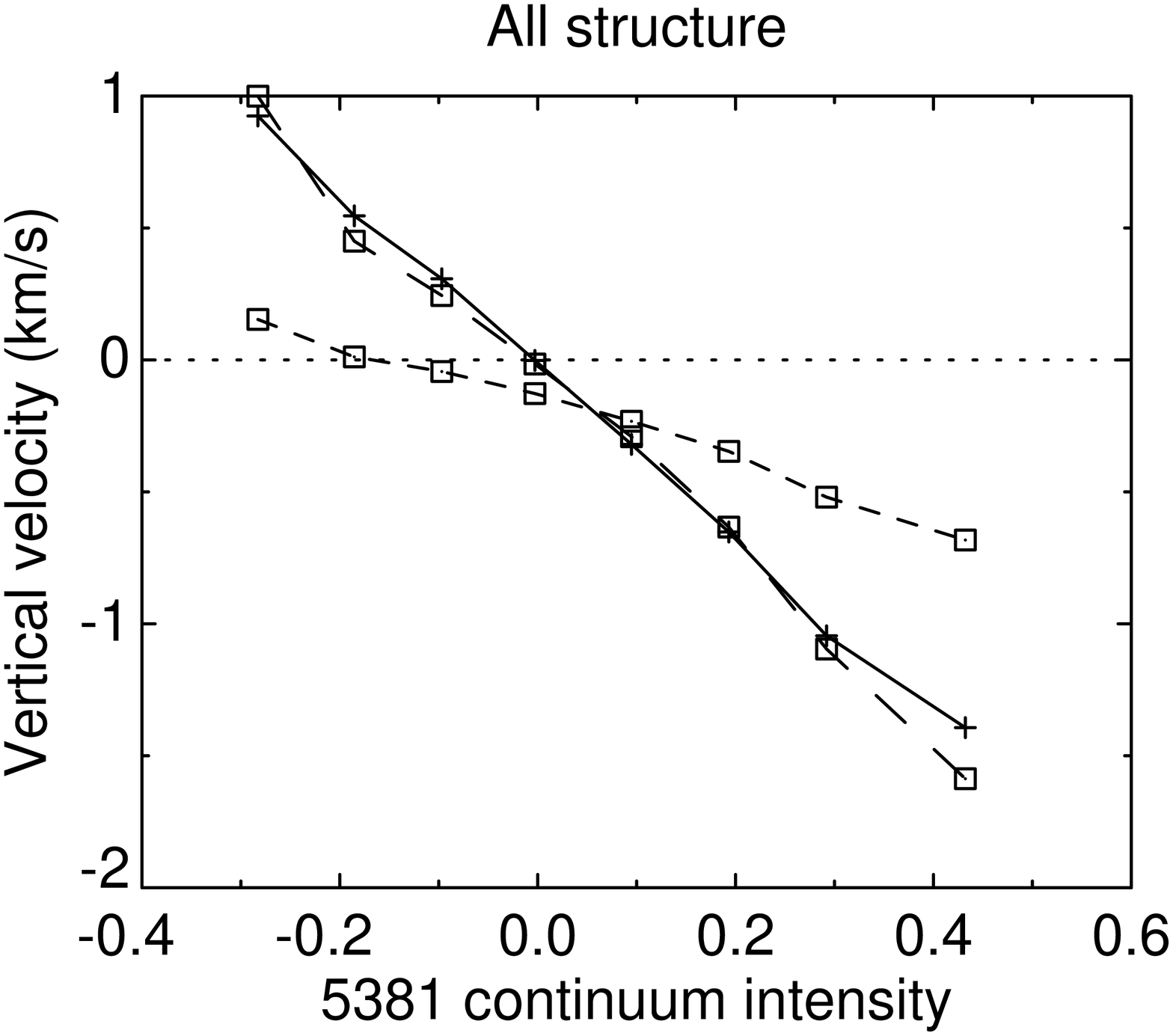}
\includegraphics[bb=18 42 578 702,angle=-90,width=0.49\linewidth]{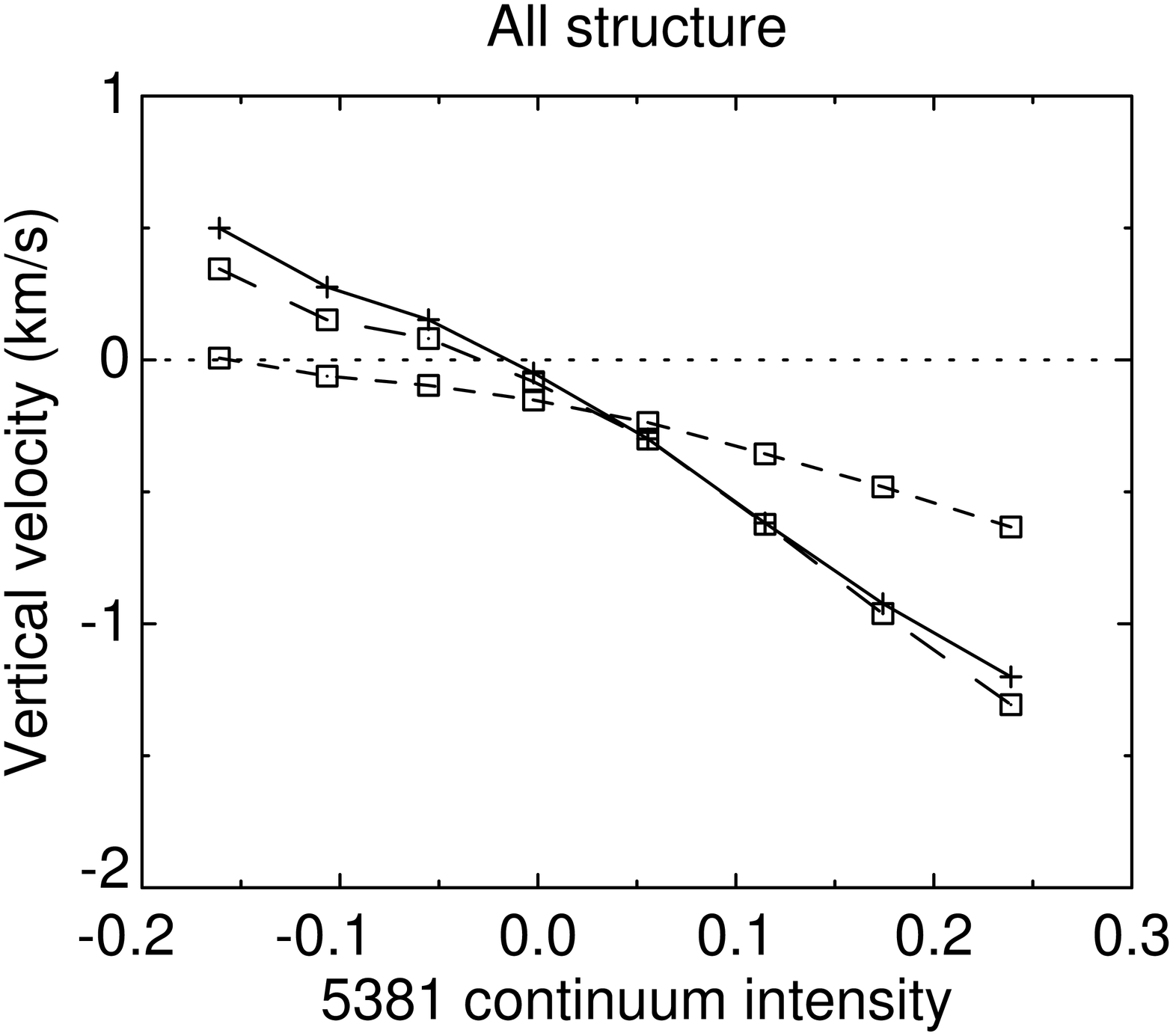}\\[2mm]
\includegraphics[bb=18 42 578 702,angle=-90,width=0.49\linewidth]{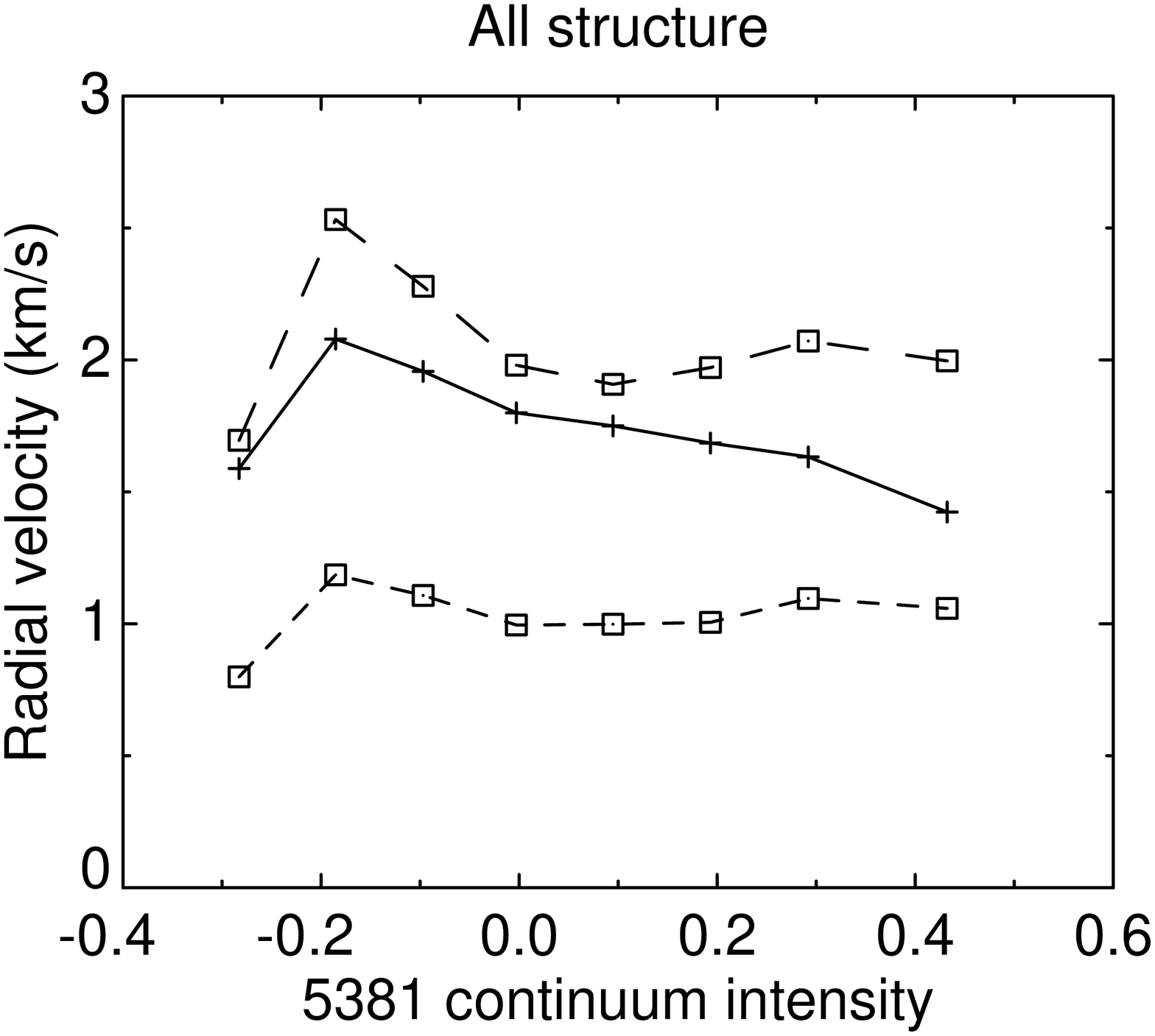}
\includegraphics[bb=18 42 578 702,angle=-90,width=0.49\linewidth]{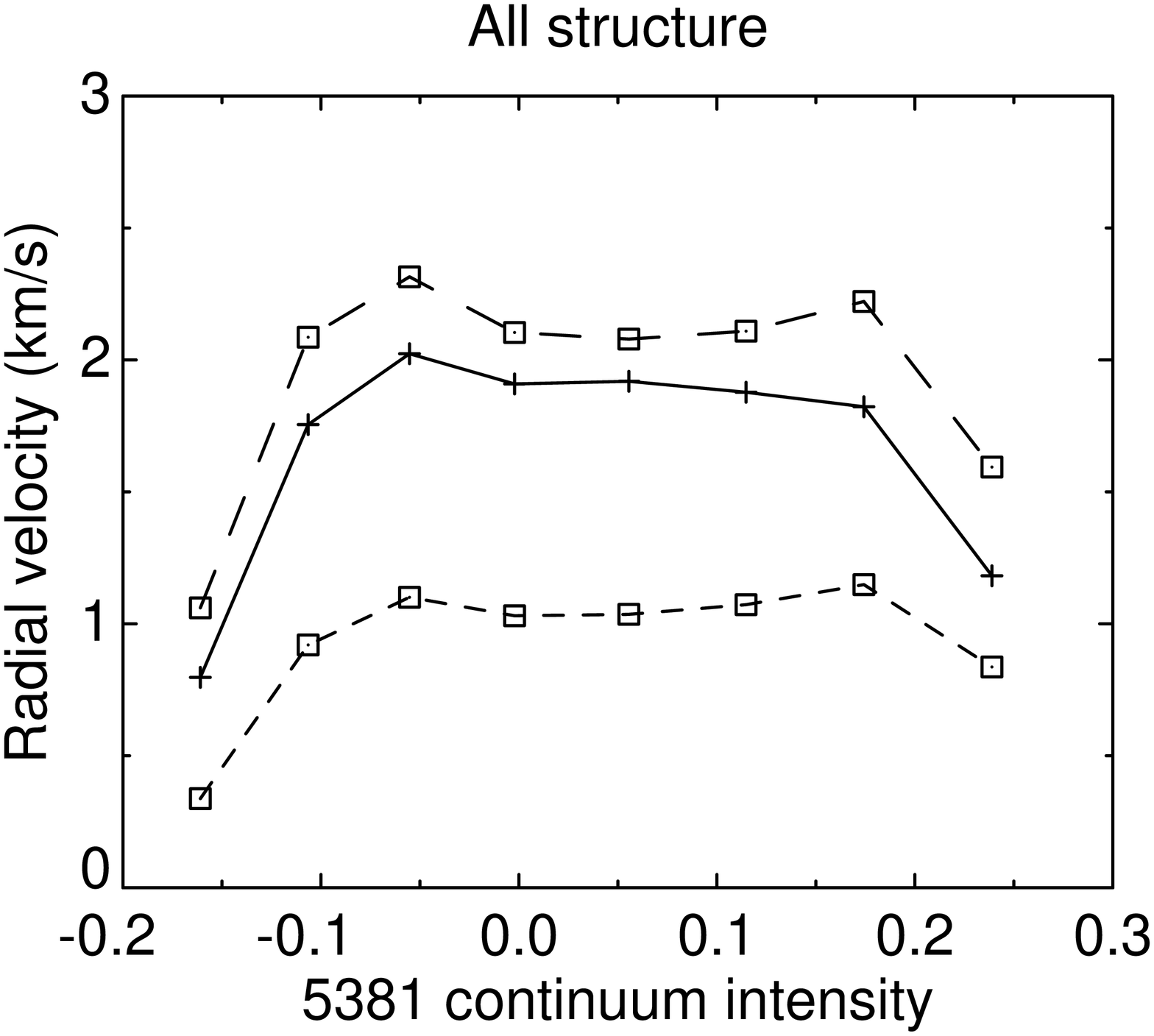}
\caption{Comparison of vertical (top) and radial (bottom) velocities measured in the 6301 line, with (left column) and without (right column) straylight compensation. All structure in the interior penumbra (spines, inter-spines and the intermediate population) were included in the fits. Lines drawn full correspond to COG measurements, short dashes to the line core, and long dashes to the 70\% bisector LOS velocities.}
\label{fig:fig_e3}
\end{figure}

\section{Conclusions}
\label{sec:conclusions}

We have combined the measurements of penumbral LOS velocities in the 5380 line core reported on recently \citep{2011Sci...333..316S} with LOS velocity and magnetic field measurements in the 6301 line as well as COG LOS velocity measurements in the 5380 line. These observations were obtained with SST/CRISP at the highest spatial resolution (0\farcs14 at 538~nm and 0\farcs16 at 630~nm) attainable today for spectropolarimetric data. By comparing observed quiet Sun RMS continuum intensities and velocities to values established from simulations of convection, we estimate the level of straylight in our data. Constraints on the FWHM of the straylight PSF are obtained from the minimum umbral intensity, which is low (15--18\%) in the observed data. By using these constraints, we compensate the data for spatial straylight \citep{2011Sci...333..316S}.

We assume that the measured small-scale fluctuations in intensity and LOS velocity across penumbral filaments do not vary systematically with azimuth angle at constant radial distance from the inner boundary of the penumbra. Azimuthal fits of the measured LOS velocities can then be used to determine statistical properties of vertical and radial flows separately \citep{1952MNRAS.112..414P}.
We have extended this widely used method \citep[e.g.,][]{1964ApNr....8..205M,1993ApJ...403..780T,2000A&A...358.1122S,2004A&A...415..717T,2005A&A...436.1087L,2006A&A...453.1117B,2007ApJ...658.1357S} by making azimuthal fits \emph{separately} for different (spatially filtered) continuum intensities \citep{2011Sci...333..316S}. These fits are made for the \emph{interior} penumbra (radial zones 2--4), excluding its innermost and two outermost radial zones \citep[note that the previous analysis included radial zones 2--5, c.f. SOM in ][]{2011Sci...333..316S}. {\gs We note that \citet{2007ApJ...658.1357S} made azimuthal fits by assuming linear relations between local intensity fluctuations and vertical/horizontal velocities. These fits demonstrated a clear correlation between intensity and the \emph{vertical} velocities in the penumbra, in the sense expected for convection. However, this method cannot be used to draw conclusions about whether or not the dark component of the penumbra on the average is associated with actual downflows ($v_z > 0$).} 

{\gs Our} recent analysis of line core measurements in the 5380 line demonstrated a roughly linear relation between the penumbral continuum intensity and the vertical velocity (SOM, Fig. S10). We also found that {\gs \emph{the darkest structures on the average show downflows of about 1~km\,s$^{-1}$}}, whereas the brightest structures show upflows of up to 3~km\,s$^{-1}$ in the layers corresponding to the formation height of the 5380 line \citep{2011Sci...333..316S}. Here, we find similar downflow velocities in the dark structure but only about 1.5~km\,s$^{-1}$ bright upflows in the wings of the 6301 line. This analysis corroborates the results obtained, and conclusions drawn from the 5380 line core measurements \citep{2011Sci...333..316S}, but indicates that the strong bright upflows seen in the 5380 line do not extend to the heights of the 6301 line. %the relation found between intensity and vertical velocity constitutes clear evidence of penumbral convection. 
We also find that the convective signature for vertical flows is the same irrespective of whether the magnetic field is relatively vertical and strong (in the spines) or weaker and more horizontal (in the inter-spines). However, strong radial outflows are found only in the inter-spines. {\gs This suggests the existence of either two distinct modes of (magneto-) convection simultaneously in the penumbra, where the strong Evershed flow is related to convection only in the more horizontal and weaker magnetic field structures in the inter-spines, or a more gradual change of the horizontal flow topology with magnetic field inclination.}

{\gs Correlations between local fluctuations in intensity and LOS velocity in the penumbra in the sense expected for convection have been reported repeatedly in the literature since 1969 \citep[e.g.,][]{1969SoPh...10..384B, 1993ASPC...46..192S, 1999A&A...349L..37S, 2000A&A...364..829S, 2007ApJ...658.1357S, 2009A&A...508.1453F, 2011arXiv1107.2586F}. However, the interpretation of these observations in terms of convective flows is not without ambiguity. At low spatial resolution, the large-scale structure of these flows appears to be that of an upflow of bright structure in in the inner penumbra and a downflow of dark structure in the outer parts. This flow pattern has been interpreted as support for flux tube models \citep{1999A&A...349L..37S, 2000A&A...364..829S, 2009A&A...508.1453F, 2011arXiv1107.2586F}. 

A major objection to the interpretation of these flows as being of convective origin is the absence of evidence of dark \emph{downflows} in the interior body of the penumbra \citep{2009A&A...508.1453F, 2011arXiv1107.2586F}. In our analysis, we therefore exclude the outermost penumbra, where dark downflows have been reported repeatedly, from our analysis. We also exclude the innermost part of the penumbra, where bright upflows are omnipresent and have been interpreted as the footpoints of penumbral flux tubes.} The discovery of dark (convective) \emph{downflows} in the penumbra reported recently \citep{2011Sci...333..316S,2011ApJ...734L..18J} and the correlation between intensity and vertical flows found from the 5380 line \citep{2011Sci...333..316S} in the {\gs \emph{ interior}} penumbra, is confirmed here for the 6301 line and for COG velocities in the 5380 line. The presence of intensity correlated upflows and downflows at roughly the same radial distance in the \emph{interior} penumbra contradicts earlier interpretations in terms of embedded flux tubes. The present observations, combined with {\gs earlier past and recent indications from correlations between penumbral brightness and LOS velocities \citep[e.g.,][]{1969SoPh...10..384B, 2007ApJ...658.1357S}}, and observed dynamics of penumbral filaments \citep{2007Sci...318.1597I,2008A&A...488L..17Z,2010A&A...521A..72S}, provides overwhelming observational evidence for interpreting the filamentary structure, dynamics, heat flux, magnetic field topology and Evershed flow of the penumbra as due to convection, rather than in terms of embedded flux tubes. This interpretation is consistent with theoretical arguments \citep{2006A&A...447..343S,2006A&A...460..605S} as well as numerical simulations \citep{2007ApJ...669.1390H,2008ApJ...677L.149S,2009Sci...325..171R,2009ApJ...691..640R,2011ApJ...729....5R}. 

{\gs Our observations could also be interpreted as evidence for convection rolls \citep{1961ApJ...134..289D}, as has been suggested previously \citep{1969SoPh...10..384B, 2007Sci...318.1597I, 2008A&A...488L..17Z, 2010mcia.conf..210S}. However, objections against this interpretation were raised by \citet{2009SSRv..144..229S} on the following grounds: the convection rolls are expected to be associated with a nearly horizontal magnetic field, whereas e.g., the observations of \citet{2007Sci...318.1597I} refer to the inner penumbra, where the magnetic field has a strong vertical component. Also, to sustain the radiative output of a filament over its life time of on the order of an hour \citep{2007A&A...464..763L}, the convective upflows must persist to depths much larger than a few 100 km, whereas the convective roll is a shallow phenomenon \citep{1961ApJ...134..289D}. %Our data are consistent with ``Global'' properties of the inter-spines, with upflow in the inner penumbra and downflows in the outer penumbra mimic a flux tube (or collection of flux tubes) only when the spatial resolution is inadequate for us to observe the small-scale intensity correlated vertical flows and associated filament dynamics. 
%The approximately linear relations found between continuum intensity and vertical velocities suggest that continuum images are meaningful proxies for vertical flows and that downflows are

The combination of high spatial resolution and straylight compensation of our data results in RMS velocities for the penumbra that are on the order of 65--70\% of those for the quiet Sun, such that they suffice to explain the penumbral heat flux \citep[c.f.,][for a more detailed discussion]{2011Sci...333..316S}. This also strongly suggests that the fundamental scales at which this convection occurs is resolved in our data.}

Our data also contain a wealth of information about the toplogy of the convective flows and their magnetic fields. In future work, we will investigate this in more detail. We will also compare the present data with simultaneously obtained even more highly resolved images in the Ca~H line (Henriques, in prep.), with emphasis on exploring properties of bright penumbral grains. We also intend to investigate properties of penumbral dark cores \citep{2002Natur.420..151S}, which are obvious mostly in the innermost penumbra (radial zone 1), and thus were {\gs to a large extent} excluded in the present analysis. {\gs We do not exclude the possibility, however, that a small fraction of pixels within our mask defining the interior penumbra actually corresponds to dark cores, and thus are included in our azimuthal fits.}
%including th difficult! Proxy for vertical flows: the continuum image!

\begin{acknowledgements}       

  Mats L\"ofdahl, Dan Kiselman and Jorge S{\'a}nchez Almeida are thanked for valuable comments on the manuscript. The Swedish 1-m Solar Telescope is operated on the island of La Palma by the Institute for Solar Physics of the Royal Swedish Academy of Sciences in the Spanish Observatorio del Roque de los Muchachos of the Instituto de Astrof\'isica de Canarias.

\end{acknowledgements}

%\bibliography{bib-strings_aa,%
%  svst,%
%  scharmer,%
%}

\end{document}